\documentclass[journal]{IEEEtran}
\usepackage[sort]{cite}
\usepackage{tikz}
\usepackage[cmex10]{amsmath}
\usepackage{float}
\usepackage{framed}
\usepackage{graphics} 			
\usepackage{epsfig} 				
\usepackage{amssymb}  			
\usepackage{url}

\usepackage[caption=false,font=footnotesize]{subfig}

\usepackage[ruled]{algorithm}
\usepackage{algorithmicx}
\usepackage{algpseudocode}

\providecommand{\e}[1]{\ensuremath{\times 10^{#1}}}

\makeatletter
\algnewcommand{\LineComment}[1]{\Statex \hskip\ALG@thistlm \parbox[t]{\dimexpr\linewidth-\ALG@thistlm}{\(\triangleright\)\textit{#1}\strut}}
\makeatother



\begin{document}
\title{Efficient estimation of probability of conflict between air traffic using Subset Simulation}


\author{
				Chinmaya Mishra, Simon Maskell, Siu-Kui Au and Jason F. Ralph\\
				\{c.mishra, smaskell, siukuiau, jfralph\}@liv.ac.uk\\
				Department of Electrical Engineering and Electronics, University of Liverpool, United Kingdom
				}


%
%

\markboth{}%
{Shell \MakeLowercase{\textit{et al.}}: Bare Demo of IEEEtran.cls for IEEE Journals}
%



\maketitle


\begin{abstract}
This paper presents an efficient method for estimating the probability of conflict between air traffic within a block of airspace. Autonomous Sense-and-Avoid is an essential safety feature to enable Unmanned Air Systems to operate alongside other (manned or unmanned) air traffic. The ability to estimate probability of conflict between traffic is an essential part of Sense-and-Avoid. Such probabilities are typically very low. Evaluating low probabilities using naive Direct Monte Carlo generates a significant computational load. This paper applies a technique called Subset Simulation. The small failure probabilities are computed as a product of larger conditional failure probabilities, reducing the computational load whilst improving the accuracy of the probability estimates. The reduction in the number of samples required can be one or more orders of magnitude. The utility of the approach is demonstrated by modeling a series of conflicting and potentially conflicting scenarios based on the standard Rules of the Air.
\end{abstract}

\begin{IEEEkeywords}
Probability of conflict, air traffic, Subset Simulation, Direct Monte Carlo, Metropolis Hastings, Sense-and-Avoid
\end{IEEEkeywords}

%

%
%
%
%

\section{Introduction}
\IEEEPARstart{F}{uture} autonomous operations of Unmanned Air Systems (UAS) within densely populated airspace require an automated Sense-and-Avoid (SAA) system~\cite{angelov2012sense}. A key element within the Sense-and-Avoid (SAA) topic is Conflict Detection and Resolution (CD\&R)~\cite{angelov2012sense}. A conflict occurs when the separation between any aircraft or obstacle reduces below a minimum distance. Such a situation could~$-$ in the worst case~$-$ generate a collision between air vehicles but even in the absence of an actual collision it will violate the mandated Rules of the Air, and may give rise to an air ‘incident’. Such incidents must be reported as soon as possible to the local Air Traffic Service Unit (ATSU)~\cite{ICAOannex13}. 

Initial work on CD\&R can be found in robotics where the collision avoidance problem has been treated as a path planning task~\cite{moravec1980obstacle} and an early approach to the collision avoidance problem involved using artificial potential fields~\cite{khatib1986real}. Such methods are suitable for scenarios where movement of the vehicles may be relatively slow, restricted in space or in scope. However, over the following decades the increased use of UAS has created demand for autonomous CD\&R solutions which are suitable for the more dynamic aerospace environment. A large number of CD\&R methods have been proposed during this period and comprehensive surveys have been conducted by Kuchar and Yang~\cite{kuchar2000review}, Krozel et al.~\cite{krozel1997conflict}, Warren~\cite{warren1997medium} and Zeghal~\cite{zeghal1998review}. Kuchar and Yang have proposed a taxonomy of methods useful in identifying gaps and directing future efforts within the SAA community~\cite{kuchar2000review}. More recently, Albaker and Rahim have presented an up to date survey of CD\&R methods for UAS~\cite{albaker2009survey}. The work presented in this paper can be categorized as a Conflict Detection method that assumes non-cooperative sensor technology.

The CD\&R methods are broadly categorized as cooperative and non-cooperative. Cooperative methods assume that traffic shares relevant information via radio, data link or by contacting ground based ATSU. These methods are dependent on cooperative equipment such as Transponders and/or Automatic Dependent Surveillance-Broadcast (ADS-B) that are carried on-board the aircraft. This equipment declares the current state of the aircraft to nearby traffic. If the potential for a conflict is identified the situation will be resolved by coordinating maneuvers between the traffic, often via two-way radio communications. The maneuvers are dictated by following a set of customary rules that determine the ‘right-of-way’ for each aircraft. These are based on existing Visual Flight Rules (VFR) within the civil aviation domain~\cite{ICAOannex2}. In VFR, it is the flight crew's responsibility to maintain safe separation with traffic. In the absence of visual information (due to limited visibility caused by bad weather), the flight crew must rely on external information. In such situations, Instrument Flight Rules (IFR) are used with the ATSU monitoring traffic separation using Radar and then directing the flight crew so as to maintain safe separation. Alternatively, on larger aircraft, a Traffic Alert Collision Avoidance System (TCAS)~\cite{kuchar2007traffic} can be used. The TCAS system provides Resolution Advisories (RA) to flight crews of conflicting traffic in the form of maneuvers to be followed to resolve the conflict. In each case, a potential conflict is resolved in accordance with the rules given by the local aviation authority for the airspace within which the aircraft are operating; such as the Federal Aviation Authority (FAA) in the US~\cite{authority2015FAA} or the Civil Aviation Authority (CAA) in the UK~\cite{authority2015cap}. The rules stated by most aviation authorities are based on the rules outlined by the International Civil Aviation Organization (ICAO)~\cite{ICAOannex1to18}. When a conflict type is identified the appropriate resolution maneuver is executed. For example, when aircraft are approaching each other head-on the rules will say that both aircraft maneuver to their right. All traffic involved with the conflict must cooperate for a successful resolution~\cite{mishra2013doing}. Each of these methods assumes that all aircraft involved in the potential conflict are sharing information and behaving in accordance with the accepted Rules of the Air. 

In contrast, non-cooperative methods assume that no information related to the current state or future intent of traffic has been shared (i.e. there is no flight plan exchange or radio/data link). This is a far more challenging problem since information related to traffic state and intentions must be measured or inferred from the behavior of non-cooperative aircraft. Normally, this will be due to the lack of appropriate technology on-board the aircraft: for example, a lightweight commercial of-the-shelf (COTS) UAS, obtained by the general public and used for recreational purposes. Problems occur when these aircraft are operated within non-segregated airspace. This type of airspace contains aircraft (manned or unmanned) that adhere to the Rules of the Air and expect traffic to do so as well. The lack of cooperative technology on-board a lightweight UAS prevents awareness of traffic and increases the risk of a midair collision. This problem needs to be addressed due to the increased number of near miss incidents involving such UAS operating within non-segregated airspace~\cite{WashpostAug2015NearMiss}. The problem of the lack of information is addressed by using on-board sensors. Information related to state of traffic is obtained from observations using sensors such as Radar, Lidar and/or cameras. For example, Mcfadyen et al. have considered using visual predictive control with a spherical camera model to create a collision avoidance controller~\cite{mcfadyen2013aircraft}. Recently, Huh et al. have proposed a vision based Sense-and-Avoid framework that utilizes a camera to detect and avoid approaching airborne intruders~\cite{HUH2010}. A collision avoidance system that uses a combination of Radar and electro-optical sensors have been prototyped and tested by Accardo et al~\cite{Accardo2013}. Measurement data obtained from sensors are inherently noisy. This gives rise to uncertainties in the observed state and predicted motion of the non-cooperative aircraft. In an environment where future trajectories are uncertain, the likelihood of a conflict is an essential metric. Obtaining an accurate estimate for the Probability of Conflict ($P_{c}$), given the sensor data, is a key parameter required to resolve traffic conflicts. This paper provides a method to calculate the $P_{c}$ metric that is more efficient than the standard approach of using Direct Monte Carlo (DMC) methods.

Probabilistic methods for conflict resolution requiring the calculation of metrics like the Probability of Conflict ($P_{c}$) have been discussed in~\cite{kuchar2000review}. Nordlund and Gustafsson~\cite{Nordlund2011} noted the huge number of simulations required to get sufficient reliability for small risks and suggested an approach that reduced the three dimensional problem to a one dimensional integral along piecewise straight paths~\cite{nordlund2008probabilistic,Lindsten200965}.
More recently, Jilkov et al. have extended a method developed by Blom and Bakker~\cite{blom2002conflict} and estimated $P_{c}$ using multiple models for aircraft trajectory prediction~\cite{jilkov2014improved}. Many probabilistic methods involve the use of Monte Carlo methods where uncertainties exist and Monte Carlo methods can be found in existing CD\&R methods~\cite{jilkov2014improved,chryssanthacopoulos2011accounting,wolf2011aircraft,belkhouche2013modeling,blom2006free,Watkins2003Stochastic,prandini2000probabilistic,Krozel1997Decision}. Unfortunately, for scenarios where the expected $P_{c}$ is low, a Monte Carlo method will require a very large number of simulations to estimate $P_{c}$ with any accuracy. To reduce the computational cost associated with Monte Carlo methods, Prandini et al. have estimated the risk of conflict using the Interacting Particle System (IPS) method~\cite{prandini2011air}. This method fixes a set of initial conditions of the aircraft and alters reducing subsets of the propagated trajectories to satisfy the intermediate thresholds; this assumes that the predicted trajectories are non-deterministic with the probability of conflict being associated with outliers in the propagation, not outliers in the initial conditions. If, however, the trajectory is deterministic (or near-deterministic), then IPS is unable to provide improved computational efficiency relative to direct (Monte Carlo) sampling. This paper proposes the use of the Subset Simulation method~\cite{au2001estimation} to avoid this problem and allows the initial conditions to be adjusted as the subsets are navigated. Subset Simulation approaches the problem of reducing the computational load associated with calculating low probabilities by focusing the simulation towards the rare regions of interest within the probability distribution function (pdf). The regions of interest correspond to the events which may lead to conflict between traffic. 

Originally, Au and Beck proposed Subset Simulation as a method for computing small failure probabilities as a result of (larger) conditional failure probabilities~\cite{au2001estimation}. The method was proposed in Civil Engineering to compute probabilities of structural failure and identify associated failure scenarios~\cite{au2003subset}. The focus of their work was on understanding the risk to structures posed by seismic activity. This paper modifies the methods developed by Au and Beck~\cite{IVAN} and demonstrates that they can significantly reduce the computational load required to estimate the value of $P_{c}$ for air traffic within a block of airspace by reducing the number of samples required. The proposed method is applied to a set of conflicting and potentially conflicting test scenarios based on the Rules of the Air specified by aviation authorities. Since these scenarios are standard engagements considered by aviation authorities, they could also be used as a benchmark for comparison against future methods. The $P_{c}$ during some scenarios is low; despite this, it is essential to provide an approximation this metric due to the catastrophic nature of a collision.

The paper is structured as follows: sections~\ref{sec:DMC} and~\ref{sec:mh} describe the Direct Monte Carlo (DMC) and Metropolis Hastings (MH) methods respectively. The Subset Simulation theory is based on a combination of DMC and MH methods. Section~\ref{sec:SS} describes Subset Simulation. Section~\ref{sec:SS_app} then describes the application of Subset Simulation to the estimation of $P_{c}$ between air traffic in non-cooperative scenarios. Section~\ref{sec:results} presents simulation results of estimating $P_{c}$ between air traffic for conflicting and potentially conflicting non-cooperative scenarios. Section~\ref{sec:acc_eff} analyzes the efficiency and accuracy of estimating the $P_{c}$ using Subset Simulation and Direct Monte Carlo. Finally, section~\ref{sec:conclusion} concludes the paper.

\section{Direct Monte Carlo}
\label{sec:DMC}

The Direct Monte Carlo (DMC) method is a sampling method that can be used to characterize a distribution of interest. The objective of this section is to estimate the probability of a type of event to occur. Therefore the DMC method is used as a `statistical averaging' tool, where the probability of failure $P_{F}$ is estimated as the ratio of failure responses to the total number of trials~\cite{IVAN}.

A set of $N$ independent identically distributed (i.i.d) inputs $\{X_{n}: n = 1,...,N\}$ are drawn from the proposal distribution $q(X|\mu,\sigma^{2})$ of the input parameter space. The proposal distribution can be any known distribution that can be sampled from. We choose a Normal distribution that is centered at the mean $\mu$ and has a variance of $\sigma^{2}$. A set of system responses are observed $\{Y_{n} = h(X_{n}): n = 1,...,N\}$, where $h(...)$ is the system process. The occurrence of a failure event $F$ is indicated when a scalar quantity $b_{F}$ (threshold) is exceeded. The number of samples that exceed the threshold is $Y_{F}$. Therefore the probability of failure is estimated as $P_{F} = P(Y \geq b_{F}) = \frac{Y_{F}}{N}$. Such an approach is suitable for large probabilities (such as $P > 0.1$) where a small number of samples can be used to estimate the probability. However for small probabilities (such as the tail region of the pdf, where $P \leq 10^{-3}$) a significantly large number of samples must be drawn to accurately estimate the probability. This is illustrated by the following example.

\subsection{Estimating probability of drawing samples from region $F$}
\begin{algorithm}[!t]
\caption{Determine distance between samples X and C}
\label{alg:h}
\begin{algorithmic}[1]
	\Function{h}{$X$,$C$}
        \State $V = X - C$
				\State $R = \sqrt{V_{x}^{2} + V_{y}^{2}}$
				
	\State \textbf{return $R$}	
	\EndFunction
\end{algorithmic}
\end{algorithm}

\begin{algorithm}[!t]
\caption{Direct Monte Carlo}
\label{alg:dmc}
\begin{algorithmic}[1]
	\Function{DMC}{$N$, $C$, $r_{c}$}
	
	\State $D = 0$
	
		\For{\texttt{$n = 1:N$}}
				\State $x \sim{~} \mathcal{N}(0,1)$
        \State $y \sim{~} \mathcal{N}(0,1)$
        \State $X_{n} = [x,y]^{T}$
				
				\State $R_{n} = $ \Call{H}{$X_{n}$, $C$}
				\If {$R_{n} \leq r_{c}$}
					\State $D = D + 1$
				\EndIf
      \EndFor
			
			\State $P_{F} = \frac{D}{N}$
	\State \textbf{return $P_{F}$}	
	\EndFunction
\end{algorithmic}
\end{algorithm}

\begin{figure}[!t]
	\centering
	\subfloat[Direct Monte Carlo with 100 samples]{%
	\includegraphics[width=\columnwidth]{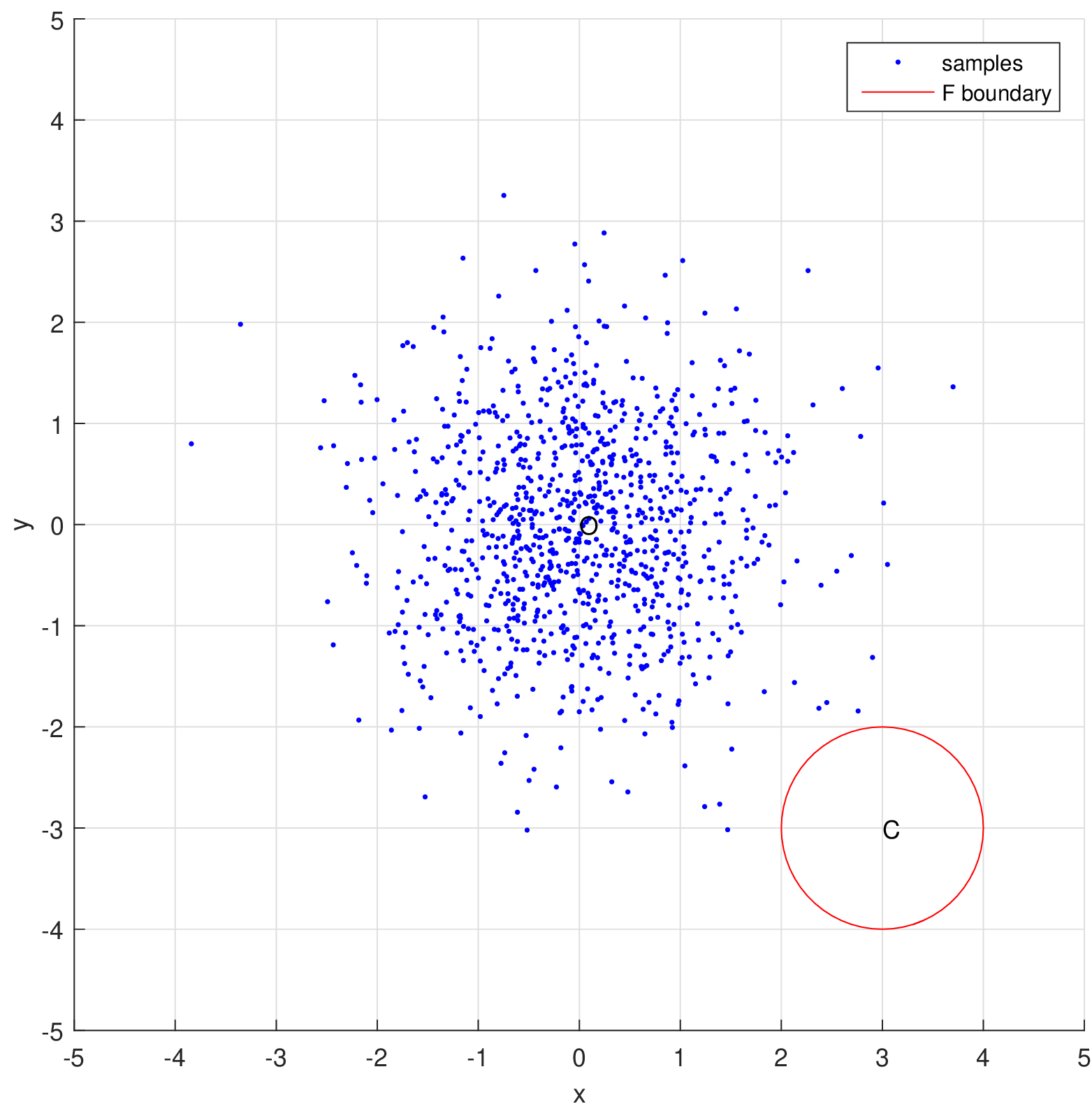}
	\label{fig:dmc_100}}
	\\
	\subfloat[Direct Monte Carlo with $10^{5}$ samples]{%
	\includegraphics[width=\columnwidth]{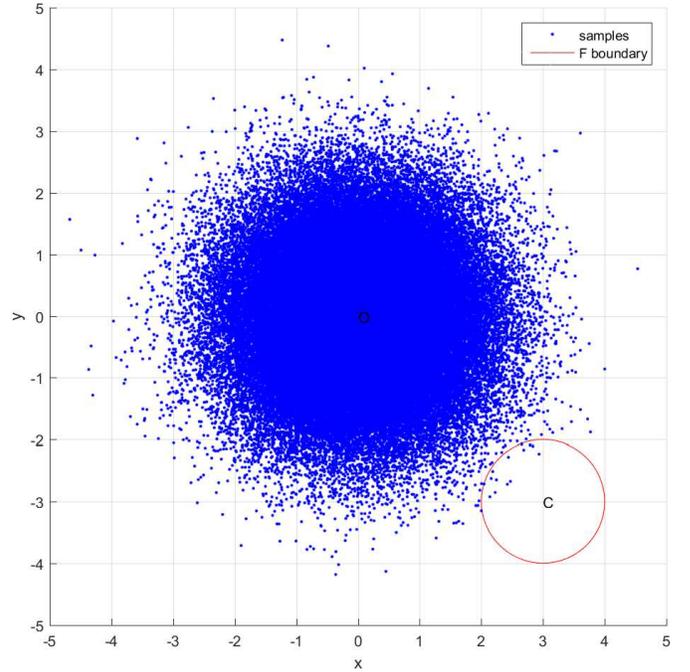}
	\label{fig:dmc_10e4}}
	\caption{The probability of drawing samples from the region $F$ is estimated using Direct Monte Carlo. Fig.~\ref{fig:dmc_100} estimates the $P_{F} = 0$ with 100 samples. Fig.~\ref{fig:dmc_10e4} estimates the $P_{F} = 1.5\e{-4}$ with $10^{5}$ samples.}
	\label{fig:DMC_100_10e4}
\end{figure}

Fig.~\ref{fig:DMC_100_10e4} shows a $10 \times 10$ square centered at $O = [0,0]^{T}$. The region $F$ is a circle with radius $r_{c} = 1$, centered at $C = [3,-3]^{T}$ within this square. The objective is to estimate the probability of drawing samples from this region. The probability distribution of the overall area is represented by a Gaussian distribution centered at $O = [0,0]^{T}$. A set of $N$ samples $\{X_{n}: n = 1,...,N\}$ are drawn where each sample is a vector; $X_{n} = [x_{n},y_{n}]^{T}$. The $x$ and $y$ values of each sample are the x-coordinate and y-coordinates of the position respectively. To clarify, $X_{1} = [x_{1},x_{2}]^{T}$ where $x_{1} \sim{~} \mathcal{N}(0,1)$ and $y_{1} \sim{~} \mathcal{N}(0,1)$. The distance between the position of each sample and center of circle $C$ is $\{R_{n} = H(X_{n},C): n = 1,...N\}$ as defined by Algorithm~\ref{alg:h}. To clarify, the distance between sample $X_{1}$ and $C$ is $R_{1} = H(X_{1},C)$. Algorithm~\ref{alg:dmc} is used to estimate the probability of drawing samples from the region $F$.

Fig.~\ref{fig:dmc_100} shows $100$ samples drawn from the distribution. Note no samples are drawn from the area $F$. The probability is estimated $P_{F} = 0$. The number of samples are increased to $N = 10^{5}$. Fig.~\ref{fig:dmc_10e4} shows some samples are drawn from the region $F$ and the probability is estimated $P_{F} = 1.5\e{-4}$ This illustrates that Direct Monte Carlo requires a significantly large number of samples to estimate the probability of drawing samples from the region $F$.

This method estimates $P_{F}$ by attempting to realize the entire pdf centered at $O$ that includes the area F. As the area $F$ reduces the number of samples required to estimate $P_{F}$ increases making such an approach computationally demanding. A different algorithm is needed. 

\section{Metropolis Hastings}
\label{sec:mh}

Metropolis-Hastings (MH) is a Markov Chain Monte Carlo (MCMC) method used to characterize a distribution of interest by sampling from a known distribution. We refer to this distribution of interest as the target distribution. The MH algorithm originates from the Metropolis algorithm first used in statistical Physics by Metropolis and co-workers (Metropolis et al, 1953)\cite{metropolis1953equation}. Hastings proposed a generalized form of this algorithm leading to the Metropolis Hastings (MH) algorithm~\cite{hastings1970monte}.

The MH method generates samples from the proposal distribution $q(X|x_{0},\sigma^{2})$ by starting from a seed value $x_{0}$. A chain of $n$ samples is then generated, starting with $x_{0}$. The sample $x_{k+1}$ is generated from the current sample $x_{k}$ using the following steps~\cite{IVAN}:

\begin{enumerate}
  \item Generate a candidate sample $x^{*} \sim{~} q(x^{*}|x_{k}, \sigma^{2})$.
  \item Calculate an acceptance ratio: $\alpha = \frac{q(x_{k}|x^{*},\sigma^{2})f(x^{*})}{q(x^{*}|x_{k},\sigma^{2})f(x_{k})}$ 
	\item Draw a sample $e$ from a uniform distribution [0,1]	
  \item Set $ x_{k+1} = 
												\left \{ \begin{array}{l l}
																		x^{*} & \text{if} \ e < \alpha \\
																		x_{k} & \text{otherwise} 
																	\end{array} \right .$
						
	\item Repeat steps 1 to 4 until $n$ samples have been generated.
\end{enumerate}

\noindent The function $f(...)$ defines the target density for the input sample. While, $n \rightarrow \infty$, this process is guaranteed to accept samples from $q$ that leads to the realization of the target distribution~\cite{robert2004metropolis}. To help ensure that all regions of the target density are explored, multiple seeds can be used to generate multiple chains of samples in parallel~\cite{IVAN}.

\subsection{Drawing samples from the region $F$}

\begin{figure}[h]
	\centering
	\includegraphics[width=\columnwidth]{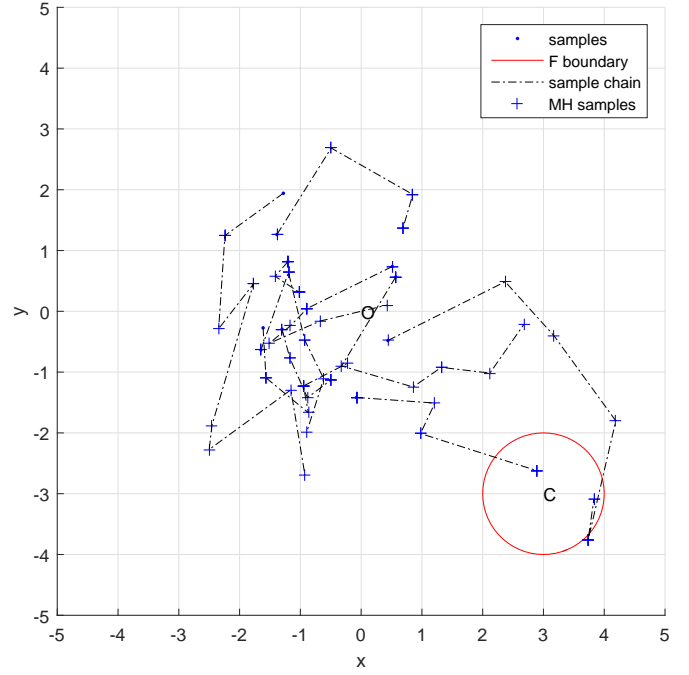}
	\caption{Drawing samples from the region F using Metropolis Hastings algorithm to generate chains of conditional samples. The initial samples used as seeds are drawn using Direct Monte Carlo.}
	\label{fig:mh}
\end{figure}

The Metropolis Hastings method is defined in algorithm~\ref{alg:mh} and it is applied to the example of estimating the probability of drawing samples from region $F$ as shown in the previous section. The covariance of the proposal $\sigma^{2}$ is a $2 \times 2$ identity matrix $I_{2 \times 2}$ and the covariance of the distribution of interest $\sigma_{r_{c}}^{2} = r_{c}^{2} \times I_{2 \times 2}$ where $r_{c}$ is the radius of the region $F$. For this example $r_{c} = 1$, therefore $\sigma_{r_{c}}^{2} = I_{2 \times 2}$.

Fig.~\ref{fig:mh} illustrates the chains of samples generated by the Metropolis Hastings algorithm. This figure shows 10 samples drawn from the proposal distribution using the DMC method. These samples are seeds $s =\{X_{1},...,X_{10}\}$. The MH algorithm is applied using the seeds $s$. Each seed generates a chain of 10 samples. Note that many sample chains do not reach the region $F$. It is clear that it might be more efficient to generate more samples for chains with seeds that are closer to the region $F$ since they have higher likelihood of generating samples that are within the region $F$ or closer to the region $F$. Subset Simulation achieves this and is described in the next section.

\begin{algorithm}
\caption{Generate conditional chains of samples using Metropolis Hastings algorithm}
\label{alg:mh}
\begin{algorithmic}[1]
\Function{MH}{$s$, $n$, $C$, $r_{c}$}
	\State $\sigma_{r_{c}}^{2} = r_{c}^{2} I_{2 \times 2}$
	\For{\texttt{$j = 1:|s|$}} \textit{$\triangleright$ For each seed}	
		\State $X_{0} = s_{j}$	\textit{$\triangleright$Select seed sample}
		\For{\texttt{$k = 0:n-1$}}
			
			\LineComment{Generate Candidate sample $X^{*}$}
			\State $g \sim{~} \mathcal{N}(0,1)$
			\State $X^{*} = X_{k} + g$
			
			\LineComment{Calculate acceptance ratio}
			
			\State $\beta = \frac{q(X^{*}|X_{k}, \sigma^{2})}{q(X_{k}|X^{*},\sigma^{2})}\frac{p(X^{*}|C, \sigma_{r_{c}}^{2})}{p(X_{k}|C, \sigma_{r_{c}}^{2})}$
			\State $\alpha = min \left \{ 1, \beta\} \right. $
			\State $e \sim{~} [0,1]$
			
			\State $X_{k+1}^{(j)} = \left \{ \begin{array}{l l}
												X^{*} & \text{if} \ e < \alpha \\
												X_{k} & \text{if} \ e \geq \alpha \\
										  \end{array} \right .$
		\EndFor	
	\EndFor

\State \textbf{return $X^{(j)}$}

\EndFunction
\end{algorithmic}
\end{algorithm}

\section{Subset Simulation}
\label{sec:SS}
Subset Simulation (SS) is based on a combination of Direct Monte Carlo (DMC) and Metropolis Hastings (MH) methods as described in sections~\ref{sec:DMC} and~\ref{sec:mh} respectively. It calculates the probability of rare events occurring as the product of the probabilities of less-rare events. Such an approach is less computationally expensive than either DMC or MH alone. A general outline of the SS method is presented in this paper and the interested reader is referred to~\cite{IVAN} for more details.

Subset Simulation generates a Complimentary Cumulative Distribution Function (CCDF) of the response quantity of interest $Y$. The probability of failure $P_{F}$ can be directly estimated from the CCDF. This CCDF is constructed by generating samples that satisfy a series of intermediate thresholds $b_{1} > b_{2} > b_{3} > ... > b_{m-1}$ that divide the space into $m$ nested regions. These thresholds are adaptively defined as the simulation progresses. This is described later on in this section. The threshold $b_{m-1}$ is the required failure threshold $b_{F}$ ($b_{m-1} = b_{F}$). The intermediate thresholds allow the probability of failure to be estimated using a classical conditional structure given by

\begin{equation}	 
	P_{F} = P(Y < b_{m-1} | Y < b_{m-2})P(Y < b_{m-2})
	\label{eq:cond_struct}
\end{equation}

Samples are generated to satisfy the threshold for each level. The total number of levels $m$ is dependent on the magnitude of the target probability $P_{F}$. Subset Simulation uses `level probability' $p_{0} \in (0,1)$ to control how quickly the simulation reaches the target event of interest~\cite{IVAN}. The target probability is used to approximate the number of levels $m$ required by evaluating $P_{F} = (p_{0})^{m}$. To clarify, if the target probability is $P_{F} = 10^{-5}$ and $p_{0} = 0.1$ then the total number of levels required will be $m = 5$.

\subsection{Level 0}

Subset Simulation begins at level $i = 0$ with Direct Monte Carlo (DMC) sampling from the entire region of interest. A set of $N$ samples $\{X_{n}^{(0)}: n = 1,...,N\}$ are drawn from a proposal distribution $q(X_{n}^{(0)}|\mu,\sigma^{2})$ (as described in section~\ref{sec:DMC}). The set of output responses $Y_{n}^{(0)}$ are evaluated $\{Y_{n}^{(0)} = h(X_{n}^{(0)}): n = 1,...,N\}$. The function $h(...)$ defines the system response to the input sample. In the context of SS, the responses $Y_{n}^{(0)}$ are also known as the quantity of interest. The set $Y_{n}^{(0)}$ is sorted in descending order to create the set $\{B_{n}^{(0)}: n = 1,...,N\}$. The input samples $X_{n}^{(0)}$ are reordered $\tilde{X}_{n}^{(0)}$ and correspond to the sorted quantity of interest $B_{n}^{(0)}$. To clarify, $\tilde{X}_{1}^{(0)}$ is the input sample that generates the largest output $B_{1}^{(0)}$. A CCDF is generated by plotting $B_{n}^{(0)}$ against the probability intervals $P_{n}^{(0)}$. The probability intervals $P_{n}^{(0)}$ are generated using the following equation:

\begin{equation}	
	P_{n}^{(i)} = p_{0}^i\frac{N - n}{N}	\quad n = 1......N
	\label{eq:prob_intervals}
\end{equation}

\begin{table*}[!t] 
\centering
\resizebox{2\columnwidth}{!}{%
\begin{tabular}{cccccccccc}
\multicolumn{1}{c}{} & \multicolumn{1}{c}{Level 0} & \multicolumn{1}{c}{} & & Level $i$ & & & & Level $m - 1$ & \\ \cline{1-10}
\multicolumn{1}{|c|}{$\textbf{P}_{n}^{(0)}$} & \multicolumn{1}{c|}{$\textbf{B}_{n}^{(0)}$} & \multicolumn{1}{c|}{$\tilde{\textbf{X}}_{n}^{(0)}$} & \multicolumn{1}{|c|}{$\textbf{P}_{n}^{(i)}$}& \multicolumn{1}{c|}{$\textbf{B}_{n}^{(i)}$}& \multicolumn{1}{|c|}{$\tilde{\textbf{X}}_{n}^{(i)}$} & \multicolumn{1}{c|}{.....} & \multicolumn{1}{|c|}{$\textbf{P}_{n}^{(m-1)}$}& \multicolumn{1}{c|}{$\textbf{B}_{n}^{(m-1)}$} & \multicolumn{1}{c|}{$\tilde{\textbf{X}}_{n}^{(m-1)}$}\\ \cline{1-10}
\multicolumn{1}{|c|}{$P_{1}^{(0)}$} & \multicolumn{1}{c|}{$B_{1}^{(0)}$} & \multicolumn{1}{c|}{$\tilde{X}_{1}^{(0)}$} & & & & & & & \\ \cline{1-3}
\multicolumn{1}{|c|}{\vdots} & \multicolumn{1}{c|}{\vdots} & \multicolumn{1}{c|}{\vdots} & & & & & &                       &                       \\ \cline{1-3}
\multicolumn{1}{|c|}{$P_{N-N_{c}}^{(0)}$} & \multicolumn{1}{c|}{$B_{N-N_{c}}^{(0)}$} & \multicolumn{1}{c|}{$\tilde{X}_{N-N_{c}}^{(0)}$}   	 & & & & & & & \\ \cline{1-6}
\multicolumn{1}{|c|}{$P_{N-N_{c}+1}^{(0)}$} & \multicolumn{1}{c|}{$B_{N-N_{c}+1}^{(0)}$}    & \multicolumn{1}{c|}{$\tilde{X}_{N-N_{c}+1}^{(0)}$}   & \multicolumn{1}{c|}{$P_{1}^{(i)}$} & \multicolumn{1}{c|}{$B_{1}^{(i)}$} & \multicolumn{1}{c|}{$\tilde{X}_{1}^{(i)}$} &  & & & \\ \cline{1-6}
\multicolumn{1}{|c|}{\vdots}    & \multicolumn{1}{c|}{\vdots}    & \multicolumn{1}{c|}{\vdots}   & \multicolumn{1}{c|}{\vdots} & \multicolumn{1}{c|}{\vdots} & \multicolumn{1}{c|}{\vdots} &  &  & &                       \\ \cline{1-6}
\multicolumn{1}{|c|}{$P_{N}^{(0)}$}    & \multicolumn{1}{c|}{$B_{N}^{(0)}$}    & \multicolumn{1}{c|}{$\tilde{X}_{N}^{(0)}$}   & \multicolumn{1}{c|}{$P_{N-N_{c}}^{(i)}$} & \multicolumn{1}{l|}{$B_{N-N_{c}}^{(i)}$} & \multicolumn{1}{l|}{$\tilde{X}_{N-N_{c}}^{(i)}$} & & & & \\ \cline{1-10}
                             &                          & \multicolumn{1}{c|}{}   & \multicolumn{1}{c|}{$P_{N-N_{c}+1}^{(i)}$} & \multicolumn{1}{c|}{$B_{N-N_{c}+1}^{(i)}$} & \multicolumn{1}{c|}{$\tilde{X}_{N-N_{c}+1}^{(i)}$} &  \multicolumn{1}{c|}{.....}&  \multicolumn{1}{c|}{$P_{1}^{(m-1)}$} & \multicolumn{1}{c|}{$B_{1}^{(m-1)}$} & \multicolumn{1}{c|}{$\tilde{X}_{1}^{(m-1)}$}\\ \cline{4-10}
                             &                          & \multicolumn{1}{c|}{}   & \multicolumn{1}{c|}{\vdots} & \multicolumn{1}{c|}{\vdots} & \multicolumn{1}{c|}{\vdots} & \multicolumn{1}{c|}{.....} & \multicolumn{1}{c|}{\vdots} & \multicolumn{1}{c|}{\vdots} &  \multicolumn{1}{c|}{\vdots}      \\ \cline{4-10}
                             &                          & \multicolumn{1}{c|}{} & \multicolumn{1}{c|}{$P_{N}^{(i)}$} & \multicolumn{1}{c|}{$B_{N}^{(i)}$} & \multicolumn{1}{c|}{$\tilde{X}_{N}^{(i)}$} &   \multicolumn{1}{c|}{.....} & \multicolumn{1}{c|}{$P_{N-N_{c}}^{(m-1)}$} & \multicolumn{1}{c|}{$B_{N-N_{c}}^{(m-1)}$} & \multicolumn{1}{c|}{$\tilde{X}_{N-N_{c}}^{(m-1)}$}\\ \cline{4-10}
                             &                          &                         &                       &                       & \multicolumn{1}{c|}{} & \multicolumn{1}{c|}{.....} & \multicolumn{1}{c|}{$P_{N-N_{c}+1}^{(m-1)}$} & \multicolumn{1}{c|}{$B_{N-N_{c}+1}^{(m-1)}$} & \multicolumn{1}{c|}{$\tilde{X}_{N-N_{c}+1}^{(m-1)}$} \\ \cline{7-10} 
                             &                          &                         &                       &                       &                       & \multicolumn{1}{c|}{} & \multicolumn{1}{c|}{\vdots} & \multicolumn{1}{c|}{\vdots} & \multicolumn{1}{c|}{\vdots} \\ \cline{8-10}
                             &                          &                         &                       &                       &                       & \multicolumn{1}{c|}{} & \multicolumn{1}{c|}{$P_{N}^{(m-1)}$} & \multicolumn{1}{c|}{$B_{N}^{(m-1)}$} & \multicolumn{1}{c|}{$\tilde{X}_{N}^{(m-1)}$} \\ \cline{8-10} 
														
\end{tabular}
}
\caption{}
\label{table:general_subset_multilevel}
\end{table*}

\noindent The vector of probability intervals $P_{n}^{(0)}$ is concatenated with the sorted quantity of interest $B_{n}^{(0)}$ and their respective samples $\tilde{X}_{n}^{(0)}$ as illustrated in table~\ref{table:general_subset_multilevel} by the column titled `Level 0'. 

The set of probability intervals $P_{n}^{(0)}$ are plotted against $B_{n}^{(0)}$ to generate the CCDF. Level 0 makes it possible to accurately approximate CCDF values from $1-N^{-1}$ to $p_{0}$. Typically the region of interest within the pdf is outside this range (since SS is typically used to realize rare events). To explore probabilities below $p_{0}$, further levels of simulation must be conducted.

\subsection{Level $i > 0$}

The subsequent levels of SS where, $i > 0 $ explore the rarer regions of the probability distribution. This is achieved by generating multiple chains of conditional samples using the MH method as discussed in the previous section. The number of chains and number of samples per chain are $N_{c}$ and $N_s$ respectively. They are determined as

\begin{equation}	
	N_{c} = p_{0}N
\label{eq:n_c}
\end{equation}

\begin{equation}
	N_{s} = p_{0}^{-1}
\label{eq:n_s}
\end{equation}

\noindent Each level of subset simulation maintains $N$ samples ($N = N_{c} N_{s}$). The response values of conditional samples generated for the current level $i$ must not exceed the intermediate threshold $b_{i}$ for this level. This threshold is determined by

\begin{equation} 
	b_{i} = B_{N-N_c}^{(i-1)} \quad i \text{ is the current subset level}
	\label{eq:thresholds}
\end{equation}

\noindent The intermediate threshold for level $i = 1$ is $b_{1} = B_{N-N_{c}}^{(0)}$. To clarify the intermediate threshold is the $(N-N_{c})^{\text{th}}$ element of the sorted set of response values $B_{n}^{(0)}$. The set of seeds $s_{j}^{(i)}$ are used to generate samples for the current level $i$ are samples generated from the previous level ($i-1$) are defined by

\begin {equation}
	s_{j}^{(i)} = \tilde{X}_{n}^{(i-1)}
\label{eq:seeds}
\end{equation}

\noindent where $1 \leq j \leq N_{c}$, $(N-N_{c}+1) \leq n \leq N$ and $i > 0$. 

The set of seeds used to generate conditional samples for level $i = 1$ is $s^{(1)} = \{\tilde{X}_{N-N_{c}+1}^{(0)},...,\tilde{X}_{N}^{(0)}\}$. The $N$ conditional samples $X_{n}^{(1)}$ are generated using the MH method. The quantities of interest for $X_{n}^{(1)}$ are determined $\{Y_{n}^{(1)} = h(X_{n}^{(1)}): n = 1,...,N\}$ and are sorted in the same manner as the previous level $B_{n}^{(1)}$. The set $B_{n}^{(1)}$ and respective samples $\tilde{X}_{n}^{(1)}$ are concatenated with the probability intervals $P_{n}^{(1)}$ as illustrated in table~\ref{table:general_subset_multilevel} by the column titled `Level $i$'. Note the samples $\{\tilde{X}_{N-N_{c}+1}^{(0)},...,\tilde{X}_{N}^{(0)}\}$ shown in the column titled `Level 0' are used as seeds to generate the conditional samples $\{\tilde{X}_{1}^{(i)},...,\tilde{X}_{N}^{(i)}\}$ in column titled `Level $i$'.

This process is continued until the target level of probability $(p_{0})^{m}$ is reached at level $i = m-1$; as shown by the column titled `Level $m-1$'. The samples used as seeds to generate samples for the consecutive level are discarded and replaced with the generated samples. This is illustrated in table~\ref{table:subset_concat_table}. The column of probability intervals $\textbf{P}_{n}$ are plotted against the respective quantities of interest $\textbf{B}_{n}$ to generate a CCDF.

This method is continued until the target level of probability $P_{F} = (p_{0})^{m}$ is reached. By generating and evaluating conditional samples, the output samples tend towards the target distribution with significantly less trials than are needed when using the DMC method. The progressive nature of the algorithm can be demonstrated in the example problem of estimating the probability of drawing samples from the region $F$.

\begin{algorithm}
\caption{Generate conditional chains of samples of Subset Simulation using Metropolis Hastings algorithm}
\label{alg:mh_ss}
\begin{algorithmic}[1]
\Function{MH\_I}{$s$, $n$, $C$, $r_{c}$}
	\State $\sigma_{r_{c}}^{2} = r_{c}^{2} I_{2 \times 2}$
	\For{\texttt{$j = 1:|s|$}} \textit{$\triangleright$ For each seed}
		\State $X_{0} = s_{j}$	\textit{$\triangleright$Select seed sample}
		\For{\texttt{$k = 0:n-1$}}
			
			\LineComment{Generate Candidate sample $X^{*}$}
			\State $g \sim{~} \mathcal{N}(0,1)$
			\State $X^{*} = X_{k} + g$

			\LineComment{Determine distance between $X^{*}$ and C}
			\State $R^{*} = $ \Call{H}{$X^{*}$, $C$}
			
			\LineComment{Determine distance between $X_{k}$ and C}
			\State $R_{k} = $ \Call{H}{$X_{k}$, $C$}
			
			\LineComment{Indicator function for range}
			\State $d = \left \{ \begin{array}{l l}
												1 & \text{if} \ R^{*} \leq r_{c}\\
												0 & \text{if} \ R^{*} > r_{c} \\
										  \end{array} \right .$
														
			\LineComment{Calculate acceptance ratio}			
			\State $\beta = \frac{q(X^{*}|X_{k}, \sigma^{2})}{q(X_{k}|X^{*},\sigma^{2})}\frac{p(X^{*}|C, \sigma_{r_{c}}^{2})}{p(X_{k}|C, \sigma_{r_{c}}^{2})}$
			
			\State $\alpha = min \left \{ 1, \beta\} \right. $
			\State $e \sim{~} [0,1]$
			
			\State $X_{k+1}^{(j)} = \left \{ \begin{array}{l l}
												X^{*} & \text{if} \ e < \alpha \\
												X_{k} & \text{if} \ e \geq \alpha \\
										  \end{array} \right .$
											
			\State $R_{k+1}^{(j)} = \left \{ \begin{array}{l l}
												R^{*} & \text{if} \ e < \alpha \\
												R_{k} & \text{if} \ e \geq \alpha \\
										  \end{array} \right. $
		\EndFor	
	\EndFor

\State \textbf{return $X^{(j)}, R^{(j)}$}

\EndFunction
\end{algorithmic}
\end{algorithm}

\begin{algorithm}[!t]
\caption{Subset Simulation}
\label{alg:SS}
\begin{algorithmic}[1]
	\Function{SS}{$C$, $N$, $p_{0}$, $m$}

	\State $N_{c} = p_{0}N$
	\State $N_{s} = p_{0}^{-1}$
	
	\State $i = 0$ \textit{ Set current level}
	
	\LineComment{Direct Monte Carlo: Draw N samples and determine quantity of interest}
	\For{\texttt{$n = 1:N$}}
		\State $X_{n}^{(i)} \sim{~}\mathcal{N}(0,1)$
		
		\LineComment{Quantity of interest: Determine distance between samples $X_{n}^{(i)}$ and $C$}
		\State $R_{n}^{(i)} = $ \Call{H}{$X_{n}^{(i)}$, $C$}
  \EndFor
		
	\State $B_{n}^{(i)} \leftarrow R_{n}^{(i)}$ \textit{Sort distances in descending order}
	
	\State $\tilde{X}_{n}^{(i)} \leftarrow X_{n}^{(i)}$ \textit{Reorder the input samples to correspond to the sorted quantity of interest $B_{n}^{(i)}$}
	
	\LineComment{Generate probability intervals; equation~\ref{eq:prob_intervals}}
	\For{\texttt{$n = 1:N$}}
		\State $P_{n}^{(i)} = p_{0}^i\frac{N - n}{N}$
  \EndFor
	
	\LineComment{CCDF: Concatenate vectors $P_{n}^{(i)}$, $B_{n}^{(i)}$ and sample $\tilde{X}_{n}^{(i)}$}
	\State $E_{n} = [P_{n}^{(i)},B_{n}^{(i)},\tilde{X}_{n}^{(i)}]$
	
	\LineComment{Begin lower levels of subset simulation}		
	\For{\texttt{$i = 1:m-1$}}
		
		\LineComment{Set threshold}
		\State $b_{i} = B_{N-N_{c}}^{(i-1)}$
		
		\LineComment{Set seeds using equation~\ref{eq:seeds}}
		\For{\texttt{$j = 1:N_{c}$}}
			\State $n = N-N_{c}+j$
			\State $s_{j}^{(i)} = \tilde{X}_{n}^{(i-1)}$
		\EndFor
	
		\LineComment{Generate conditional samples using Metropolis Hastings algorithm}
		\State $[X_{n}^{(i)}, R_{n}^{(i)}] = $ \Call{MH\_I}{$s_{j}^{(i)}$, $N_{s}$, $C$, $b_{i}$}
		
	\State $B_{n}^{(i)} \leftarrow R_{n}^{(i)}$ \textit{Sort distances in descending order}
	
	\State $\tilde{X}_{n}^{(i)} \leftarrow X_{n}^{(i)}$ \textit{Reorder the input samples to correspond to the sorted quantity of interest $B_{n}^{(i)}$}	
	
	\LineComment{Generate probability intervals; equation~\ref{eq:prob_intervals}}
	\For{\texttt{$n = 1:N$}}
		\State $P_{n}^{(i)} = p_{0}^i\frac{N - n}{N}$
  \EndFor
	
	\LineComment{CCDF: Discard all rows after $E_{i(N-N_{c})}$}
	\LineComment{Concatenate $P_{n}^{(i)}$, $B_{n}^{(i)}$, $\tilde{X}_{n}^{(i)}$ and append to $E$}
		\For{\texttt{$n = 1:N$}}
			\State $E_{i(N-N_{c}+n)} = [P_{n}^{(i)}, B_{n}^{(i)}, \tilde{X}_{n}^{(i)}]$
		\EndFor
	\EndFor
	
	\State \textbf{return $E$}	
	\EndFunction
\end{algorithmic}
\end{algorithm}

\begin{table}
\centering
	\begin{tabular}{|c|c|c|c}
	\cline{1-3}
	$\textbf{P}_{n}$ & $\textbf{B}_{n}$ & $\tilde{\textbf{X}}_{n}$ & \\ \cline{1-3} \hline
	$P_{1}^{(0)}$ & $B_{1}^{(0)}$ & $\tilde{X}_{1}^{(0)}$ & \\ \cline{1-3}
	\vdots & \vdots & \vdots & \textit{Level 0} \\ \cline{1-3}
	$P_{N-N_{c}}^{(0)}$ & $B_{N-N_{c}}^{(0)}$ & $\tilde{X}_{N-N_{c}}^{(0)}$ & \textit{samples retained} \\ \cline{1-3} \hline
	$P_{1}^{(i)}$ & $B_{1}^{(i)}$ & $\tilde{X}_{1}^{(i)}$ &  \\ \cline{1-3}
	\vdots & \vdots & \vdots & \textit{Level $i$} \\ \cline{1-3}
	$P_{N-N_{c}}^{(i)}$ & $B_{N-N_{c}}^{(i)}$ & $\tilde{X}_{N-N_{c}}^{(i)}$ & \textit{samples retained} \\ \cline{1-3} \hline
	\vdots & \vdots & \vdots &  \\ \cline{1-3} \hline
	$P_{1}^{(m-1)}$ & $B_{1}^{(m-1)}$ & $\tilde{X}_{1}^{(m-1)}$ & \\ \cline{1-3}
	\vdots & \vdots & \vdots & \\ \cline{1-3}
	$P_{N-N_{c}}^{(m-1)}$ & $B_{N-N_{c}}^{(m-1)}$ & $\tilde{X}_{N-N_{c}}^{(m-1)}$ \\ \cline{1-3}
	$P_{N-N_{c}+1}^{(m-1)}$ & $B_{N-N_{c}+1}^{(m-1)}$ & $\tilde{X}_{N-N_{c}+1}^{(m-1)}$ \\ \cline{1-3}
	\vdots & \vdots & \vdots & \textit{Level $m - 1$} \\ \cline{1-3}
	$P_{N}^{(m-1)}$ & $B_{N}^{(m-1)}$ & $\tilde{X}_{N}^{(m-1)}$ & \textit{samples retained} \\ \cline{1-3} \hline
	\end{tabular}
	\vspace{2mm}
	\caption{}
\label{table:subset_concat_table}
\end{table}

\subsection{Estimating Probability of drawing samples from region F}

\begin{figure*}[!t]
	\centering
	\subfloat[Subset Simulation Level 0]{%
	\includegraphics[width=\columnwidth]{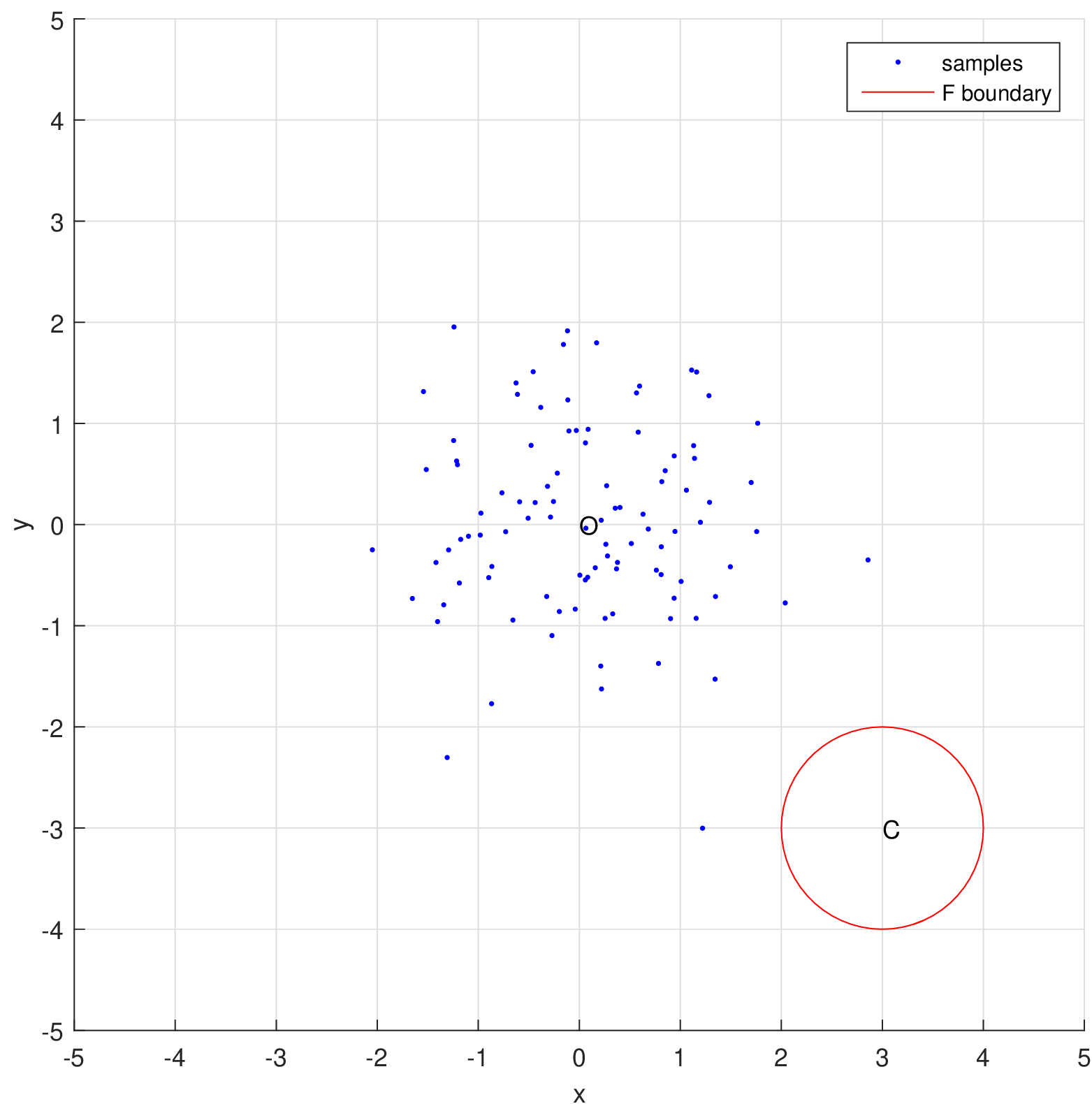}
	\label{fig:SS_0}}
  \hfill
	\subfloat[Subset Simulation Level 0 CCDF]{%
	\includegraphics[width=\columnwidth]{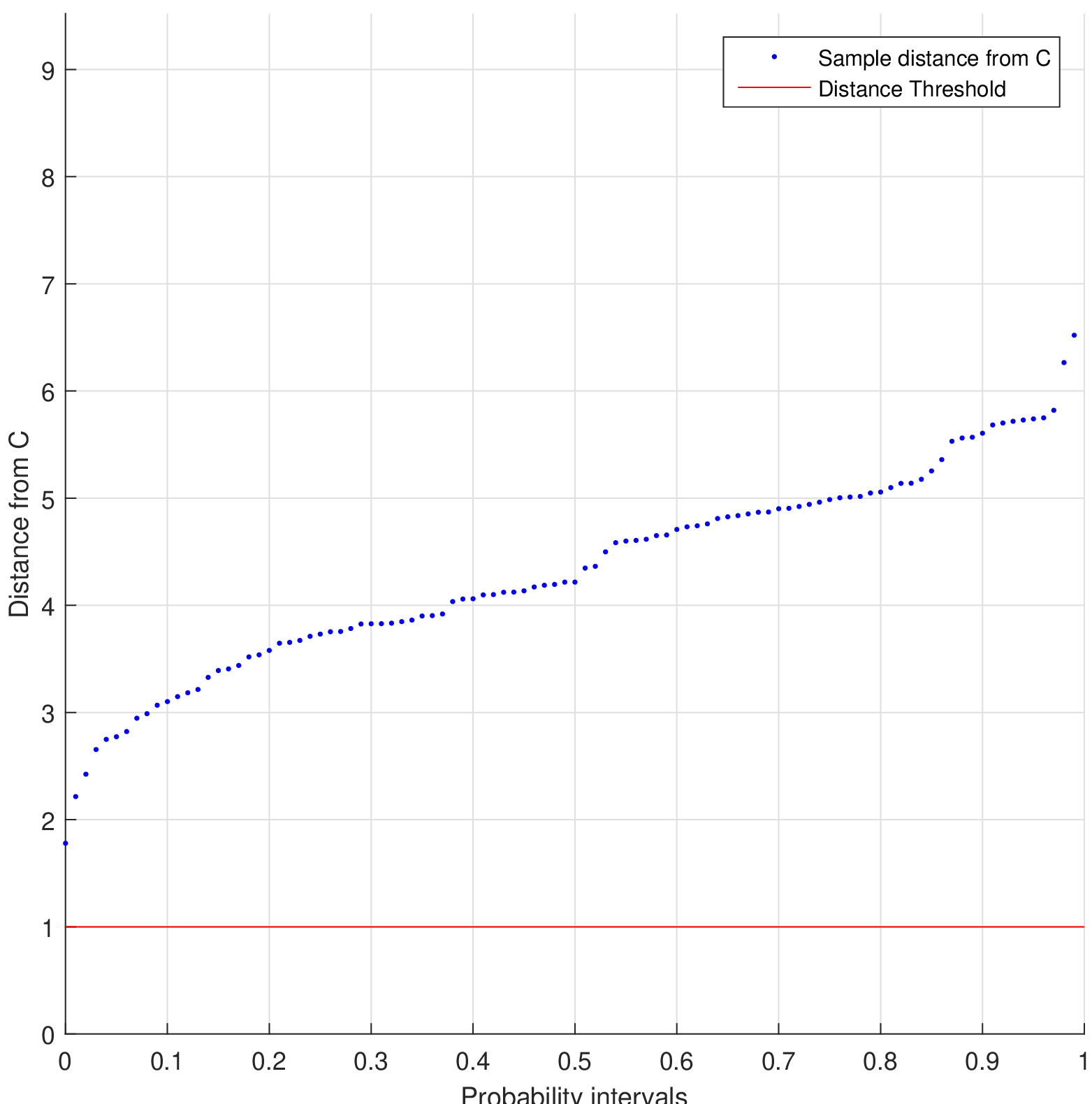}
	\label{fig:SS_0_CCDF}}  
	\hfill
	\subfloat[Subset Simulation Level 1]{%
	\includegraphics[width=\columnwidth]{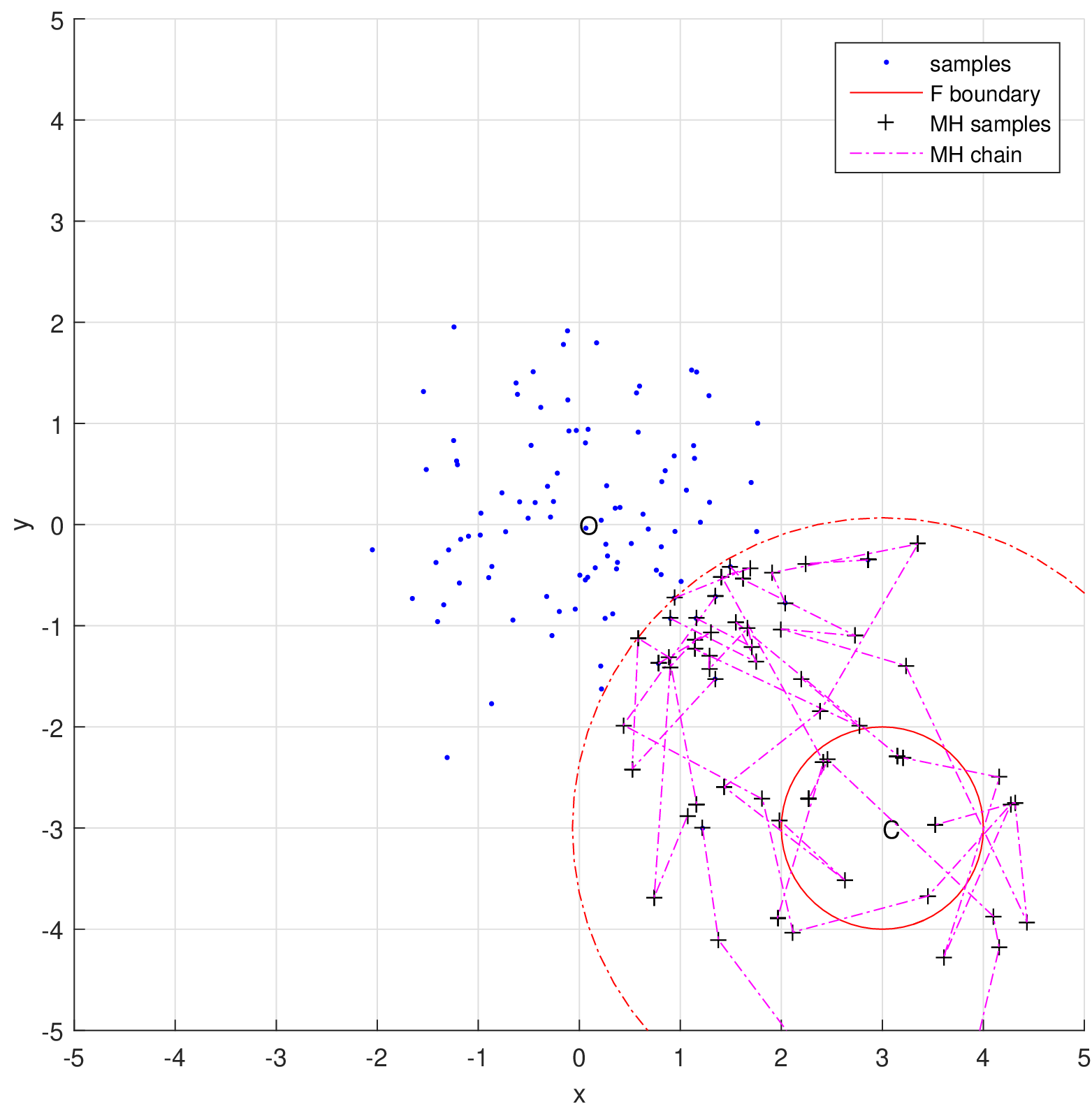}
	\label{fig:SS_1}}
	\hfill
	\subfloat[Subset Simulation Level 1 CCDF]{%
	\includegraphics[width=\columnwidth]{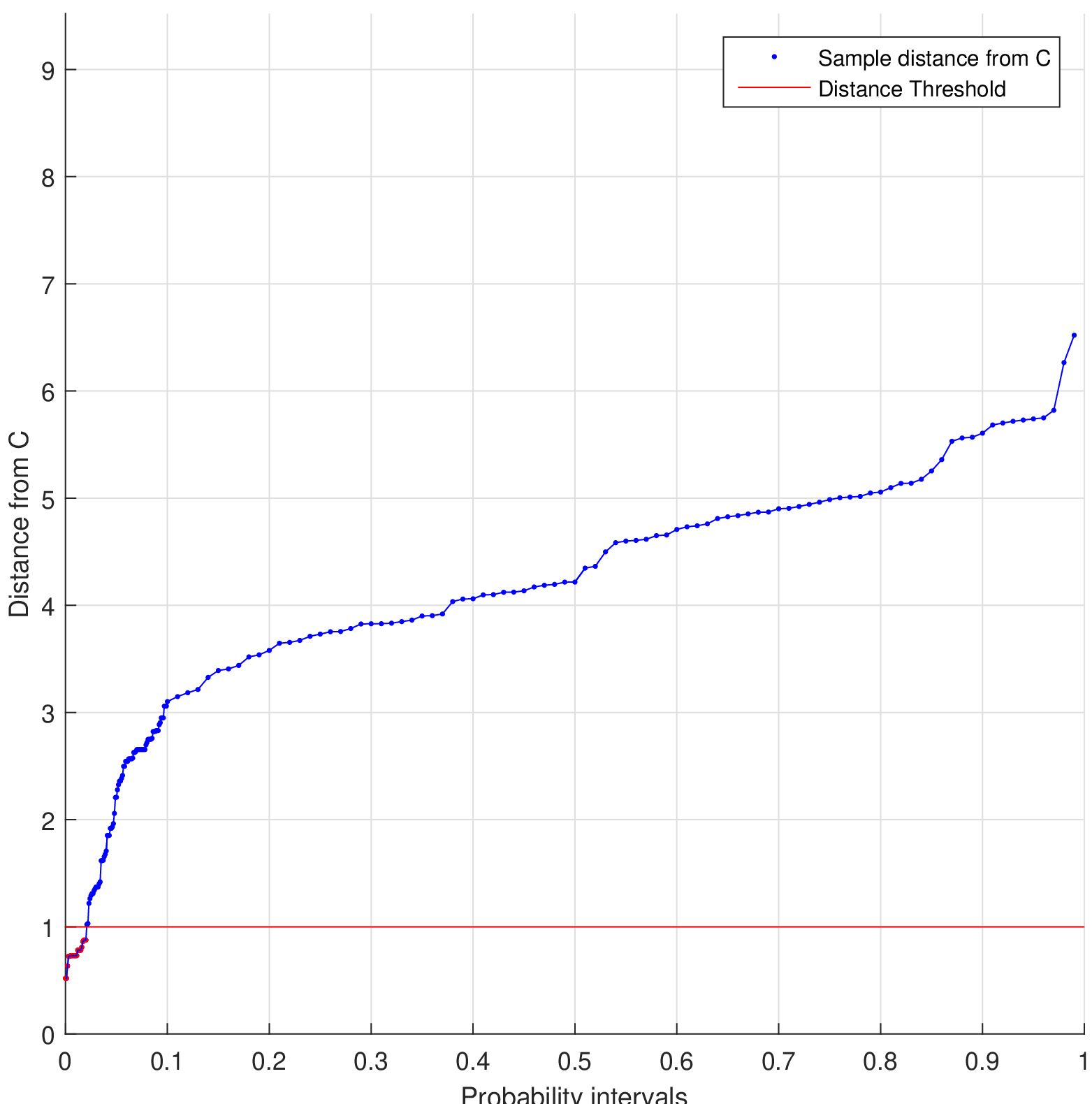}
	\label{fig:SS_1_CCDF}}
	\caption{Subset Simulation is applied to the problem of estimating the probability of drawing samples from the region $F$. Subset Simulation begins with level 0 by drawing N = 100 samples from a Gaussian distribution centered at $O = [0,0]$ using the DMC method as shown in Fig.~\ref{fig:SS_0}. The quantity of interest is the distance between each sample and $C$. These are plotted against probability intervals to generate a CCDF as shown in Fig.~\ref{fig:SS_0_CCDF}. No samples are within the region $F$. The SS method proceeds to level 1 and conditional samples are generated using the MH method. The $N_{c}$ level 0 samples are used to generate the conditional samples shown in Fig.~\ref{fig:SS_1}. These conditional samples are drawn progressively closer to the region F until some samples are drawn from the region F. This is achieved by drawing samples from intermediate thresholds closer to the boundary of $F$. The quantity of interest for the samples are determined and plotted against the probability intervals for the current level. This CCDF is appended to the previous CCDF by replacing the samples used as seeds from the previous level as shown in Fig.~\ref{fig:SS_1_CCDF}.}
	\label{fig:SS_0_1_CCDF}
\end{figure*}

The example of estimating the probability of drawing samples from the region $F$ shown in the previous sections is used to illustrate the Subset Simulation method (using algorithm~\ref{alg:SS}). The radius of the circle bounding the region $F$ is $r_{c} = 1$. The SS parameters used for this example are: $p_{0} = 0.1$, $N = 100$, $N_{s} = 10$, $N_{c} = 10$, $m = 2$. Subset Simulation is typically used to realize rare events (for $P_{F} \leq 10^{-3}$ therefore $m > 3$). However for the purpose of this example the number of levels is kept low ($m = 2$).

The simulation begins with level 0 Direct Monte Carlo where a set of $N = 100$ samples $\{X_{n}^{(0)}: n = 1,...,100\}$ are drawn from a Gaussian distribution centered at $O = [0,0]^{T}$ as shown in Fig.~\ref{fig:SS_0}. The quantity of interest $\{R_{n}^{(0)} = H(X_{n}^{(0)},C): n = 1,...,100\}$ is the distance between each sample $X_{n}^{(0)}$ and the center of the circle $C = [3,-3]^{T}$ (this is the equivalent of $Y_{n}^{(0)}$ used previously). This is determined by process $H(...)$ as defined by algorithm~\ref{alg:h} in section~\ref{sec:DMC}. If the condition $R_{n}^{(0)} \leq r_{c}^{(0)}$ is satisfied then the $n^{\text{th}}$ sample $X_{n}^{(0)}$ is within the region $F$. This condition is used to determine if a sample is within the region $F$. The quantity of interest $R_{n}^{(0)}$ is sorted in descending order $\{B_{n}^{(0)}: n = 1,...,100\}$. This is because the samples with the lowest distances will be closest to the region $F$ and have a higher likelihood of generating conditional samples closer to or within the region $F$ than other samples as the simulation progresses to higher levels ($i > 0$). The input samples $X_{n}^{(0)}$ are reordered $\tilde{X}_{n}^{(0)}$ and correspond to the sorted quantity of interest $B_{n}^{(0)}$; to clarify, the distance between the sample $\tilde{X}_{1}^{(0)}$ and $C$ is $B_{1}^{(0)}$. The probability intervals $P_{n}^{(0)}$ are determined by equation~\ref{eq:prob_intervals}. The sorted quantity of interest $B_{n}^{(0)}$ and respective samples $\tilde{X}_{n}^{(0)}$ are concatenated with the probability intervals $P_{n}^{(0)}$ as shown in the column titled `Level 0' in table~\ref{table:SS_example_lvl1}. The CCDF shown in Fig.~\ref{fig:SS_0_CCDF} is generated by plotting the probability intervals $P_{n}^{(0)}$ against $B_{n}^{(0)}$. This CCDF shows that no samples have a distance less than the radius $r_{c}$ therefore no samples have been drawn from the region $F$.

\begin{table}[h]
\centering
\subfloat[]{
\begin{tabular}{cccccc}
\multicolumn{1}{c}{} & \multicolumn{1}{c}{Level 0} & \multicolumn{1}{c}{} & & Level $1$ & \\ \cline{1-6}
\multicolumn{1}{|c|}{$\textbf{P}_{n}^{(0)}$} & \multicolumn{1}{c|}{$\textbf{B}_{n}^{(0)}$} & \multicolumn{1}{c|}{$\tilde{\textbf{X}}_{n}^{(0)}$} & \multicolumn{1}{|c|}{$\textbf{P}_{n}^{(1)}$}& \multicolumn{1}{c|}{$\textbf{B}_{n}^{(1)}$}& \multicolumn{1}{|c|}{$\tilde{\textbf{X}}_{n}^{(1)}$} \\ \cline{1-6}
\multicolumn{1}{|c|}{$P_{1}^{(0)}$} & \multicolumn{1}{c|}{$B_{1}^{(0)}$} & \multicolumn{1}{c|}{$\tilde{X}_{1}^{(0)}$} & & & \\ \cline{1-3}
\multicolumn{1}{|c|}{\vdots} & \multicolumn{1}{c|}{\vdots} & \multicolumn{1}{c|}{\vdots} & & & \\ \cline{1-3}
\multicolumn{1}{|c|}{$P_{90}^{(0)}$} & \multicolumn{1}{c|}{$B_{90}^{(0)}$} & \multicolumn{1}{c|}{$\tilde{X}_{90}^{(0)}$}   	 & & & \\ \cline{1-6}
\multicolumn{1}{|c|}{$P_{91}^{(0)}$} & \multicolumn{1}{c|}{$B_{91}^{(0)}$}    & \multicolumn{1}{c|}{$\tilde{X}_{91}^{(0)}$}   & \multicolumn{1}{c|}{$P_{1}^{(1)}$} & \multicolumn{1}{c|}{$B_{1}^{(1)}$} & \multicolumn{1}{c|}{$\tilde{X}_{1}^{(1)}$} \\ \cline{1-6}
\multicolumn{1}{|c|}{\vdots} & \multicolumn{1}{c|}{\vdots} & \multicolumn{1}{c|}{\vdots} & \multicolumn{1}{c|}{\vdots} & \multicolumn{1}{c|}{\vdots} & \multicolumn{1}{c|}{\vdots} \\ \cline{1-6}
\multicolumn{1}{|c|}{$P_{100}^{(0)}$}    & \multicolumn{1}{c|}{$B_{100}^{(0)}$}    & \multicolumn{1}{c|}{$\tilde{X}_{100}^{(0)}$}   & \multicolumn{1}{c|}{$P_{90}^{(1)}$} & \multicolumn{1}{l|}{$B_{90}^{(1)}$} & \multicolumn{1}{l|}{$\tilde{X}_{90}^{(1)}$} \\ \cline{1-6}
                             &                          & \multicolumn{1}{c|}{}   & \multicolumn{1}{c|}{$P_{91}^{(1)}$} & \multicolumn{1}{c|}{$B_{91}^{(1)}$} & \multicolumn{1}{c|}{$\tilde{X}_{91}^{(1)}$} \\ \cline{4-6}
                             &                          & \multicolumn{1}{c|}{}   & \multicolumn{1}{c|}{\vdots} & \multicolumn{1}{c|}{\vdots} & \multicolumn{1}{c|}{\vdots} \\ \cline{4-6}
                             &                          & \multicolumn{1}{c|}{} & \multicolumn{1}{c|}{$P_{100}^{(1)}$} & \multicolumn{1}{c|}{$B_{100}^{(1)}$} & \multicolumn{1}{c|}{$\tilde{X}_{100}^{(1)}$} \\ \cline{4-6}
														
\end{tabular}
\label{table:SS_example_lvl1}}
\qquad \qquad \qquad
\subfloat[]{
	\begin{tabular}{|c|c|c|}
  \multicolumn{3}{c}{\quad} \\
	\cline{1-3}
	$\textbf{P}_{n}$ & $\textbf{B}_{n}$ & $\tilde{\textbf{X}}_{n}$ \\ \cline{1-3} \hline
	$P_{1}^{(0)}$ & $B_{1}^{(0)}$ & $\tilde{X}_{1}^{(0)}$ \\ \cline{1-3}
	\vdots & \vdots & \vdots \\ \cline{1-3}
	$P_{90}^{(0)}$ & $B_{90}^{(0)}$ & $\tilde{X}_{90}^{(0)}$\\ \cline{1-3} \hline
	$P_{1}^{(1)}$ & $B_{1}^{(1)}$ & $\tilde{X}_{1}^{(1)}$\\ \cline{1-3} \hline
	\vdots & \vdots & \vdots \\ \cline{1-3}
	$P_{90}^{(1)}$ & $B_{90}^{(1)}$ & $\tilde{X}_{90}^{(1)}$\\ \cline{1-3} \hline	
	$P_{91}^{(1)}$ & $B_{91}^{(1)}$ & $\tilde{X}_{91}^{(1)}$\\ \cline{1-3} \hline
	\vdots & \vdots & \vdots \\ \cline{1-3}
	$P_{100}^{(1)}$ & $B_{100}^{(1)}$ & $\tilde{X}_{100}^{(1)}$\\ \cline{1-3} \hline
	\end{tabular}
	\label{table:subset_example_lvl1_concat}}
	\vspace{2mm}
	\caption{}
\end{table}

The SS method continues to the next level ($i = 1$) and generates $N$ conditional samples using the MH method. The conditional samples $\{X_{n}^{(1)}: n = 1,...,100\}$ are generated from a set of seeds $s_j^{(1)} = \{\tilde{X}_{91}^{(0)},...,\tilde{X}_{100}^{(0)}\}$ that correspond to the sorted distances $\{B_{n}^{(0)}: n = 91,...,100\}$ from the previous level 0. The intermediate threshold $b_{1} = B_{90}^{(0)}$ determined by equation~\ref{eq:thresholds} is used to ensure the conditional samples $X_{n}^{(1)}$ generated by each seed satisfies the condition $R_{n}^{(1)} \leq b_{1}$. The respective sample distances $R_{n}^{(1)}$ from $C$ are less than or equal to the level 1 threshold $b_{1}$. This is to enable a progressive nature of drawing samples that are closer to the region $F$. The conditional samples are genrated using algorithm~\ref{alg:mh_ss}. This will eventually lead to samples being drawn from the region $F$ as SS proceeds to higher number of levels in the future. The level 1 threshold is marked by the dotted arc in Fig.~\ref{fig:SS_1}. The figure shows chains of samples that lead to the region $F$. The distances $R_{n}^{(1)}$ of samples $X_{n}^{(1)}$ generated in level 1 are sorted in descending order $\{B_{n}^{(1)}: n = 1,...,100\}$. The input samples $X_{n}^{(1)}$ are reordered $\tilde{X}_{n}^{(1)}$ and correspond to the sorted distances $B_{n}^{(1)}$. The probability intervals $P_{n}^{(1)}$ are generated using equation~\ref{eq:prob_intervals} and concatenated with the sorted distances $B_{n}^{(1)}$ and their corresponding samples $\tilde{X}_{n}^{(1)}$. Table~\ref{table:SS_example_lvl1} illustrates the conditional samples generated in level 1 using samples from level 0. The seeds used to generate samples in level 1 are discarded and replaced with the generated level 1 samples as illustrated in table~\ref{table:subset_example_lvl1_concat}. Note the probability intervals $\{P_{n}^{(0)}: n = 91,...,100\}$, sorted distances $\{B_{n}^{(1)}: n = 91,...,100\}$ and the corresponding input samples $\{\tilde{X}_{n}^{(1)}: n = 91,...,100\}$ from level 0 that were used as seeds to generate the samples for level 1 are discarded and replaced with level 1 samples $\tilde{X}_{n}^{(1)}$ and their respective distances $B_{n}^{(1)}$ and probability intervals $P_{n}^{(1)}$. This process is repeated until the maximum number of levels $m$ is reached. This is when $i = m-1$. Fig.~\ref{fig:SS_1_CCDF} shows the overall CCDF at level 1. The overall CCDF is used to estimate the probability of drawing samples from the region $F$ as approximately $P_{F} = 0.02$.

This example demonstrates the progressive nature of Subset Simulation when used to generate conditional samples to realize the rare `tail' region of the pdf. This feature of SS results in the empirical observation that SS requires significantly less samples when compared to naive DMC to obtain estimates with the same accuracy. Subset Simulation is useful for generating samples that progress to the distribution of interest. 

The next section applies the Subset Simulation method with modifications to estimate the probability of conflict between air traffic.

\section{Application of Subset Simulation for Airborne Conflict detection}
\label{sec:SS_app}

The estimation of the probability of conflict $P_{c}$ between air traffic is a useful metric for Conflict Detection \& Resolution (CD\&R) methods. Such methods can be used in piloted aircraft but are useful for UAS where an automated method for CD\&R will be required as part of a Sense-and-Avoid system~\cite{kuchar2000review}.

\begin{figure*}
	\centering	
	\subfloat[Head-on]{%
	\includegraphics[width=0.5\columnwidth]{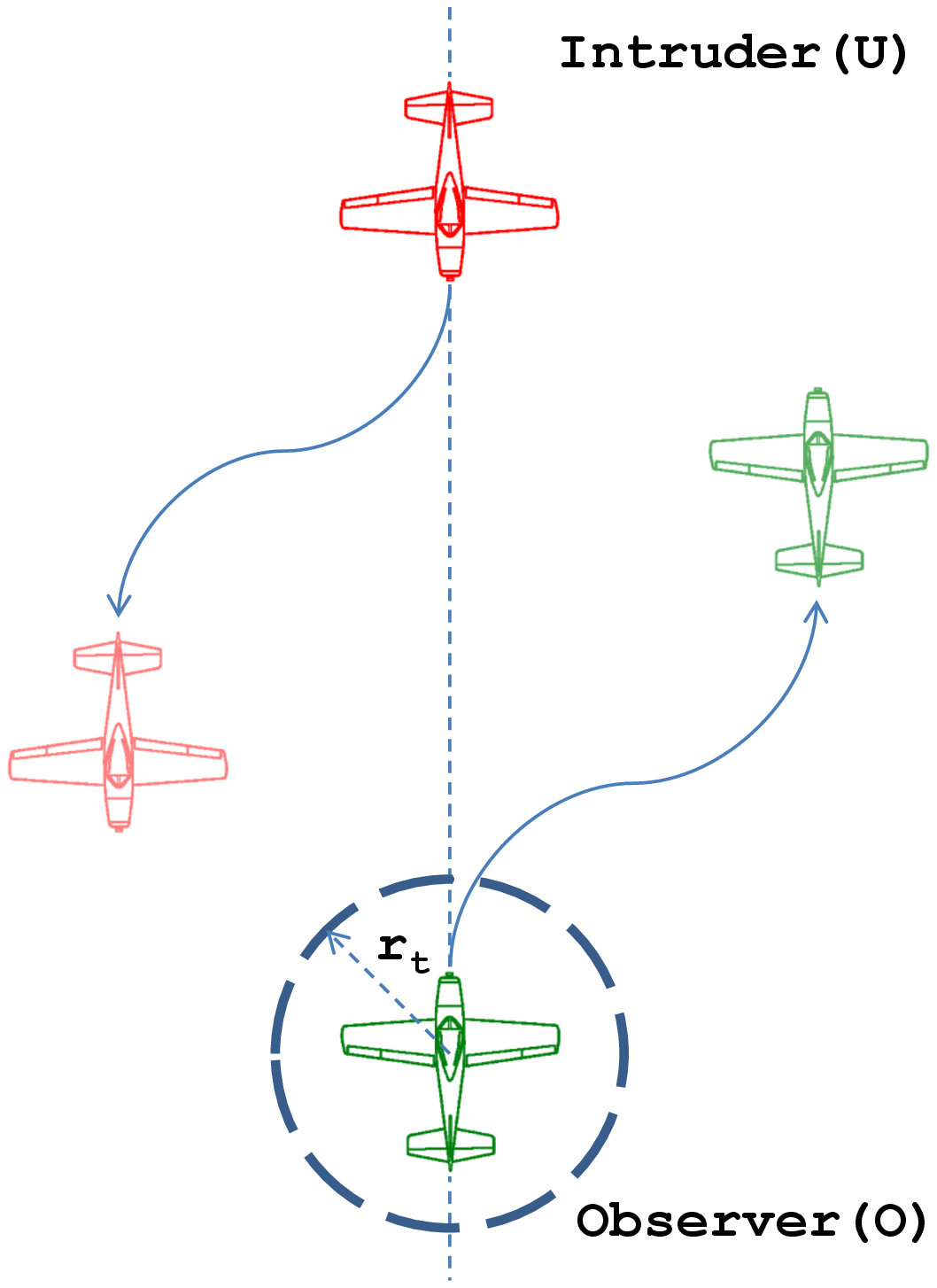}
	\label{fig:head_on}}	
	\hfill 
	\subfloat[Overtaking]{%
	\includegraphics[width=0.5\columnwidth]{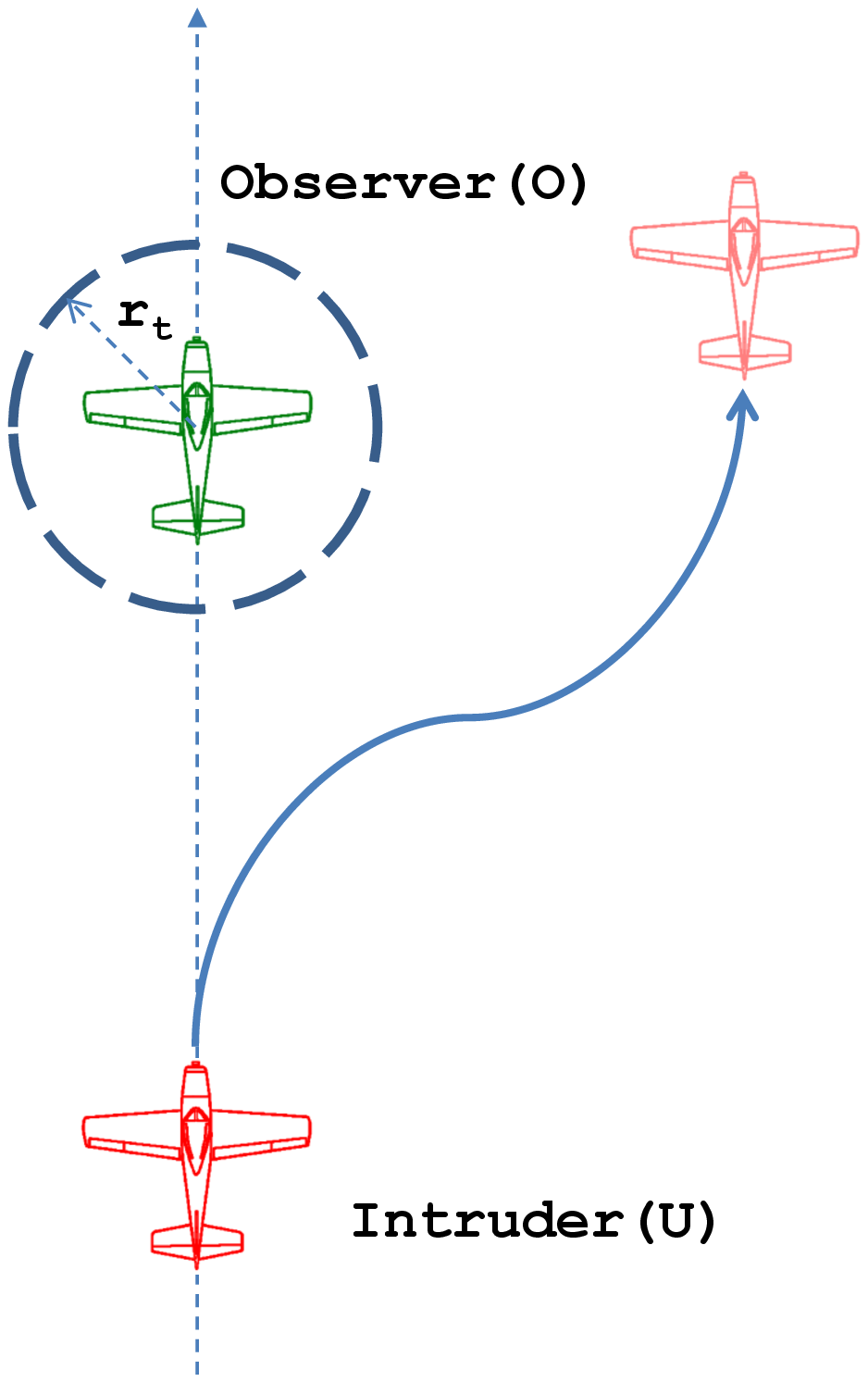}
	\label{fig:overtaking}}	
  \hfill
	\subfloat[Converging]{%
	\includegraphics[width=0.7\columnwidth]{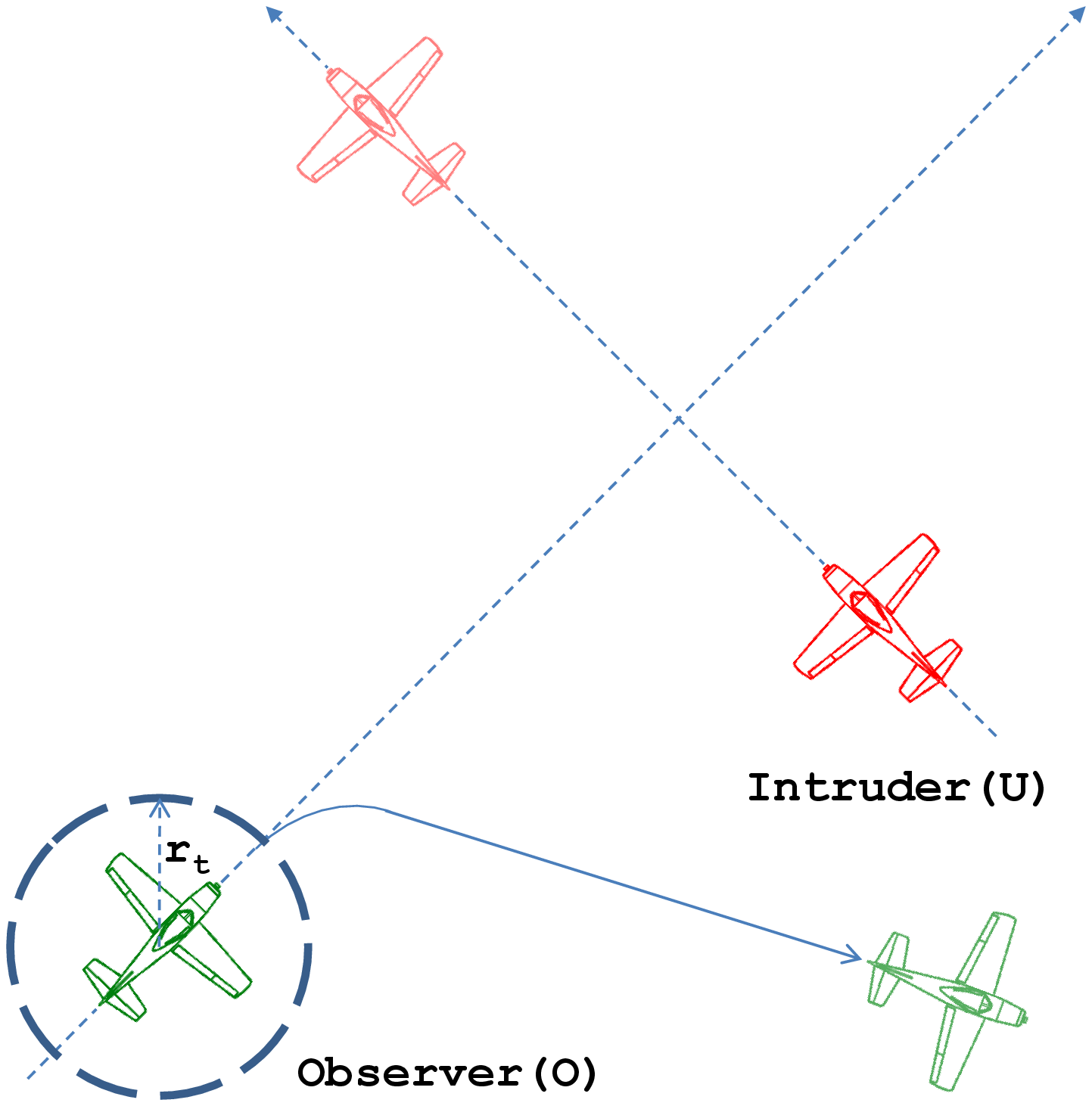}
	\label{fig:converging}}	
	\caption{These figures illustrate the geometric configuration of the different conflicts that might be encountered within a block of airspace. This includes different maneuvers required to be executed by the respective parities to resolve the conflict.}
	\label{fig:conflict_types}
\end{figure*}

According to CAA CAP 393 Rules of the Air, the minimum lateral (Horizontal) separation required between two or more aircraft at any instance is 500ft. A conflict event occurs when two or more aircraft collide or if there is a loss of this separation between them within a block of airspace. The conflict type depends on the geometry of the encounter between traffic, as defined in~\cite{authority2015cap}. These conflict types are illustrated in Fig.~\ref{fig:conflict_types} as:
\begin{itemize}
	\item A Head-on conflict scenario as shown in Fig.~\ref{fig:head_on}. In such a case each aircraft must turn right to avoid the collision.
	\item An Overtaking conflict scenario is where the aircraft being overtaken has the right of way as shown in Fig.~\ref{fig:overtaking}. The overtaking aircraft must alter course right and keep clear of the overtaken aircraft. An overtaking condition exists while the overtaking aircraft is approaching the rear of another aircraft within an angle less that 70 degrees from the extended centreline of the aircraft being overtaken.
	\item A Converging conflict scenario is where the aircraft on the right has the right of way as shown in Fig.~\ref{fig:converging}. The aircraft on the left must alter its course right to resolve the conflict.
\end{itemize}

If a conflict is detected, the conflict type needs to be identified so that the appropriate resolution maneuver can be executed by the CD\&R system to resolve the conflict. This paper addresses a key component of a detection of a conflict by estimating the probability of conflict $P_{c}$.

\begin{figure}[h]
	\centering
	\subfloat[Head-on pass]{%
	\includegraphics[width=0.48\columnwidth]{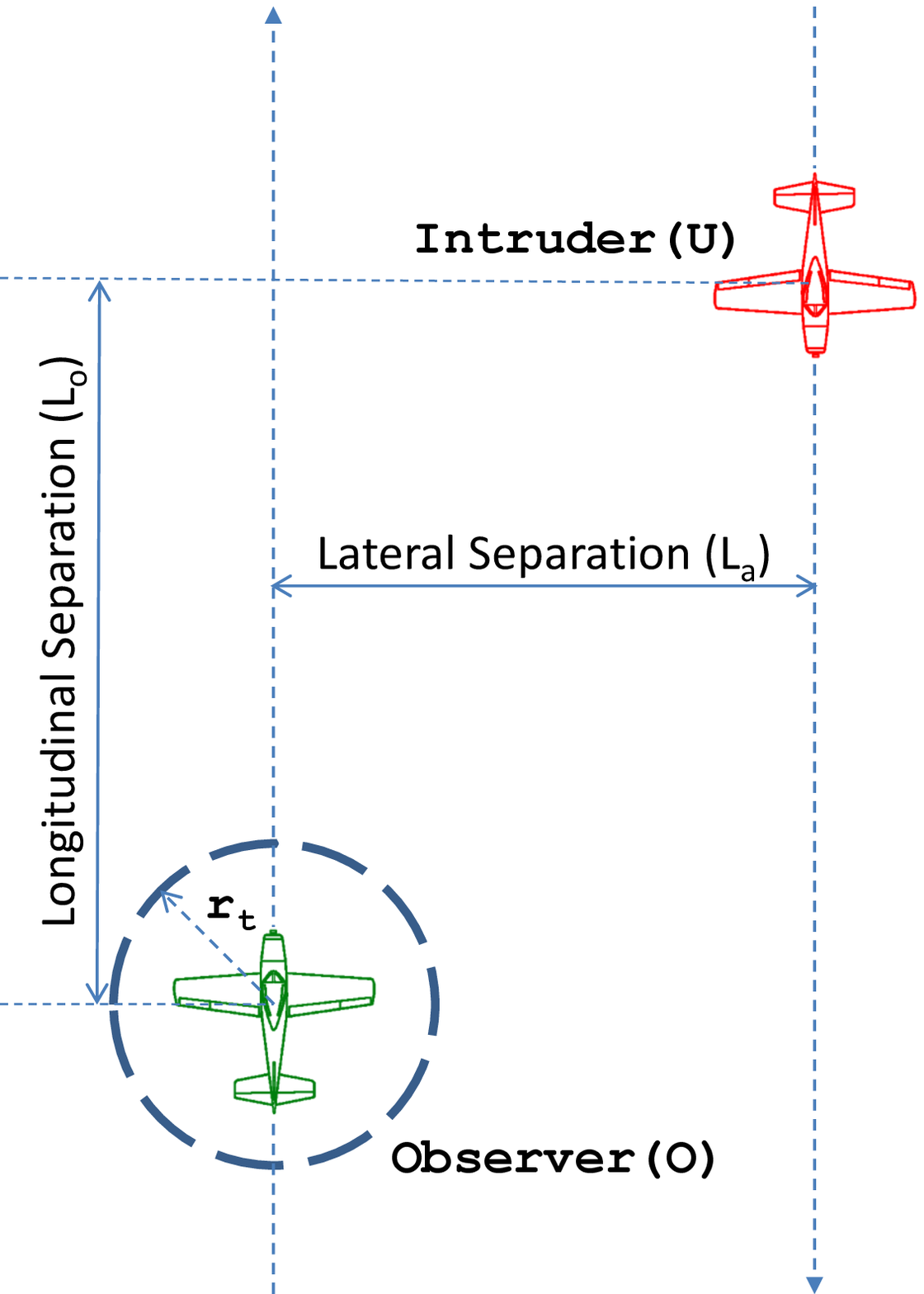}
	\label{fig:potential_headon}}
	\hfill
	\subfloat[Intruder overtaking Observer]{%
	\includegraphics[width=0.48\columnwidth]{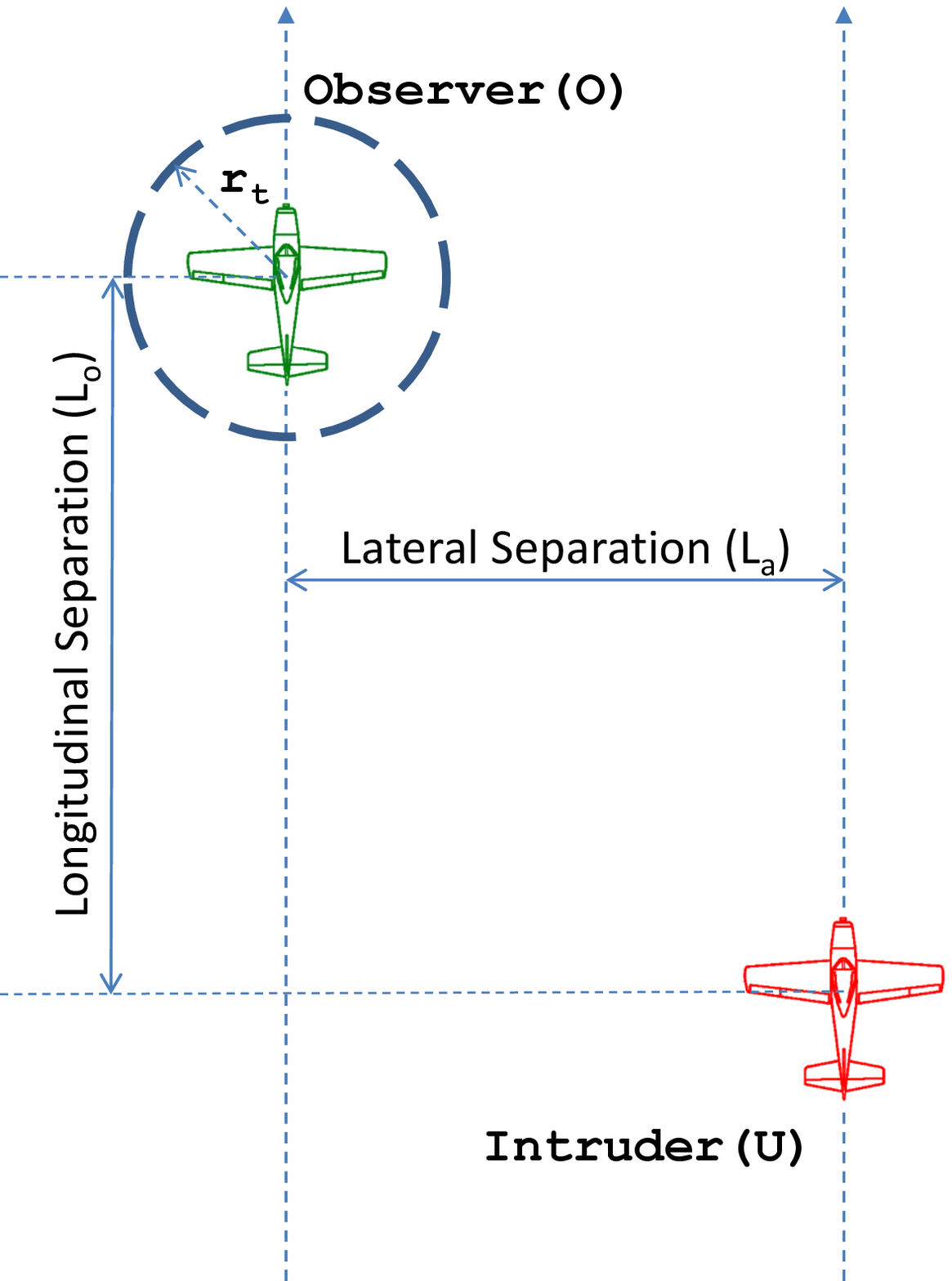}
	\label{fig:intruderovertaking_potential}}
	\caption{The potentially conflicting scenarios based on the different conflicts shown in Fig.~\ref{fig:conflict_types}}
	\label{fig:potential_conflict_scenarios}
\end{figure}

We assume a non-cooperative scenario, where the traffic does not share information. This is a challenging situation since the information related to the state and intentions of the traffic might be unknown or incorrect. The only information available regarding the state of traffic is from measurements or inference using sensors. In such a scenario, CD\&R system must allow for the possibility that the non-cooperative traffic may take inappropriate actions or may not adhere to the Rules of the Air. This type of situation requires a UAS to react and take appropriate action to ensure safe separation. To achieve this the $P_{c}$ needs to be continuously evaluated against the behavior of the observed traffic so that the likelihood of the traffic causing a conflict can be calculated. Fig.~\ref{fig:potential_conflict_scenarios} illustrates some potentially conflicting scenarios based on Fig.~\ref{fig:conflict_types}. During some phases of the scenario, the expected $P_{c}$ can be very low; such as a magnitude of $10^{-8}$ (this is demonstrated later in this section). The previous sections have demonstrated that estimating low probabilities using the Direct Monte Carlo method is inefficient and this motivates the use of Subset Simulation (SS). Assessing the full pdf may not be feasible and may not be required. Subset Simulation provides an efficient method of determining the probability associated with all predicted conflicts thereby estimating $P_{c}$. In applying SS to this problem, $P_{c}$ plays the role of the threshold of failure $P_{F}$.

The Subset Simulation method is used to estimate the probability of conflict $P_{c}$ during the simulation of the potentially conflicting scenarios of the Observer and Intruder aircraft in the Head-on and Overtaking situations as shown in figures~\ref{fig:potential_headon} and~\ref{fig:intruderovertaking_potential} respectively. Both scenarios show the Observer and Intruder in a non-conflicting a state, where the Intruder is not within the Observer's protected zone. The Observer's protected zone is marked as a circle around the Observer with radius $r_{t} = 152.4 \text{m } \text{(500ft)}$. Although the current state is non-conflicting there is a potential for future conflict. For example from the Observer's perspective the Intruder could continue on its course or turn right or turn left. The latter could cause a loss of separation or worse -- a collision between the Observer and the Intruder. Also in the situation when the lateral separation $L_{a}$ between the Observer and Intruder is lower than or equal to the radius of the Observer's protected zone $r_{t}$; ($r_{t} \leq L_{a}$) a conflict occurs due to loss of separation or collision between the Observer and the Intruder. Therefore the likelihood of such conflict needs to be realized by estimating $P_{c}$.

The Subset Simulation method is used by the Observer to determine the probability of conflict $P_{c}$ between itself and the approaching Intruder for the potentially conflicting scenarios shown in Fig.~\ref{fig:potential_conflict_scenarios}. However, since some parameters are not available this requires the method to be adapted. The order of magnitude for the target probability (conflict) region $(p_{0})^{m}$ is unknown. The solution to this problem is addressed later in this section. Therefore the number of subset levels $m$ required to reach the target probability level with a fixed $p_{0}$ is unknown. The Intruder and Observer are simulated as \textit{nearly constant acceleration} point models~\cite{RongLi2003}. This is a simple model that is used to illustrate the use of Subset Simulation. It can be augmented by more complex dynamic models such as Six-Degrees-of-Freedom (SixDoF) aircraft models as shown in~\cite{nelson1998flight}. This would not affect the use of Subset Simulation and the computational advantages that it provides. The dynamics of the Intruder and Observer are modeled in state space form as $U(K+1) = AU(K)$ and $O(K+1) = AO(K)$ respectively in two-dimensional Cartesian space, where $K$ is the time--step index. The Intruder and Observer statevectors are $U(K) = [x, u, a_{x}, y, v, a_{y}]^{T}$ and $O(K) = [x, u, a_{x}, y, v, a_{y}]^{T}$ respectively. The displacement, velocity and acceleration in the $x$-direction are represented by $x, u$ and $a_{x}$ respectively. The displacement, velocity and acceleration in the $y$ direction are represented by $y, v$ and $a_{y}$ respectively. The state transition matrix $A$ is defined as

\begin{equation}
A = \begin{bmatrix}
	1 & \Delta T & \frac{1}{2}\Delta T^{2} & 0 & 0 & 0\\	
	0 &  1 & \Delta T & 0 & 0 & 0\\	
	0 &  0 & 1 & 0 & 0 & 0 \\
	0 &  0 & 0 & 1 & \Delta T & \frac{1}{2}\Delta T^{2} \\
	0 &  0 & 0 & 0 & 1 & \Delta T \\
	0 &  0 & 0 & 0 & 0 & 1 
\end{bmatrix}
\label{eq:transition_matrix}
\end{equation}

\noindent where $\Delta T$ is the period of discretized time-step. The sampling frequency $f = \frac{1}{\Delta T}$. The Observer estimates the state of the Intruder $\hat{U}(K)$ using a Kalman Filter~\cite{bar2004estimation}. The periodic measurements of the Intruder's position $Z = [x, y]$ is defined by the measurement equation as 

\begin{equation}
	Z = HU(K) + [w_{x},w_{y}]'
	\label{eq:measurement}
\end{equation}

\noindent where $H$ is the measurement matrix.

\begin{equation}
H = \begin{bmatrix}
1 & 0 & 0 & 0 & 0 & 0\\	
0 & 0 & 0 & 1 & 0 & 0\\	
\end{bmatrix}
 \label{eq:measurement_matrix} 
\end{equation}

\begin{equation}
w_{x} \sim{~} \mathcal{N}(0,\sigma_{x})
\label{eq:noise_x}
\end{equation}

\begin{equation}
w_{y} \sim{~} \mathcal{N}(0,\sigma_{y})
\label{eq:noise_y}
\end{equation}

\noindent The periodic position measurements are simulated by adding noise as $w_{x}$ and $w_{y}$ to the $x$ and $y$ directions respectively. The standard deviation of the of the measurement error in the $x$ and $y$ directions are $\sigma_{x}$ and $\sigma_{y}$ respectively. For the sake of simplicity the measurement noise is uncorrelated. The instantaneous state estimate of the Intruder is determined using a Kalman Filter. The Intruder's state estimate $\hat{U}(K+1)$ and covariance $\hat{S}(K+1)$ is predicted using equations

\begin{equation}
\hat{U}(K+1) = A\hat{U}(K)
\label{eq:kalman_dyn}
\end{equation}

\begin{equation}
\hat{S}(K+1) = A\hat{S}(K)A^{T} + Q
\label{eq:kalman_error_predict}
\end{equation}

\noindent The process noise covariance is $Q$. This is the \textit{white-noise jerk} version of the \textit{Wiener-Process Acceleration} model~\cite{RongLi2003}.

\begin{equation}
	Q = \begin{bmatrix}		
			 Q_{\sigma}\frac{\sigma_{a_{x}}^{2}}{\Delta T} & 0 \\
			 0 & Q_{\sigma}\frac{\sigma_{a_{y}}^{2}}{\Delta T}
		\end{bmatrix}
\end{equation}

\begin{equation}
	Q_{\sigma} = \begin{bmatrix}
	\frac{1}{20}\Delta T^{5} & \frac{1}{8}\Delta T^{4} & \frac{1}{6}\Delta T^{3} \\	
	\frac{1}{8}\Delta T^{4} & \frac{1}{3}\Delta T^{3} & \frac{1}{2}\Delta T^{2} \\	
	\frac{1}{6}\Delta T^{3} & \frac{1}{2}\Delta T^{2} & \Delta T
	\end{bmatrix} 
\end{equation}

\noindent The parameters $\sigma_{a_{x}}^{2}$ and $\sigma_{a_{y}}^{2}$ are the variance of acceleration parameters in the $x$ and $y$ directions respectively. The Kalman gain $G$ is evaluated during the update stage:

\begin{equation}
G = \hat{S}(K+1)H^{T}([H\hat{S}(K+1)H^{T}]+R)^{-1}
\label{eq:kalman_gain}
\end{equation}

\noindent where $R$ is the measurement covariance.

\begin{equation}
R = \begin{bmatrix}
\sigma_{x}^{2} & 0 \\	
0 & \sigma_{y}^{2} 
\end{bmatrix} 
\end{equation}

\noindent This is followed by updating the Intruder estimate $\hat{U}(K+1)$ and error covariance $\hat{S}(K+1)$ respectively.

\begin{equation}
\hat{U}(K+1) = \hat{U}(K+1)+G\{Z(K)-[H\hat{U}(K+1)]\}
\label{eq:kalman_est}
\end{equation}

\begin{equation}
\hat{S}(K+1) = [I - GH]\hat{S}(K+1)
\label{eq:kalman_cov}
\end{equation}

\begin{algorithm}
\caption{Kalman Filter}
	\label{alg:kalman_filter}
	\begin{algorithmic}[1]	
	\Function {KF}{$\hat{U}(K)$, $\hat{S}(K)$, $Z$, $H$, $Q$, $R$, $M_{Z}$}

		\LineComment{Predict}
		\State $\hat{U}(K+1) = A\hat{U}(K)$
		\State $\hat{S}(K+1) = A\hat{S}(K)A^{T}+Q$
	
		\LineComment{Update if new measurement is available}
		\If{$M_{Z} = \text{true}$}
			\State $G = \hat{S}(K+1)H^{T} \{[H\hat{S}(K+1)H^{T}]+R\}^{-1}$
			\State $\hat{U}(K+1) = \hat{U}(K+1)+G\{Z-[H\hat{U}(K+1)]\}$
			\State $\hat{S}(K+1) = [I - GH]\hat{S}(K+1)$
		\EndIf
		
		\State \textbf{return $\hat{U}(K+1), \hat{S}(K+1)$}
	\EndFunction	
	\end{algorithmic}
\end{algorithm}

\subsection{Example}
The Subset Simulation method is applied to the Head-on pass scenario with lateral separation $L_{a} = 1000\text{m}$ and longitudinal separation $L_{o} = 2000\text{m}$. The duration of the simulation $t = 20\text{s}$ with sampling frequency $f = 20\text{Hz}$ and the measurement frequency $f_{M} = 2\text{Hz}$. The initial conditions of the Intruder and Observer are $O(0) = [0,77.2 \text{ms}^{-1},0,0,0,0]^{T}$ and $U(0) = [2000 \text{m},-77.2 \text{ms}^{-1},0,1000 \text{m},0,0]^{T}$. The Observer's protected zone radius $r_{t} = 152.4\text{m}$.

\textbf{Kalman Filter parameters:}
\begin{itemize}
	\item $\sigma_{x} = 0.1\text{m}$
	\item $\sigma_{y} = 0.1\text{m}$
	\item $\sigma_{a_{x}}^{2} = 0.01\text{m}^{2}\text{s}^{-4}$
	\item $\sigma_{a_{y}}^{2} = 0.01\text{m}^{2}\text{s}^{-4}$
\end{itemize}

\textbf{Subset Simulation parameters:}
\begin{itemize}	
	\item $N = 100$
	\item $p_{0} = 0.1$
	\item $N_{c} = 10$
	\item $N_{s} = 10$
	\item $m = 7$ 
\end{itemize}

Ideally the SS method should continue to higher levels of simulation until conflicting samples are encountered and $P_{c}$ can be estimated using the CCDF. This is assuming infinite simulation resources are available. This is impractical for implementation since simulation capacity is limited due to limited resources available. Therefore the SS method implemented requires a limited number of levels\negthinspace \footnote{An alternative implementation: During the process of SS estimating the $P_{c}$; the SS method continues to higher levels until conflicting samples are found. If new information is received (such as a new Intruder measurement that updates the Intruder state estimate) and the SS method has not found conflicting samples, then the calculation for the current time-step should be abandoned and restarted with the new information. Restarting is necessary since the information used to calculate $P_{c}$ becomes obsolete once more recent information is obtained. This approach would be useful for situations where real-time computation is enforced. Note that this paper does not enforce constraints associated with real-time computation.} to be defined $m$.


\begin{figure}
\centering
\includegraphics[trim={0 0.5cm 0 0.5cm},clip,width=\columnwidth]{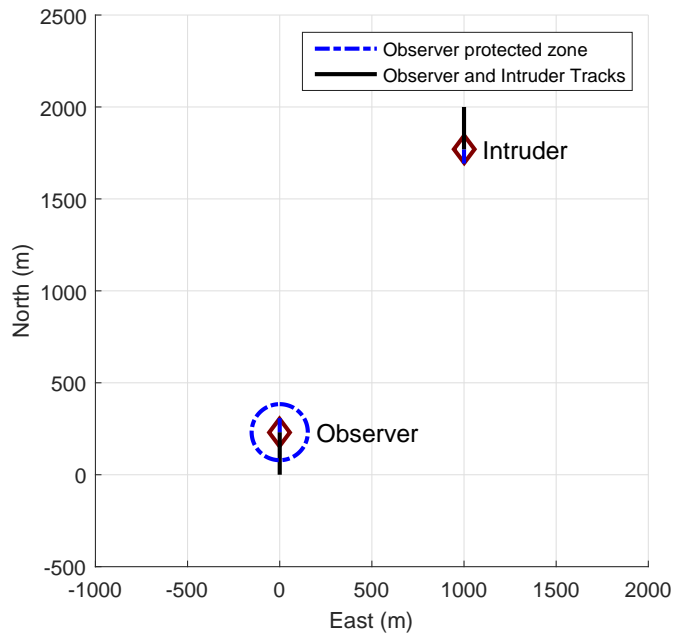}
\caption{Head-on pass scenario with 1000m Lateral Separation}
\label{fig:headon_TUT}
\end{figure}

Subset Simulation estimates $P_{c}(K+1)$ where $K+1$ is the time-step of an instance during the simulation as shown in Fig.~\ref{fig:headon_TUT}. Subset Simulation begins with level 0 Direct Monte Carlo sampling. A set of 100 samples $\{U_{n}^{(0)}: n = 1,...,100 \}$ representing the Intruder's pdf are drawn from the distribution that is centered at the Intruder's mean $\hat{U}(K+1)$ and covariance $\hat{S}(K+1)$. The mean and covariance are obtained from the Kalman filter defined in algorithm~\ref{alg:kalman_filter}.

\begin{figure*}[!t]
	\centering
	\subfloat[Level 0 DMC sample trajectories with highlighted trajectories of samples used as seeds]{%
	\includegraphics[width=\columnwidth]{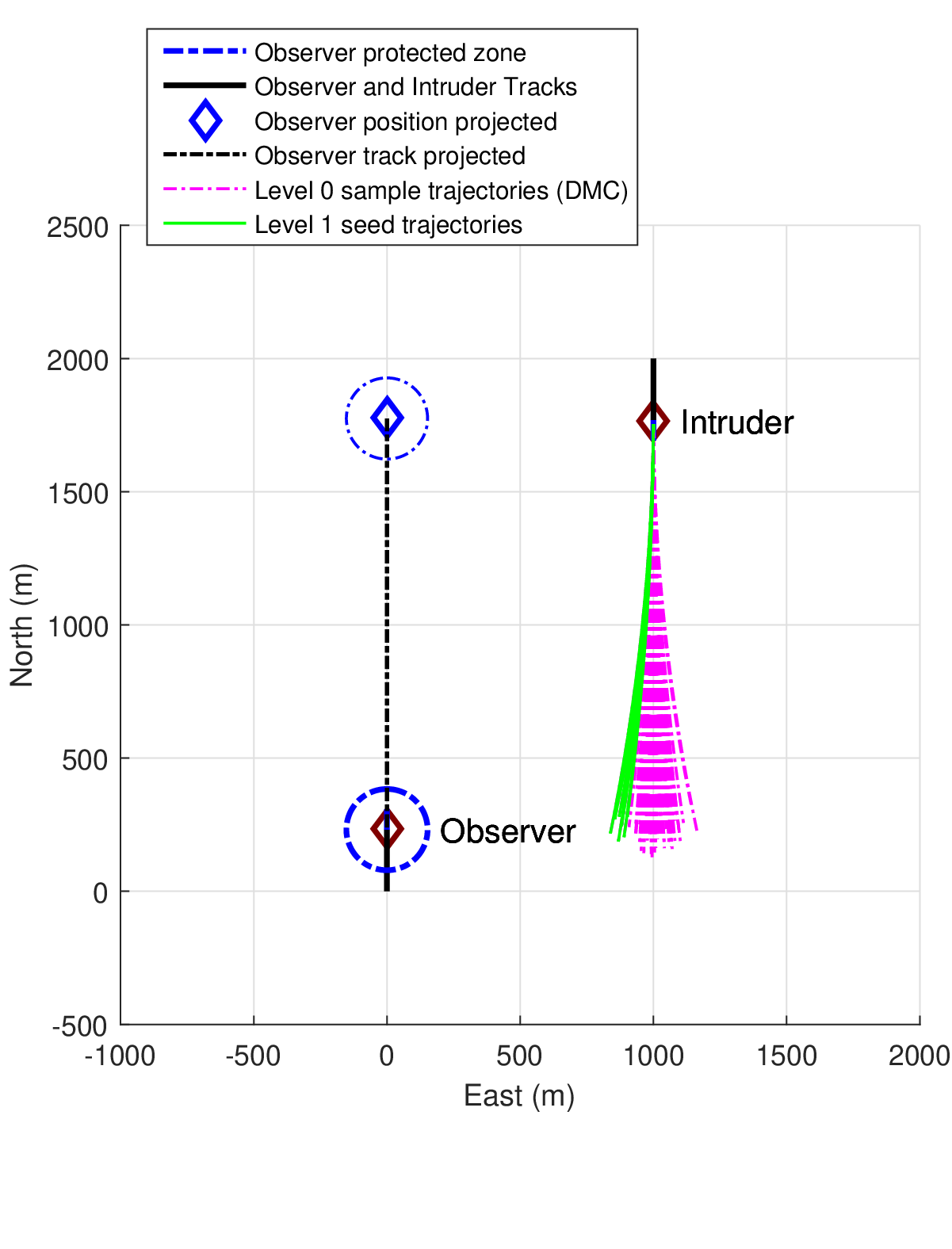}
	\label{fig:headon_level0_seeds}}
	\hfill
	\subfloat[Trajectories for Level 1 samples generated using $N_{c}$ samples from level 0]{%
	\includegraphics[width=\columnwidth]{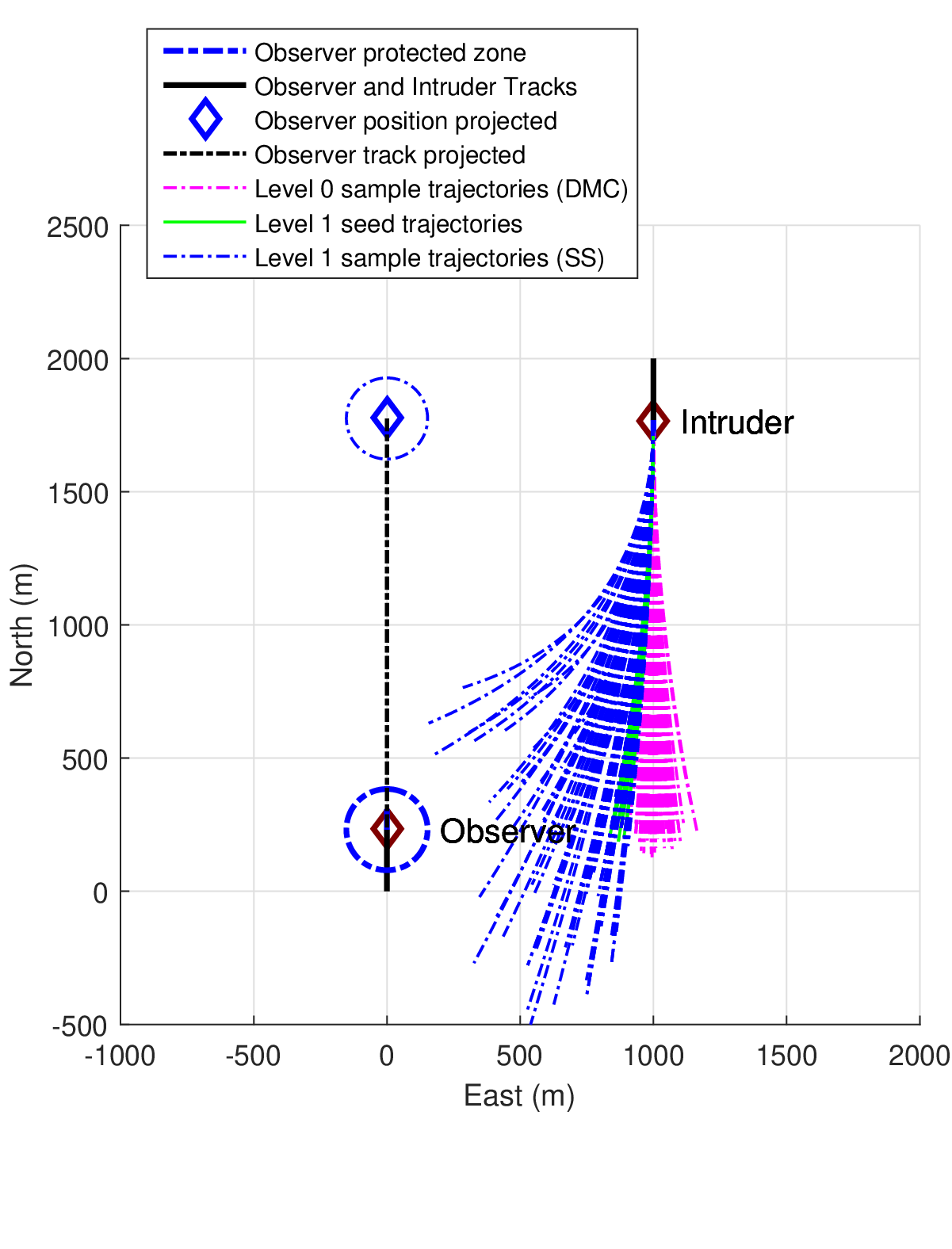}
	\label{fig:headon_level1}}
	\hfill	
	\subfloat[Level 0 CCDF with miss-distance of Level 1 seeds highlighted ]{%
	\includegraphics[width=\columnwidth]{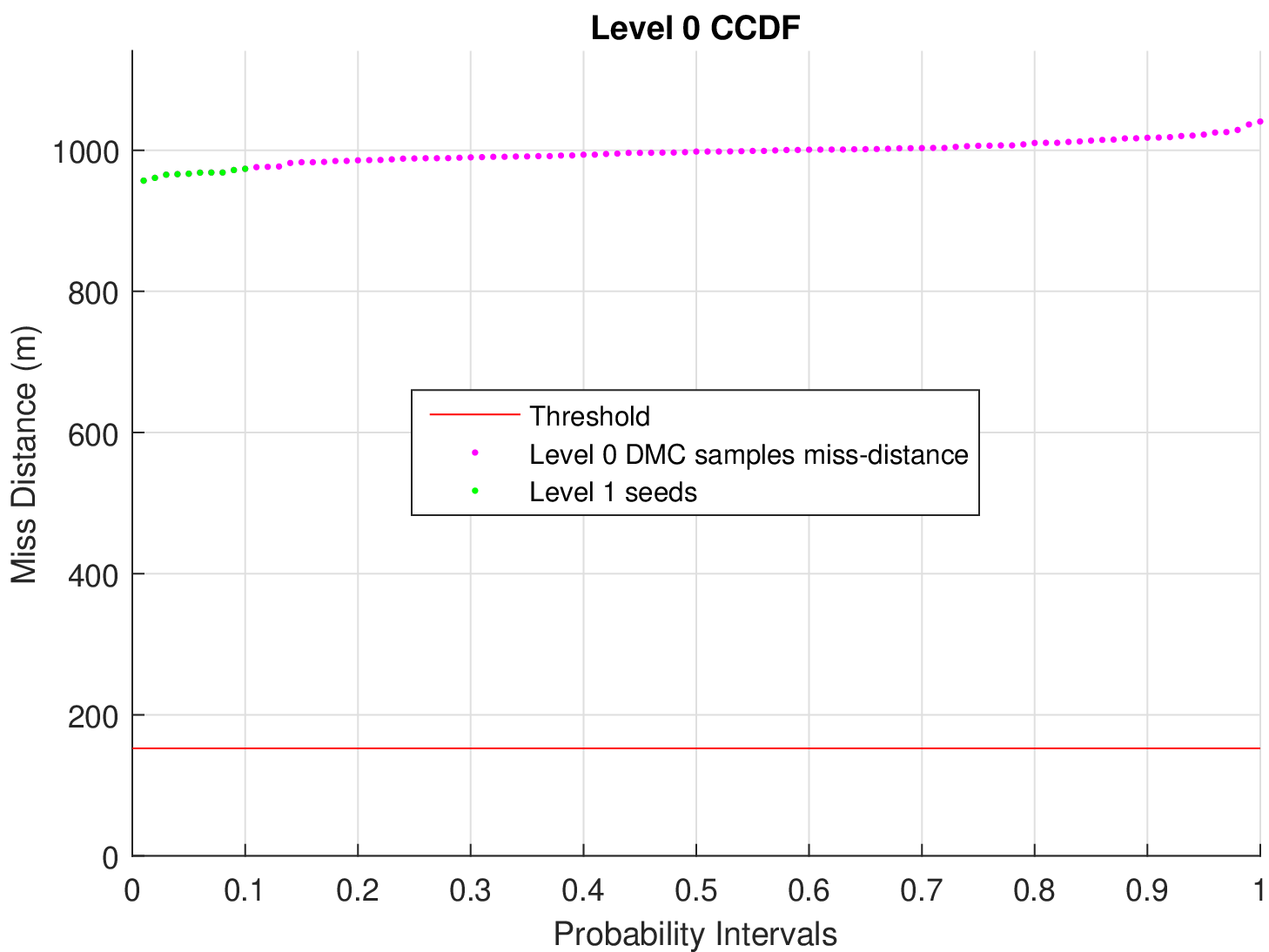}
	\label{fig:headon_level0_seeds_CCDF}}
	\hfill
	\subfloat[Level 1 CCDF]{%
	\includegraphics[width=\columnwidth]{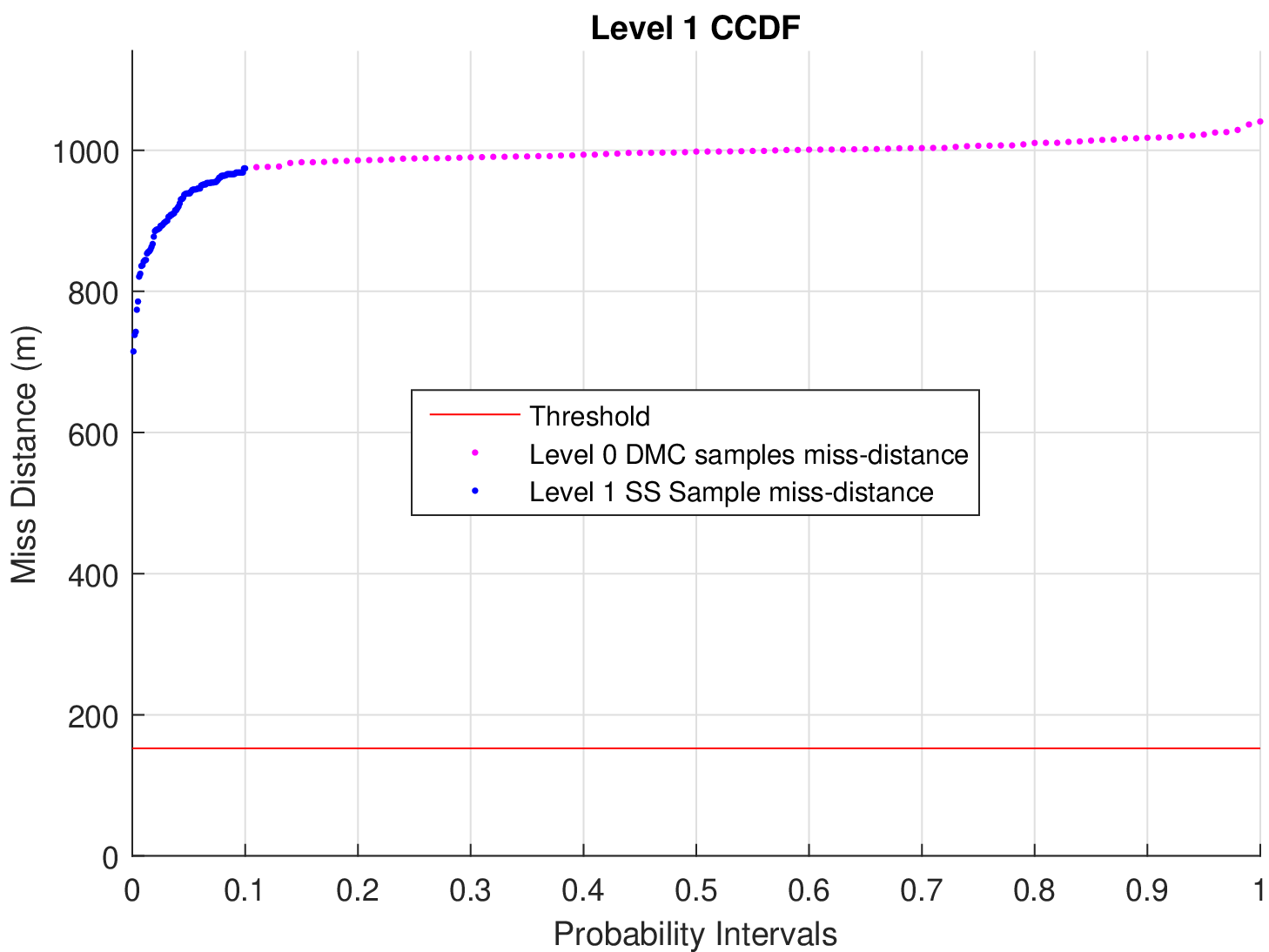}
	\label{fig:headon_level1_CCDF}}
	\caption{These figures illustrate the application of SS to estimate the $P_{c}(K+1)$ at the time-step $K+1$ during a Head-on pass between an Observer $O(K)$ and Intruder $U(K)$ with lateral separation of 1000m. SS begins with level 0 (DMC) where $N = 100$ samples are drawn from a distribution centered at the Intruder's state estimate $\hat{U}(K+1)$ with a covariance of $\hat{S}(K+1)$ obtained from the Kalman Filter. Fig.~\ref{fig:headon_level0_seeds} shows trajectories generated by level 0 samples, no conflicting samples have been encountered. The simulation proceeds to level 1 where conditional samples are generated using $N_{c}$ samples from level 0 as seeds. The trajectories of the level 0 samples used as seeds are highlighted in Fig.~\ref{fig:headon_level0_seeds}. The MH method is applied to generate conditional samples from the seeds. The trajectories of generated samples for level 1 are shown in Fig.~\ref{fig:headon_level1}. This process is continued to generates more trajectories as the number of levels increase. The method continues until conflicting samples are encountered at higher levels as shown in Fig.~\ref{fig:headon_level3_4}.}
	\label{fig:headon_level0_1}
\end{figure*}

The set of samples $U_{n}^{(0)}$ and the intended vector of the Observer $O(K)$ are propagated to generate trajectories $J_{U_{n}}^{(0)}$ and $J_{O}$ respectively. A trajectory $J$ is a set of consecutive state vectors indexed by the time-step $k$ where $k = 1,...,tf = 1,...,400$ and $f = 20\text{Hz}$ is the sampling frequency (as defined in algorithm~\ref{alg:sample_traj}). For example the Observer trajectory $J_{O} = [O(1),...,O(tf)] = [O(1),...,O(400)]$, where $O(1)$ is the state vector of the Observer at time-step $k = 1$. The propagation time $t = 20\text{s}$. This is also the period of the simulation. Fig.~\ref{fig:headon_level0_seeds} shows the Intruder samples and the respective trajectories generated with the projected position of the Observer during level 0 for a Head-on pass scenario with lateral separation $L_{a} = 1000\text{m}$. No conflicting samples have been encountered yet. A conflicting sample is an Intruder sample $U_{n}^{(i)}$ generated in level $i$ with a trajectory $J_{n}^{(i)}$ that has a miss-distance $r_{n}^{(i)}$ between the Observer trajectory $J_{O}$ and satisfies the conflict condition $r_{n}^{(i)} \leq r_{t}$. The number of conflicting samples encountered in a level is $D$.



\begin{algorithm}[!t]
\caption{Propagate State to generate trajectory}
\label{alg:sample_traj}
	\begin{algorithmic}[1]
			\Function{SampleTrajectory}{$U_{0}$, $f, t, A$}
			\State $J_{0} = U_{0}$
				\For{$k = 0: tf$}
					\State $U(k+1) = AU(k)$
					\State $J(k+1) = U(k+1)$
				\EndFor
				\State \textbf{return $J$}
			\EndFunction
	\end{algorithmic}
\end{algorithm}

\begin{algorithm}[!t]
\caption{Determine miss-distance $r$ and minimum points $\hat{U}_{xy}, O_{xy}$ between observer trajectory $J_{O}$ and Intruder trajectory $J_{\hat{U}}$}
\label{alg:miss_distance}
	\begin{algorithmic}[1]
		\Function{MinDistance}{$J_{O},J_{U}$}
		
		\LineComment{Difference between Observer and Intruder trajectory}
		\State $J_{OU} = J_{U} - J_{O}$
		
		\LineComment{Distance between each point on trajectories}		
		\State $r_{OU} = \sqrt{J_{OU_{x}}^{2} + J_{OU_{y}}^{2}}$		

		\LineComment{Minimum distance}		
		\State $r_{OU_{\text{min}}} = \text{min}(r_{OU})$
		
		\LineComment{Index of minimum distance}		
		\State $k = \{r_{OU_{n}}|n = r_{OU_{\text{min}}}\}$

		\State $J_{O_{\text{min}}} = J_{O_{xy}}(k)$

		\State $J_{U_{\text{min}}} = J_{U_{xy}}(k)$
		
		\State{\textbf{return} $r_{O\hat{U}_{\text{min}}},J_{O_{\text{min}}},J_{\hat{U}_{\text{min}}} $}
		\EndFunction
	\end{algorithmic}
\end{algorithm}

\begin{algorithm}[!t]
\caption{Estimating Probability of Conflict using Direct Monte Carlo}
\label{alg:pc_dmc}
	\begin{algorithmic}[1]
	\Function{PC\_DMC}{$f, t, A, O, \hat{U}, \hat{S}, N, r_{t}$}
		
		\State $D = 0$
		
		\LineComment{Propagate Observer for $t$ seconds}
		\State $J_{O} = $ \Call{SampleTrajectory}{$O$, $f$, $t$, $A$}		

		\For{$n = 1:N$}
			\LineComment{Draw sample}
			\State $U_{n} \sim{~}\mathcal{N}(\hat{U},\hat{S})$

			\LineComment{Propagate Intruder Samples for $t$ seconds}
			\State $J_{n} = $ \Call{SampleTrajectory}{$U_{n}$, $f, t, A$}

			\LineComment{Determine miss-distance between Observer and Sample Trajectories}
			\State $r_{n} = $ \Call{MinDistance}{$J_{O}, J_{n}$}

			\If {$r_{n} \leq r_{s}$}
					\State $D=D+1$
			\EndIf
		\EndFor
		
		\State $P_{c} = \frac{D}{N}$
		
		\State \textbf{return $P_{c}, D, U_{n}, J_{O}, J_{n}, r_{n}$}
	\EndFunction
	\end{algorithmic}
\end{algorithm}

The quantities of interest are the miss-distances $\{r_{n}^{(0)}: n = 1,...,100\}$. These are the minimum distances between the Intruder samples' trajectories $\{J_{U_{n}}^{(0)}: n = 1,...,100\}$ and the Observer trajectory $J_{O}$. Algorithm~\ref{alg:miss_distance} defines the procedure to determine the miss-distances between the Observer and Intruder trajectories. A conflict is projected to occur when there is a loss of minimum separation between any sample in set $J_{U_{n}}$ and the Observer trajectory $J_{O}$ at any instance. The set of miss-distances $r_{n}^{(0)}$ are sorted in descending order $\{B_{n}^{(0)}: n = 1,...,100\}$. The input samples $U_{n}^{(0)}$ are reordered $\tilde{U}_{n}^{(0)}$ to correspond to the sorted miss-distances $B_{n}^{(0)}$. To clarify, the sample $\tilde{U}_{1}^{(0)}$ produces a trajectory $J_{\tilde{U}_{1}}$ that has the largest miss-distance $B_{1}^{(0)}$ between itself the trajectory produced by the Observer $J_{O}$. The samples with lower miss-distances in the current level have a higher likelihood of generating conditional samples that satisfy the conflict condition than other samples in the current level. The vector of probability intervals $P_{n}^{(0)}$ are generated by

\begin{equation}	
	P_{n+1}^{(i)} = p_{0}^i\frac{N - n}{N}	\quad n = 0,...,(N-1)
	\label{eq:prob_intervals_mod}
\end{equation}

\noindent Note the range of $n$ in this equation is different to equation~\ref{eq:prob_intervals}. This is due to the maximum number of levels limit $m$. In the event that SS reaches the maximum number of levels without encountering conflicting samples the probability of conflict will be estimated $P_{c} = P_{N}^{(m-1)} = P_{100}^{(m-1)} = 0$ (the last probability interval in the $P_{n}^{(m-1)}$ vector that is generated by equation~\ref{eq:prob_intervals}) and this does not reflect the low magnitude of the probability. In contrast, the probability interval generated by equation~\ref{eq:prob_intervals_mod} allows the probability of conflict to be estimated $P_{c} < P_{100}^{(m-1)}$; $P_{100}^{(m-1)} = 1\e{-8}$. This information means that although no conflicting samples have been encountered despite exhausting all levels of SS the expected $P_{c}$ is estimated to be lower than $(p_{0})^{m}$, the lowest probability level realizable due to the maximum number of levels limit reached by SS. Such information is more useful than the estimate $P_{c} = 0$ evaluated by equation~\ref{eq:prob_intervals}. The level 0 CCDF is constructed by plotting the probabilities $P_{n}^{(0)}$ against $B_{n}^{(0)}$ as shown in Fig.~\ref{fig:headon_level0_seeds_CCDF}. No conflicting samples have been drawn in level 0 since no miss-distances satisfy the conflict condition. If the number of conflicting samples $D > N_{c}$ then the probability of conflict is estimated $P_{c} = P_{(N-D+1)}^{(i)}$. This also applies for the situation where the maximum number of levels has been reached $i = m - 1$ and some conflicts have been encountered where the number of conflicts encountered is less than or equal to $N_{c}$; ($N_{c} \geq D > 0$). 
The DMC method estimates the probability of conflict $P_{c} = \frac{D}{N}$ as defined by algorithm~\ref{alg:pc_dmc}.

\begin{algorithm}
\caption{Generate conditional samples using Metropolis Hastings}
\label{alg:conflict_samples_MH}
\begin{algorithmic}[1]
\Function{MH\_ConflictSamples}{$f$, $t$, $A$, $O$, $\hat{U}$, $\hat{S}$, $s_{j}$, $N_{s}$, $r_{t}$}
\State $\sigma_{r_{t}}^{2} = r_{t}^{2} I_{2 \times 2}$
\State $J_{O} = $ \Call{SampleTrajectory}{$O$, $f, t, A$}
	
	\For{\texttt{$j = 1:N_{c}$}}
		\State $U_{0} = s_{j}$ \textit{$\triangleright$Select seed sample}
		
		\LineComment{For each seed generate $N_{s}$ samples}
		\For{\texttt{$k = 0:N_{s}-1$}}
			
			\LineComment{Draw acceleration sample from mean}
			\State $a_{x}^{*} \sim{~} \mathcal{N}(0,1)$
			\State $a_{y}^{*} \sim{~} \mathcal{N}(0,1)$
			\State $g = [0, 0, a_{x}^{*},0, 0, a_{y}^{*}]^{T}$
			\LineComment{Generate Candidate sample $U^{*}$}
			\State $U^{*} = U_{k} + g$
			
			\LineComment{Propagate Samples for $t$ seconds} 
			\State $J_{U}^{*} = $ \Call{SampleTrajectory}{$U^{*}$, $f, t, A$}
			\State $J_{U_{k}} = $ \Call{SampleTrajectory}{$U_{k}$, $f, t, A$}

			\LineComment{Determine minimum miss-distance and $(x,y)$ coordinates of minimum points between Observer and Sample Trajectories}
			\State $[r_{k}, J_{O_{\text{min}}}, J_{U_{k_{\text{min}}}}] = $ \Call{MinDistance}{$J_{O}, J_{U_{k}}$}			
			\State $[r^{*}, J_{O_{\text{min}}}^{*}, J_{U_{\text{min}}}^{*}] = $ \Call{MinDistance}{$J_{O}, J_{U}^{*}$}
			
			\LineComment{Indicator function for miss-distance}
			\State $d = \left \{  \begin{array}{l l}
												1 & \text{if} \ r^{*} < r_{t}\\
												0 & \text{if} \ r^{*} \geq r_{t} \\
										  \end{array} \right .$			
			
			\LineComment{Calculate acceptance ratio}
			\State $\beta = \frac{p(J_{\hat{U}_{\text{min}}}^{*}|J_{O_{\text{min}}}^{*}, \sigma_{r_{t}}^{2})q(U^{*}|\hat{U}, \hat{S})}{p(J_{U_{k_{\text{min}}}}|J_{O_{\text{min}}}, \sigma_{r_{t}}^{2})q(U_{k}|\hat{U}, \hat{S})}d$
			
			\State $\alpha = \text{min} \left \{1,\beta \right \}$			
			\State $e \sim{~} [0,1]$
			
			\LineComment{Accept candidate sample, trajectory and miss-distance if $e<a$}
			\State $U_{k+1}^{(j)} = \left \{ \begin{array}{l l}
												U^{*} & \text{if} \ e < \alpha \\
												U_{k} & \text{if} \ e \geq \alpha \\
									  \end{array} \right .$			
			\State $J_{k+1}^{(j)} = \left \{ \begin{array}{l l}
												J^{*} & \text{if} \ e < \alpha \\
												J_{k} & \text{if} \ e \geq \alpha \\
										  \end{array} \right .$
			\State $r_{k+1}^{(j)} = \left \{ \begin{array}{l l}
												r^{*} & \text{if} \ e < \alpha \\
												r_{k} & \text{if} \ e \geq \alpha \\
										  \end{array} \right .$
			
		\EndFor	
	\EndFor
\State \textbf{return $U^{(j)}$, $J^{(j)}$, $r^{(j)}$}	
\EndFunction
\end{algorithmic}
\end{algorithm}

\begin{algorithm}[!t]
\caption{Estimate Probability of Conflict Using Subset Simulation}
\label{alg:SS_pc}
\begin{algorithmic}[1]
	\Function{PC\_SS}{$f$, $t$, $A$, $O$, $\hat{U}$, $\hat{S}$, $N$, $r_{t}$, $p_{0}$, $m$}
	
	\State $N_{c} = p_{0}N$
	\State $N_{s} = p_{0}^{-1}$
	
	\State $i = 0$ \textit{$\triangleright$Set current level}
	
	\LineComment{Direct Monte Carlo}
	\State $[D, U_{n}^{(i)}, r_{n}^{(i)}] = $ \Call{PC\_DMC}{$f, t, A, O, \hat{U}, \hat{S}, N, r_{t}$}
	
	\State $B_{n}^{(i)} \leftarrow r_{n}^{(i)}$ \textit{Sort distances in descending order}
	
	\State $\tilde{U}_{n}^{(i)} \leftarrow U_{n}^{(i)}$ \textit{Reorder the input samples to correspond to the sorted quantity of interest $B_{n}^{(i)}$}
	
	\LineComment{Generate probability intervals; equation~\ref{eq:prob_intervals_mod}}
		\For{\texttt{$n = 0:N-1$}}
			\State $P_{n+1}^{(i)} = p_{0}^i\frac{N - n}{N}$
		\EndFor

	\LineComment{CCDF: Concatenate vectors $P_{n}^{(i)}$, $B_{n}^{(i)}$ and samples $\tilde{U}_{n}^{(i)}$}
	\State $E_{n} = [P_{n}^{(i)},B_{n}^{(i)},\tilde{U}_{n}^{(i)}]$

  \While{$D < N_{c}$ \textbf{and} $i < m$}
		\State $i = i+1$
		
		\State $b_{i} = B_{N-N_{c}}^{(i-1)}$ \textit{$\triangleright$Set threshold}
		
		\LineComment{Set seeds using equation~\ref{eq:seeds}}
		\For{\texttt{$j = 1:N_{c}$}}
			\State $n = N-N_{c}+j$
			\State $s_{j}^{(i)} = \tilde{U}_{n}^{(i-1)}$
		\EndFor
		
		\LineComment{Metropolis Hastings to obtain conflicting samples}
		\State $[U_{n}^{(i)}, r_{n}^{(i)}] = $ \Call{MH\_ConflictSamples}{$f$, $t$, $A$, $O$, $\hat{U}$, $\hat{S}$, $s_{j}$, $N_{s}$, $b_{i}$}
	\State $B_{n}^{(i)} \leftarrow r_{n}^{(i)}$ \textit{Sort distances in descending order}
	
	\State $\tilde{U}_{n}^{(i)} \leftarrow U_{n}^{(i)}$ \textit{Reorder the input samples to correspond to the sorted quantity of interest $B_{n}^{(i)}$}
	
		\LineComment{Generate probability intervals; equation~\ref{eq:prob_intervals_mod}}
		\For{\texttt{$n = 0:N-1$}}
			\State $P_{n+1}^{(i)} = p_{0}^i\frac{N - n}{N}$
		\EndFor
	
		\LineComment{CCDF: Discard all rows after $E_{i(N-N_{c})}$}
		\LineComment{Concatenate $P_{n}^{(i)}$, $B_{n}^{(i)}$, $\tilde{U}_{n}^{(i)}$ and append to $E$}
		\For{\texttt{$n = 1:N$}}
			\State $E_{i(N-N_{c}+n)} = [P_{n}^{(i)}, B_{n}^{(i)}, \tilde{U}_{n}^{(i)}]$
		\EndFor
		
		\State $D = |B_{n}^{(i)} \leq r_{t}|$ \textit{$\triangleright$Number of conflicts $D$}
		
  \EndWhile
	
	\If {$D > 0$}
		\State $P_{c} = P^{(i)}_{(N-D+1)}$ 
	\Else
		\State $P_{c} = P^{(i)}_{N}$ \textit{$\triangleright$No conflicting samples were found select lowest probability interval}
	\EndIf

	\State \textbf{return $P_{c}$, $E$}
	\EndFunction
\end{algorithmic}
\end{algorithm}


\begin{figure*}[!t]
	\centering
	\subfloat[Sample Trajectories for levels 0 to 3]{%
	\includegraphics[width=\columnwidth]{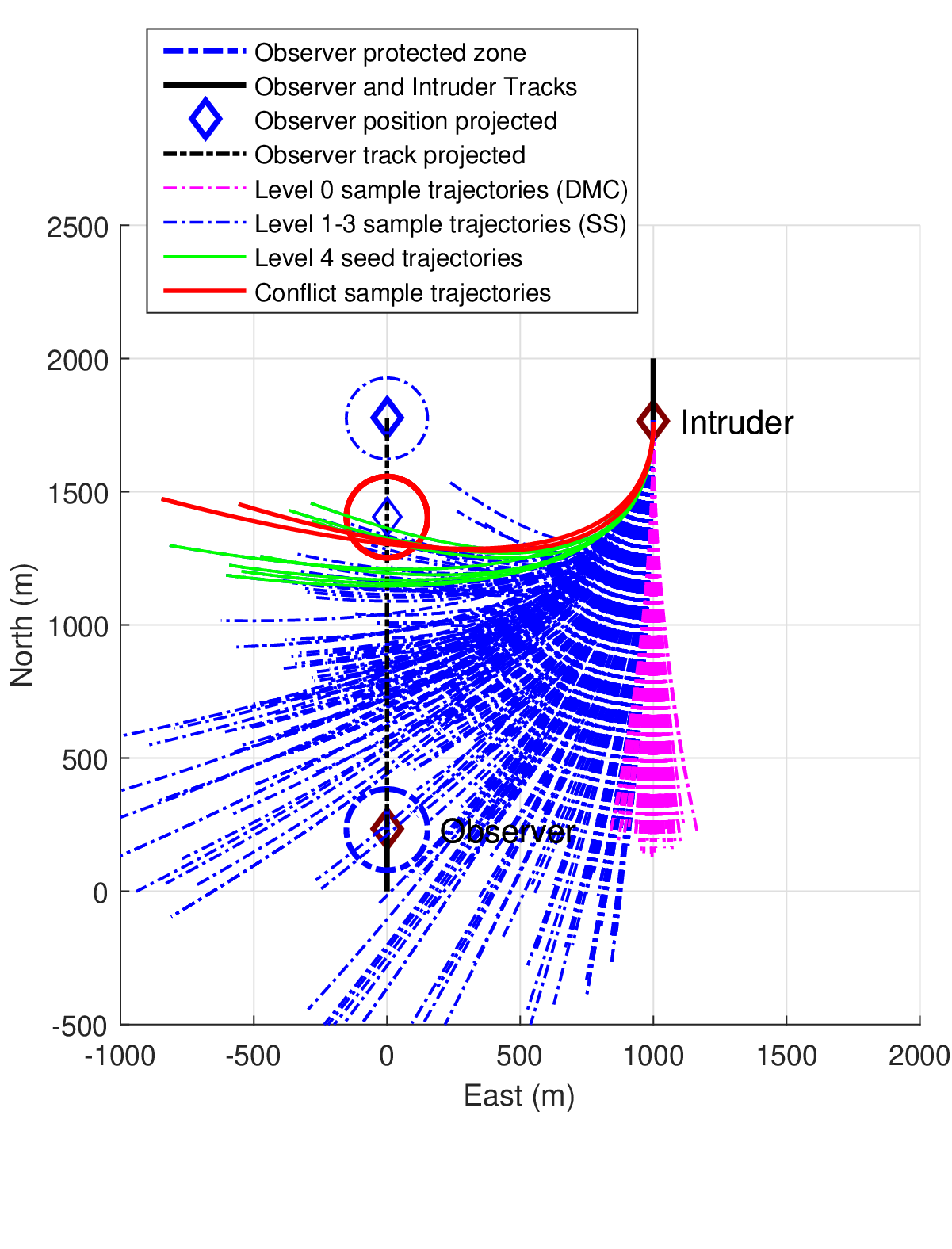}
	\label{fig:headon_level3_conflict}}
	\hfill
	\subfloat[Samples Trajectories for levels 0 to 4]{%
	\includegraphics[width=\columnwidth]{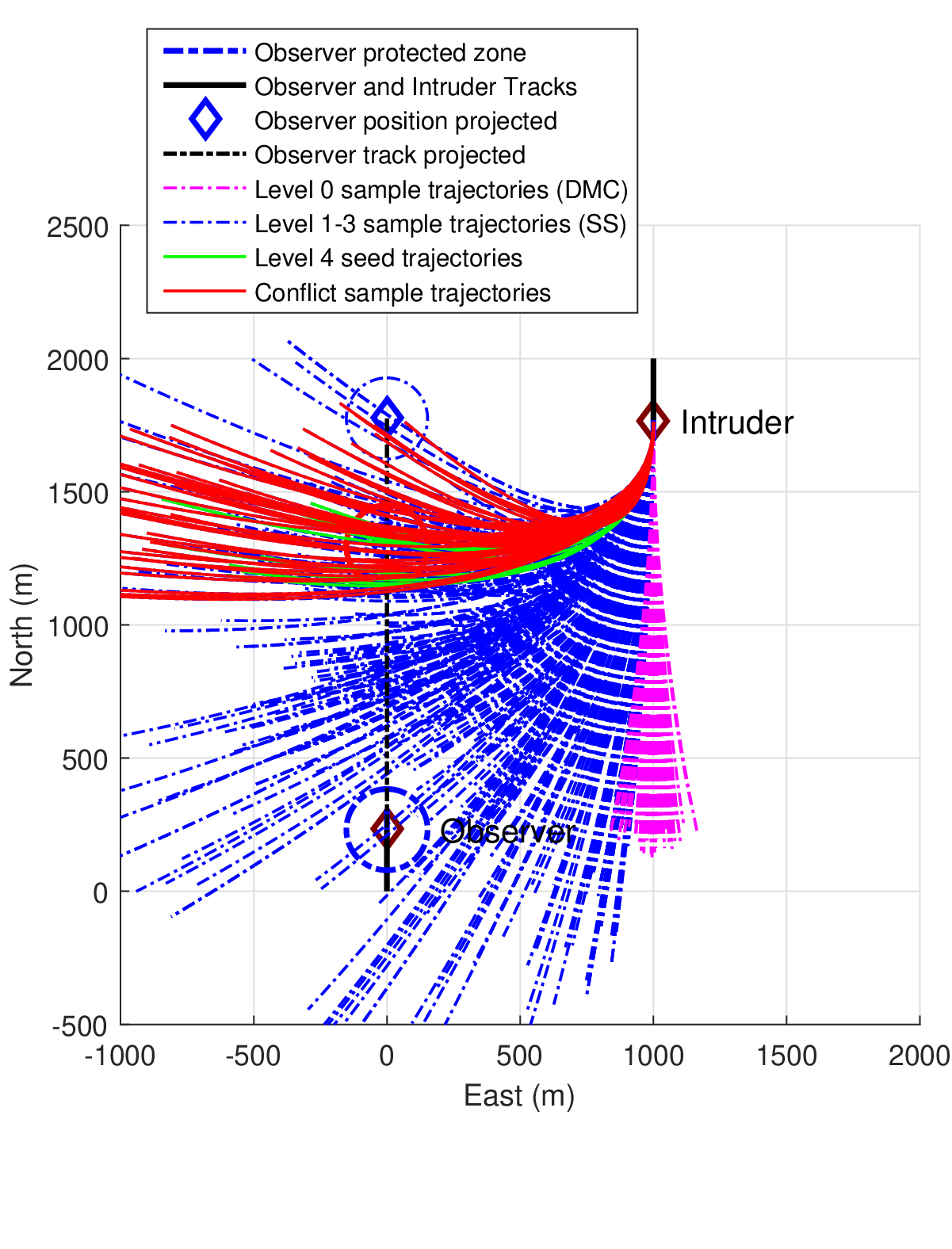}
	\label{fig:headon_level4_conflict}}
	\hfill
	\subfloat[Level 4 CCDF]{%
	\includegraphics[width=\columnwidth]{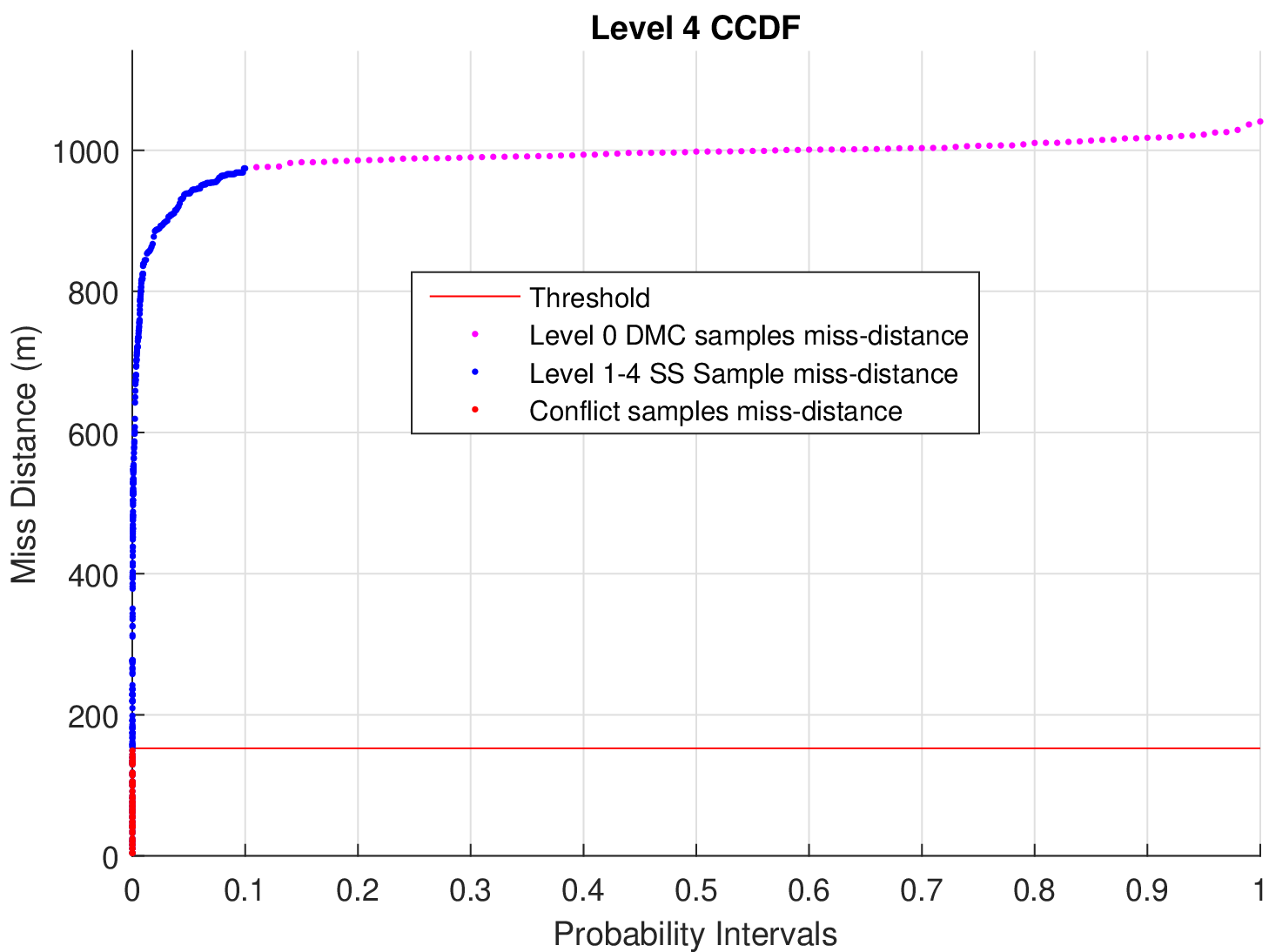}
	\label{fig:headon_level4_conflict_CCDF}}
	\hfill		
	\subfloat[Level 4 CCDF]{%
	\includegraphics[width=\columnwidth]{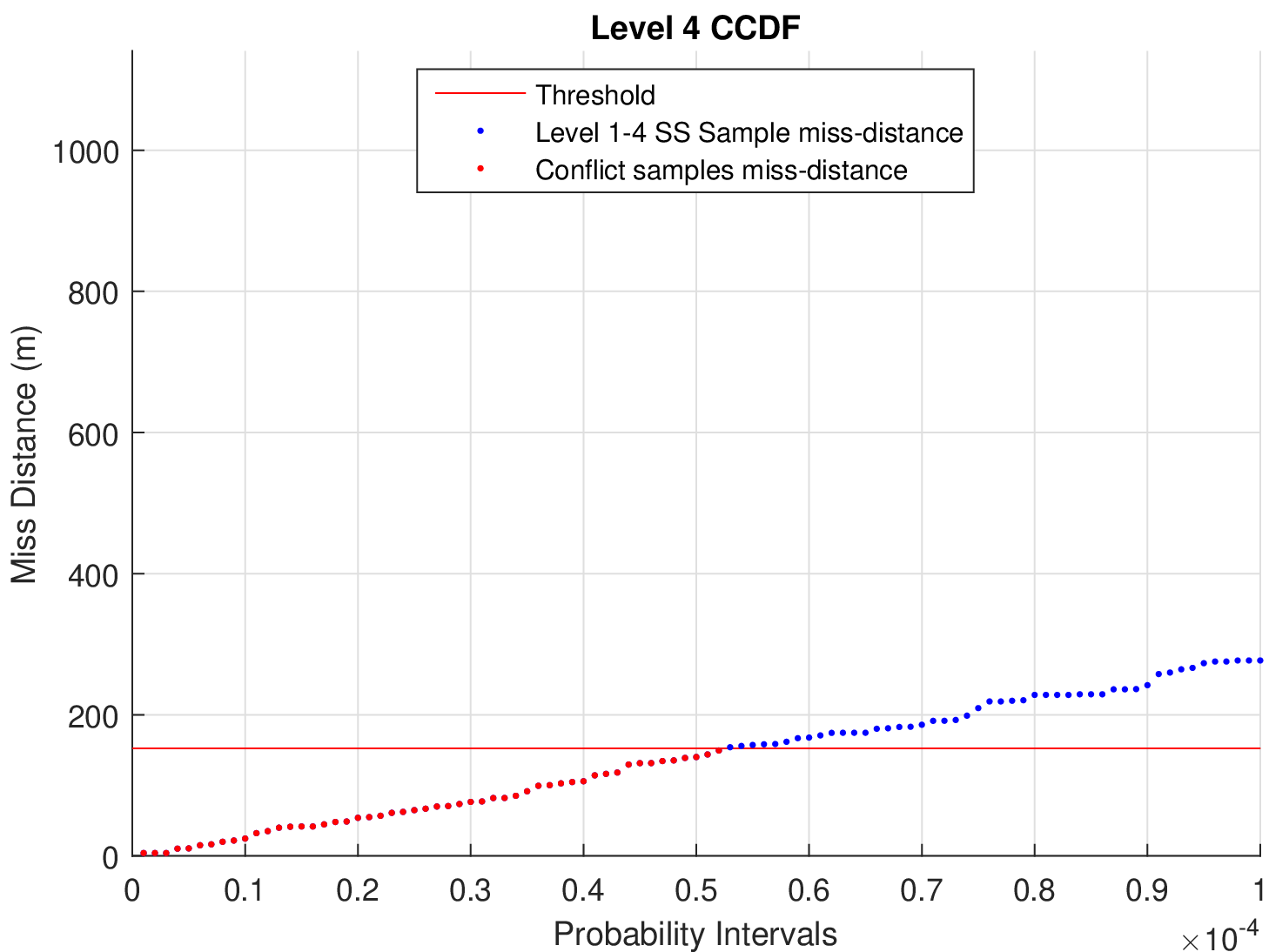}
	\label{fig:headon_level4_conflict_CCDF_zoom}}
	\caption{The above figures show trajectories of conditional samples generated as the simulation continues to higher levels. Subset Simulation continues until the number of conflicting samples $D$ found in a level is greater than $N_{c}$ within a level as shown in Fig.~\ref{fig:headon_level4_conflict}. The probability of conflict is estimated as $P_{c}(K+1) = 0.52\e{-4}$ as shown in Fig.~\ref{fig:headon_level4_conflict_CCDF_zoom}.}
	\label{fig:headon_level3_4}
\end{figure*}

However if the condition $D > N_{c}$ is not satisfied and $i < m-1$; SS proceeds to the next level $(i > 0)$ and continues until the condition is satisfied or if the maximum number of levels is reached. This is because the conflict region of the pdf is not represented accurately enough due to the lack of sufficient samples representing the conflict region in the current level. Therefore it is necessary generate more conditional samples at higher levels of SS to progress towards representing the conflict region of the pdf more accurately.

The following subset levels $(i > 0)$ generate $N$ conditional Intruder samples $\tilde{U}_{n}^{(i)}$ using the Metropolis Hastings method as defined in algorithm~\ref{alg:conflict_samples_MH}. The set of seeds $s_{j}^{(i)}$ required to generate the samples are selected from samples in the previous level using 

\begin {equation}
	s_{j}^{(i)} = \tilde{U}_{n}^{(i-1)}
\label{eq:seeds_app}
\end{equation}

\noindent where $1 \leq j \leq N_{c}$, $(N-N_{c}+1) \leq n \leq N$ and $i > 0$.

Fig.~\ref{fig:headon_level0_seeds} highlights the trajectories of level 0 samples selected as seeds to generate level 1 conditional samples. Fig.~\ref{fig:headon_level1} shows the trajectories of the conditional samples generated in level 1.
The set $s_{j}^{(i)}$ contains $N_c$ seeds; one for each chain. Each chain generates $N_s$ samples. This maintains the total number of samples as $N$ for each level. The MH method uses an indicator $d$ (as shown in algorithm~\ref{alg:conflict_samples_MH}) to ensure the miss-distance $r^{(i)*}$ between the Observer's trajectory $J_{O}$ and Intruder trajectory $J^{(i)*}$ of the proposed sample $U^{(i)*}$ is less than the intermediate threshold $b_{i}$ set by equation~\ref{eq:thresholds}. If $r^{(i)*} > b_{i}$ then the proposed sample is rejected and the current sample of the Intruder is maintained.

The miss-distances $\{r_{n}^{(1)}: n = 1,...,100\}$ of the conditional samples $U_{n}^{(1)}$ generated in level 1 are determined and sorted in descending order $B_{n}^{(1)}$ using the same method as level 0. The input samples $U_{n}^{(1)}$ are reordered $\tilde{U}_{n}^{(1)}$ to correspond to the sorted miss-distances $B_{n}^{(1)}$. The probability intervals $P_{n}^{(1)}$ for the current level are generated and plotted against $B_{n}^{(1)}$ to construct a CCDF. Fig.~\ref{fig:headon_level1_CCDF} shows the CCDF generated up to level 1. Note the miss-distances of the samples used as seeds from the previous level 0 (that are highlighted in Fig.~\ref{fig:headon_level0_seeds_CCDF}) are discarded and replaced with the miss-distances of the conditional samples generated in level 1. This illustrates that the samples used as seeds are discarded and replaced with the conditional samples generated in the current level. This process is repeated as SS progresses to higher levels until the condition $D > N_{c}$ is satisfied or the maximum number of levels is reached as defined in algorithm~\ref{alg:SS_pc}. Fig.~\ref{fig:headon_level3_conflict} shows the trajectories of the conflicting samples encountered in level 3. However the condition $D > N_{c}$ had not been satisfied. This required SS to proceed to level 4 and generate conditional samples that satisfy the condition $D > N_{c}$ as shown in Fig.~\ref{fig:headon_level4_conflict}. The CCDF generated up to level 4 is shown in Fig.~\ref{fig:headon_level4_conflict_CCDF}. The CCDF is used to estimate the $P_{c}(K+1) = 0.52\e{-4}$ as shown in Fig.~\ref{fig:headon_level4_conflict_CCDF_zoom}.
This process is repeated through out the duration of the simulation to determine the probability of conflict for each time-step using samples from the prediction of the Intruder's estimate $\hat{U}(K+1)$ and covariance $\hat{S}(K+1)$.

\begin{algorithm}[!t]
\caption{Determine Probability of Conflict using SS and DMC}
\label{alg:pc_ss_dmc}
\begin{algorithmic}[1]
\State $O(0)$ \textit{$\triangleright$Initialize Observer}
\State $U(0)$ \textit{$\triangleright$Initialize Intruder}
\State $\hat{U}(0)$ \textit{$\triangleright$Initialize Intruder Estimate}
\State $\hat{S}(0)$ \textit{$\triangleright$Initialize Intruder Covariance}

\State $M_{c} = 0$ \textit{$\triangleright$Measurement counter}

	\For{\texttt{$K = 0: tf$}}
	
		\State $O(K+1) = AO(K)$ \textit{$\triangleright$Propagate Observer}
		
		\State $U(K+1) = AU(K)$ \textit{$\triangleright$Propagate Intruder}
		
		\State $M_{Z} = false$ \textit{$\triangleright$Flag to indicate new measurement}
		\If {$M_{c} = \frac{f}{f_{M}}$} \textit{$\triangleright$Conduct Intruder position measurement}
			\State $Z = HU(K+1) + [w_{x},w_{y}]^{T}$
				\State $M_{Z} = \text{true}$ \textit{$\triangleright$Set flag to indicate that new measurement is available for Kalman filter Update}
				\State $M_{c} = 0$ \textit{$\triangleright$ Reset measurement counter}
		\EndIf
		\State $M_{c} = M_{c} + 1$ \textit{$\triangleright$Increment measurement counter}
		
		\LineComment{Predict/Update estimate of Intruder with Kalman filter}
		\State $[\hat{U}(K+1), \hat{S}(K+1)] = $ \Call{KF}{$\hat{U}(K), \hat{S}(K), Z, H, Q, R$, $M_{Z}$}
		
		\LineComment{Estimate Probability of Conflict using Subset Simulation}
		
		\State $P_{c}^{(\text{SS})}(K+1) = $
		
		\Call{PC\_SS}{$f$, $t$, $A$, $O$, $\hat{U}(K+1)$, $\hat{S}(K+1)$, $N$, $r_{t}$, $p_{0}$, $m$}
		\LineComment{Estimate Probability of Conflict using Direct Monte Carlo}
		
		\State $P_{c}^{(\text{DMC})}(K+1) = $
		
		\Call{PC\_DMC}{$f, t, A, O, \hat{U}(K+1), \hat{S}(K+1), N, r_{t}$}
	\EndFor
\end{algorithmic}
\end{algorithm}

\section{Results}
\label{sec:results}

The Subset Simulation method has been tested and compared with the Direct Monte Carlo (DMC) method to estimate the probability of conflict $P_{c}$ between the Observer and Intruder by simulating the scenarios shown in Fig.~\ref{fig:potential_conflict_scenarios}. The Observer and Intruder were modeled as points with nearly constant velocity in a geometric configuration based on the three different types of conflict shown in Fig.~\ref{fig:conflict_types}. The $P_{c}$ metric was estimated as an average of 50 Monte Carlo simulation during the Head-on and Overtaking conflicts as shown in figures~\ref{fig:potential_headon} and~\ref{fig:intruderovertaking_potential} respectively. The tests were repeated with varying lateral separations $L_{a} = \{0, 100, 152, 500, 1000, 1100\} \text{m}$. 

The following Subset Simulation parameters were used for all scenarios: $N = 100$; Level probability: $p_{0} = 0.1$;  $N_c = p_{0}N = 10$; $N_s = \frac{1}{p_{0}} = 10$; $m = 7$; Observer minimum separation threshold $r_{t} = 500 \text{ft} = 152.4 \text{m}$. Algorithm~\ref{alg:pc_ss_dmc} defines the simulation conducted.

The number of samples used for each level of SS remain constant. However the number of levels required at a given time-step vary depending on the magnitude of $P_{c}$. Therefore the total number of samples $N_{T}$ required to realize a conflict at a given time-step varies as a function of time-step. In the interest of a fair comparison of the computational effort between the two methods, an equal number of samples are evaluated for both methods. The estimation using DMC is conducted with $N_{T}$ samples, where $N_{T}$ is the number of samples that are used in the SS method at the same time-step. To clarify, if the SS method reaches level $i$ = 4 to satisfy the conflict condition for estimating the $P_{c}^{(\text{SS})}(K)$ at time-step $K$, then $N_{T} = 100 \times 5 = 500$ samples have been used by the SS method. Therefore DMC estimates the $P_{c}^{(\text{DMC})}(K)$ for the same time-step with 500 samples only.




\begin{figure*}
	\centering	
	\subfloat[$P_{c}$ during Head-on conflict with 0 m Lateral separation]{%
	\includegraphics[trim={0 0cm 0 0cm},clip,width=0.66\columnwidth]{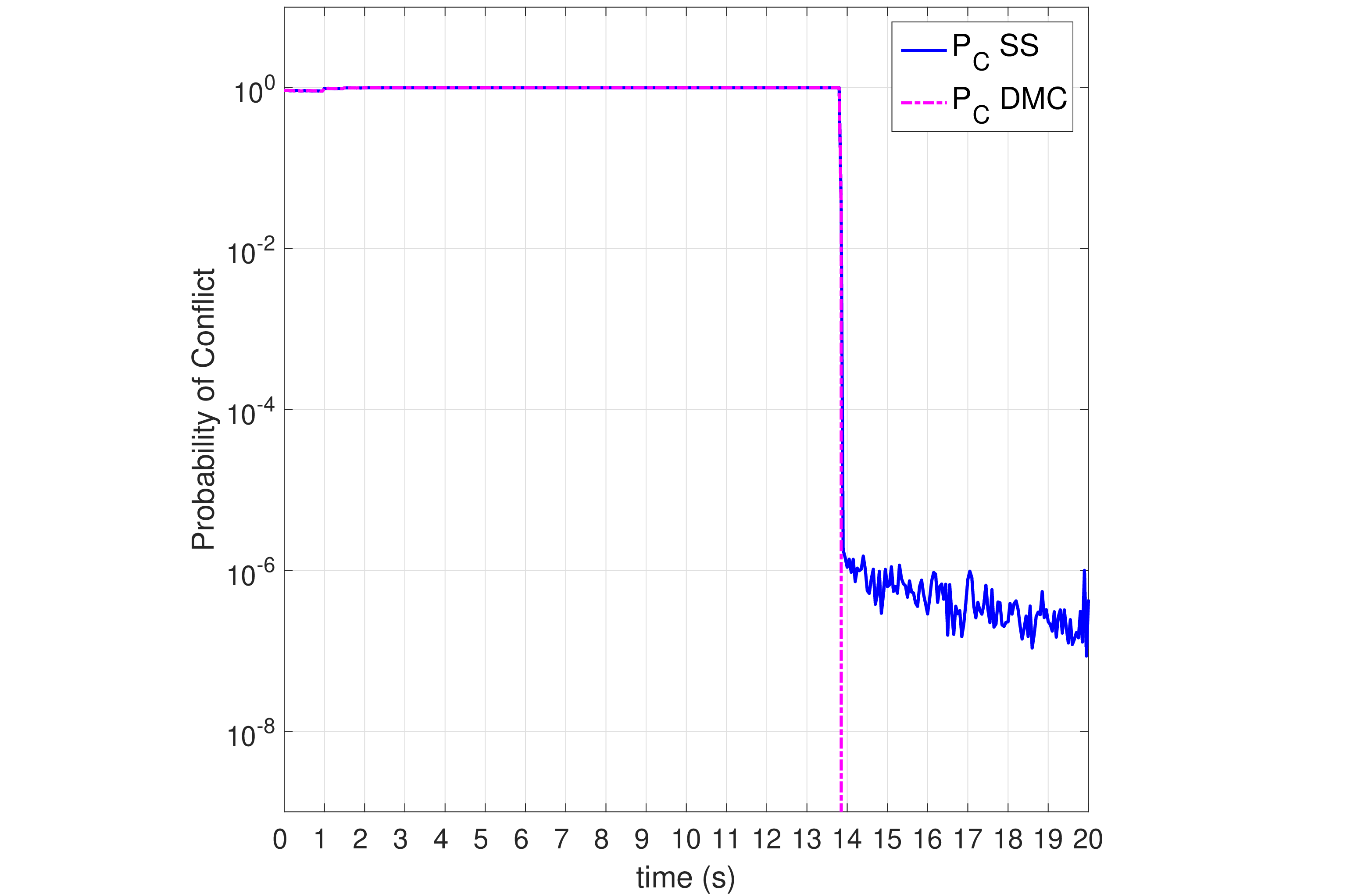}
	\label{fig:headonpass_SS_DMC_latsep_0}}
	\hfill 
	\subfloat[$P_{c}$ during Head-on conflict with 100m Lateral separation]{%
	\includegraphics[trim={0 0cm 0 0cm},clip,width=0.66\columnwidth]{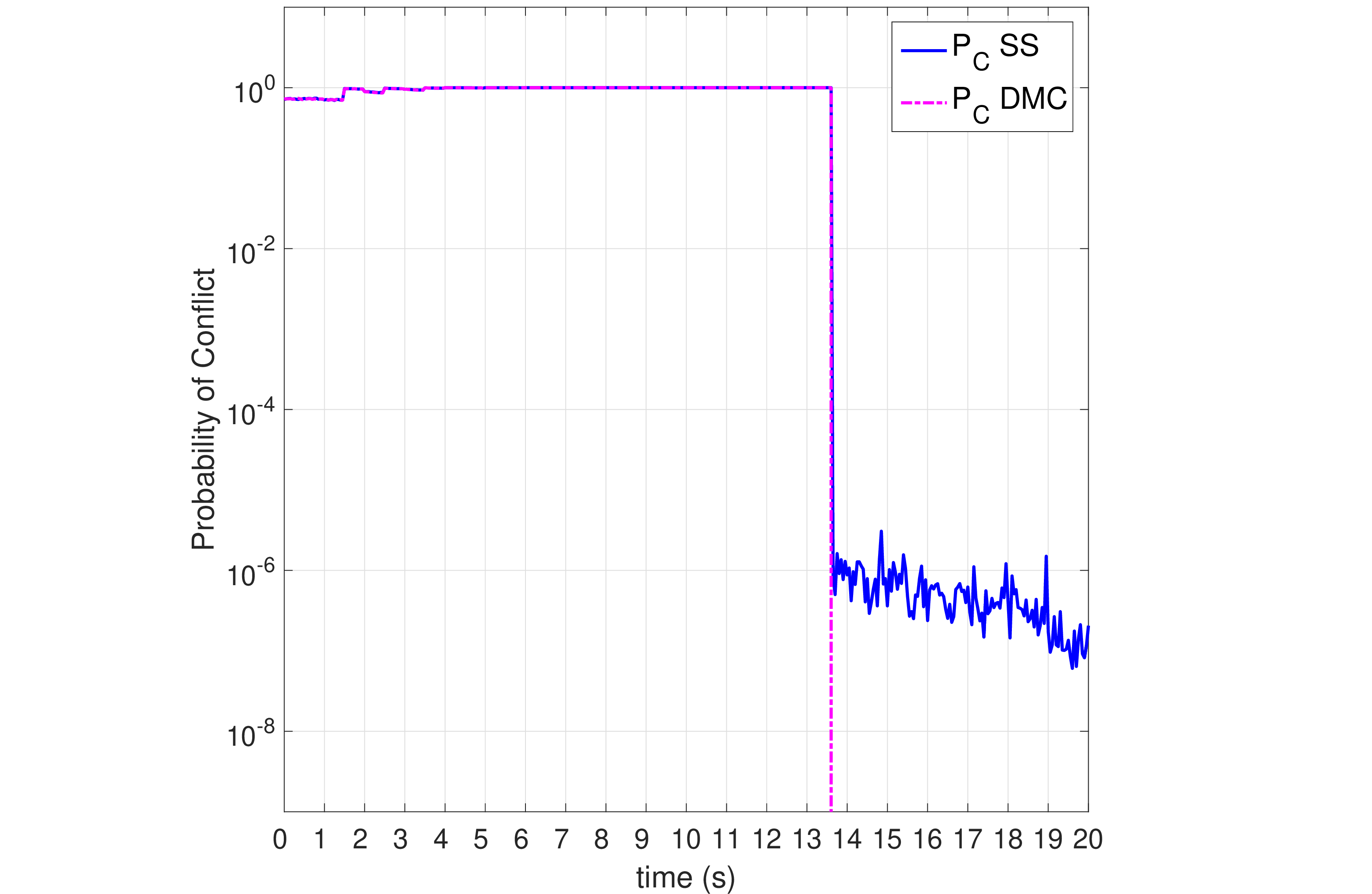}
	\label{fig:headonpass_SS_DMC_latsep_100}}	
	\hfill 
	\subfloat[$P_{c}$ during Head-on conflict with 152.4m Lateral separation]{%
	\includegraphics[trim={0 0cm 0 0cm},clip,width=0.66\columnwidth]{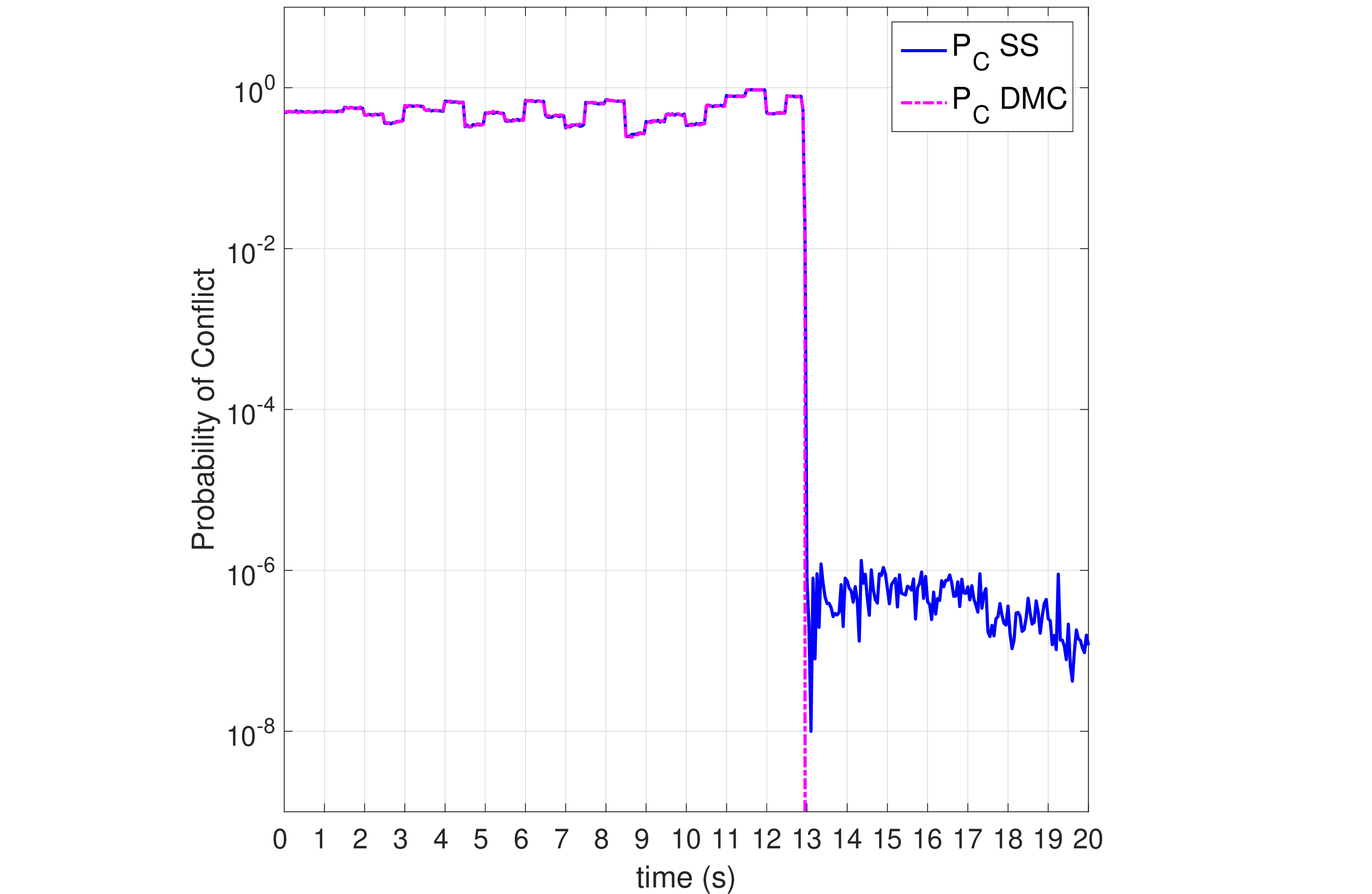}
	\label{fig:headonpass_SS_DMC_latsep_152}}
	\hfill
	\subfloat[$P_{c}$ during Head-on pass with 500m Lateral separation]{%
	\includegraphics[trim={0 0cm 0 0cm},clip,width=0.66\columnwidth]{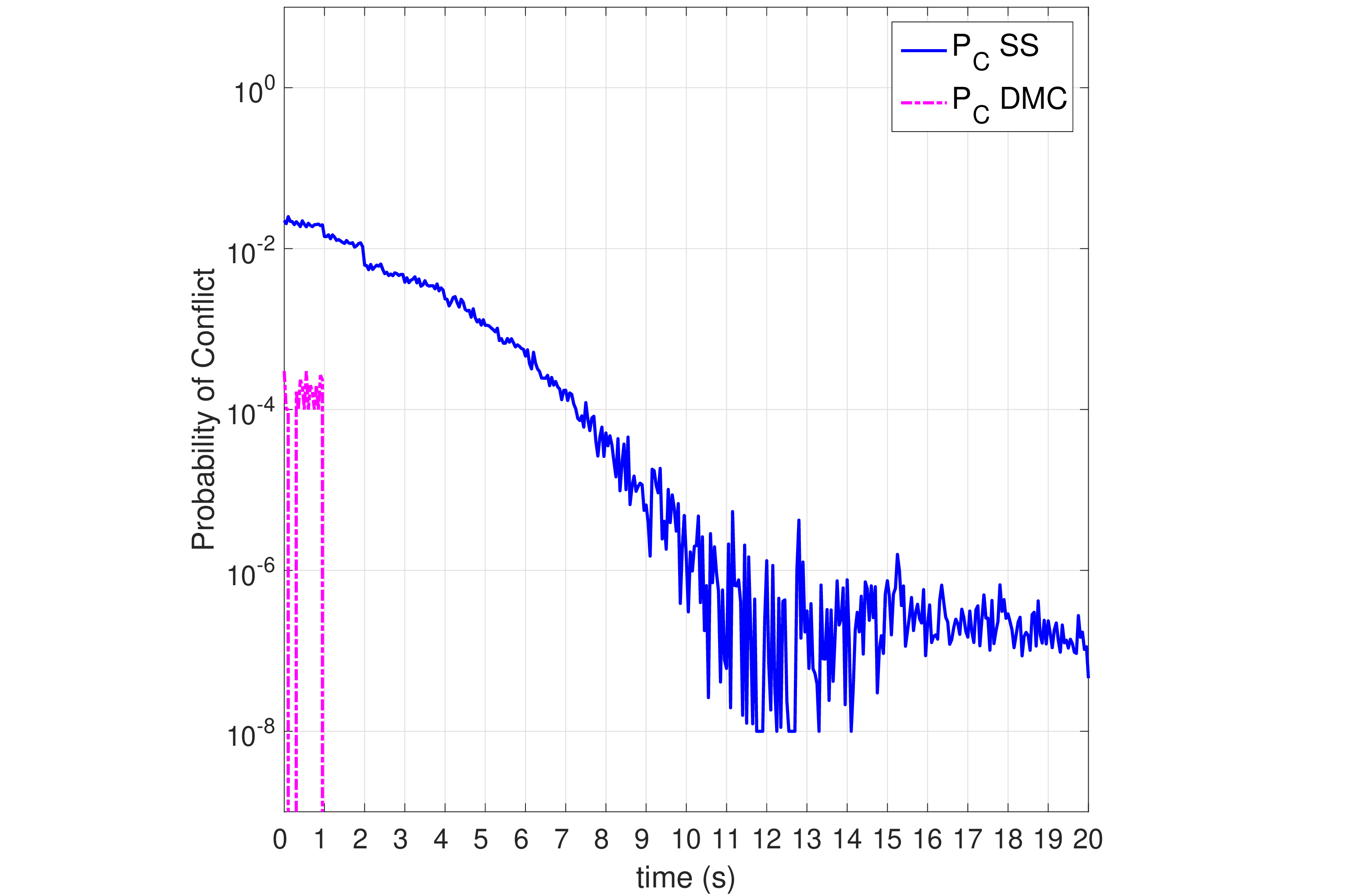}
	\label{fig:headonpass_SS_DMC_latsep_500}}
	\hfill
	\subfloat[$P_{c}$ during Head-on pass with 1000m Lateral separation]{%
	\includegraphics[trim={0 0cm 0 0cm},clip,width=0.66\columnwidth]{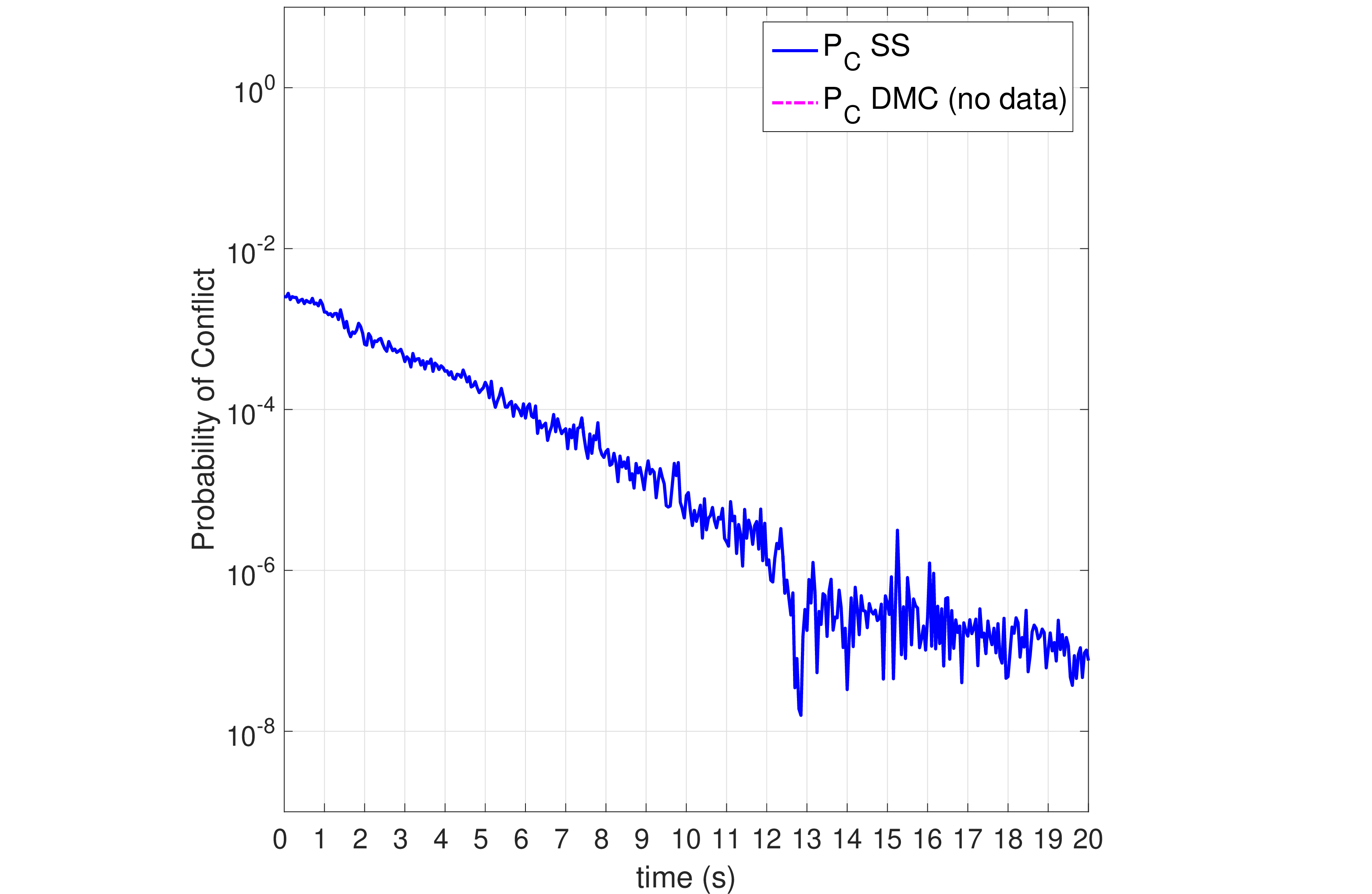}
	\label{fig:headonpass_SS_DMC_latsep_1000}}	
	\hfill
	\subfloat[$P_{c}$ during Head-on pass with 1100m Lateral separation]{%
	\includegraphics[trim={0 0cm 0 0cm},clip,width=0.66\columnwidth]{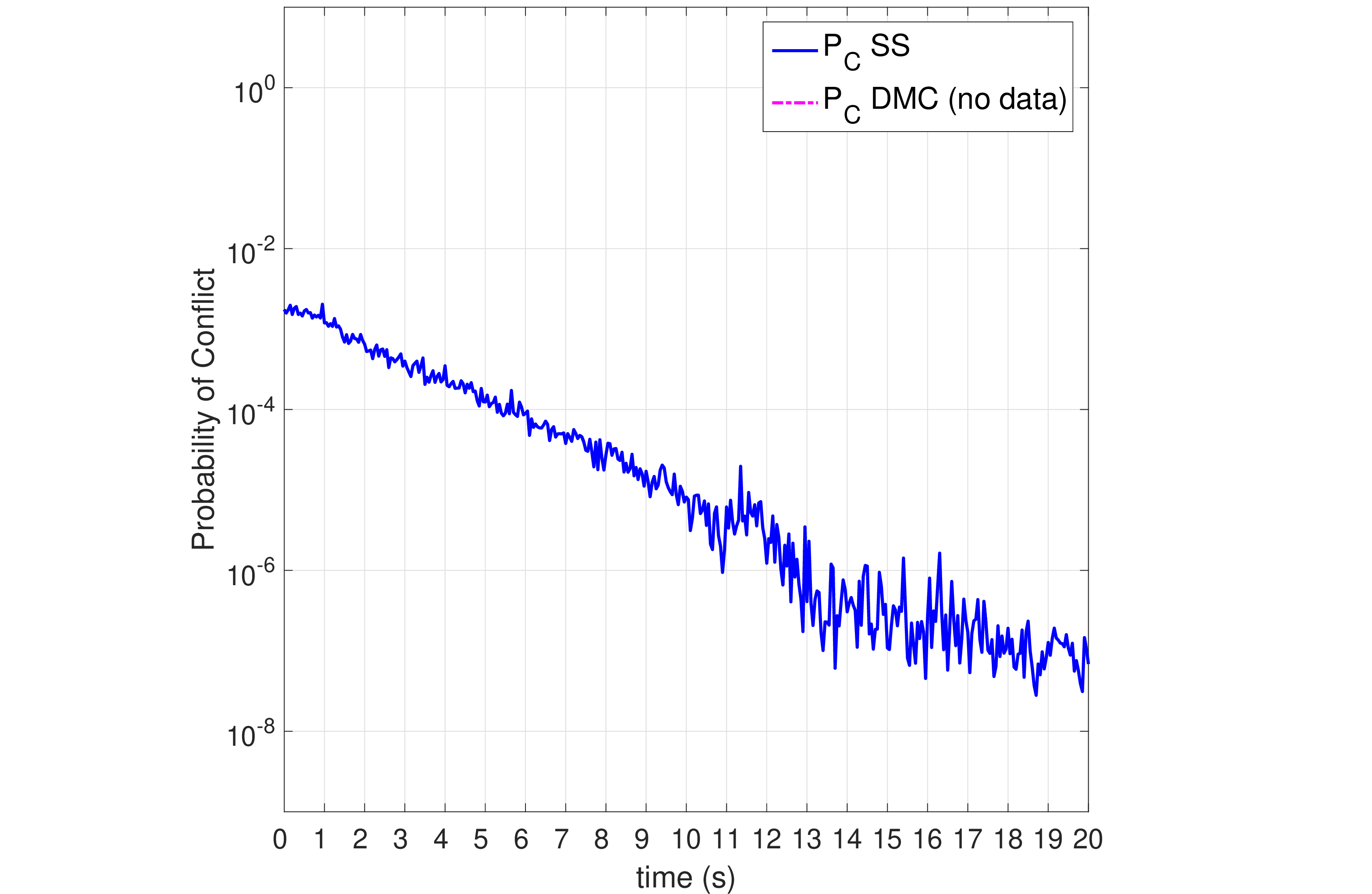}
	\label{fig:headonpass_SS_DMC_latsep_1100}}
	\caption{The estimated $P_{c}$ using the Subset Simulation and Direct Monte Carlo methods during the Head-on pass as shown in Fig.~\ref{fig:potential_headon} with varying lateral separation $L_{a}=\{0,100,152,500,1000,1100\}\text{m}$.
	}
	\label{fig:headon_set1}	
\end{figure*}

\subsection{Estimation of $P_{c}$ for Head-on Pass scenario}

The Intruder and Observer parameters used for the Head-on pass scenario are as follows: The Intruder and Observer maintain a constant speed of 150 knots (77.17ms$^{-1}$). The Observer maintains a constant heading of $0^{\circ}$; the Intruder maintains a constant heading of $180^{\circ}$. The Observer's minimum separation threshold is $r_{t} = 500 \text{ft} = 152.4 \text{m}$. The Longitudinal separation is $L_{o} = 2000 \text{m}$

Figures~\ref{fig:headonpass_SS_DMC_latsep_0},~\ref{fig:headonpass_SS_DMC_latsep_100} and~\ref{fig:headonpass_SS_DMC_latsep_152} show the estimation of $P_{c}$ for the Head-on pass scenario using SS and DMC methods with lateral separations of 0m, 100m and 152m respectively. The scenarios are conflicting because the geometric configuration and initial conditions of both the Observer and Intruder are conflicting and remain as such throughout the duration of the simulation. When $t \leq 12\text{s}$ the Intruder and Observer are approaching each other the estimated $P_{c}$ increases. This is as expected because a conflict is imminent. Both estimation methods show approximately the same $P_{c}$ as expected, since the first level of the SS method is DMC sampling. At this stage the conflict region of the pdf is large and the probability of drawing a sample which leads to a conflict is high. The conflict occurs at $t \approx 12.5$s due to the loss of separation between the Observer and Intruder. Fig.~\ref{fig:headonpass_SS_DMC_latsep_152} shows the estimation of $P_{c}$ with lateral separation $L_{a} = 152.4m = r_{t}$. This is a conflicting scenario since the Intruder skims Observer's protected boundary at $t \approx 12.5 \text{s}$ as the Observer and Intruder pass each other. The oscillations during $t \leq 12s$ are due to $L_{a} = r_{t}$. This is a borderline situation. 

The Intruder and Observer pass each other at $t \approx 13$s. The $P_{c}$ estimated by both methods is still $1$ until $t > 14$s where the Intruder has exited the Observer's protected zone. At this stage the Observer and Intruder have receding relative velocities and are moving away from each other. $P_{c}$ is expected to reduce at this stage as shown in the log-$y$ plot. The conflict region of the pdf reduces since both Intruder and Observer are moving away from each other. The SS method estimates the $P_{c}$ as being close to zero at an order of magnitude of $10^{-7}$. The lowest probability which can be realized is $P_{c} = 10^{-8}$. This is due to a maximum level restriction imposed in the simulation. In such instances the probability of conflict can be considered to be less than the order of $10^{-8}$. At this stage the DMC method draws the same number of samples as SS but is unable to find conflicting samples and estimates $P_{c} = 0$. This is because the region of conflict within the pdf has reduced and the probability of drawing a conflicting sample is rare. This requires the DMC method to draw and evaluate a larger number of samples at this stage before a conflicting sample is drawn from the rare region of conflict within the pdf. The SS method is able to obtain the conflicting samples from the rare region of the pdf by generating samples conditionally in such a way that the samples satisfy the intermediate thresholds leading to the rare region using the MH method. Each subset level corresponds to an intermediate threshold.  This progressive feature of the SS method allows a more efficient approach to reach the rare `tail' region of the pdf.

As the lateral separation of the scenario is increased, the expected $P_{c}$ decreases. The scenario is simulated with a lateral separation of 500m, 1000m and 1100m as shown in figures~\ref{fig:headonpass_SS_DMC_latsep_500},~\ref{fig:headonpass_SS_DMC_latsep_1000} and~\ref{fig:headonpass_SS_DMC_latsep_1100} respectively. These are non-conflicting scenarios. The figures show abrupt variations in $P_{c}$. These are caused by the Monte Carlo nature of our algorithm. Note that, since the sampling frequency is high relative to the thickness of the line in the figure, the variations in $P_{c}$ are particularly readily perceived. The conflict region of the pdf is smaller than the previous scenarios. The SS estimation method is able to estimate low $P_{c}$ throughout the duration of the simulation, whereas with an equivalent number of samples the DMC method is unable to find conflicting or near conflicting samples of the Intruder in most instances. Fig.~\ref{fig:headonpass_SS_DMC_latsep_500} shows abrupt variations in the $P_{c}$ estimated by the DMC method when $t < 1\text{s}$ where the estimate tends to zero. These are instances where the DMC method is unable for find any conflicting samples and estimates $P_{c} = 0$.

\begin{figure}
	\centering	
	\subfloat[Head-on conflict scenario with 1000m Lateral separation before head-on pass.]{%
	\includegraphics[width=\columnwidth]{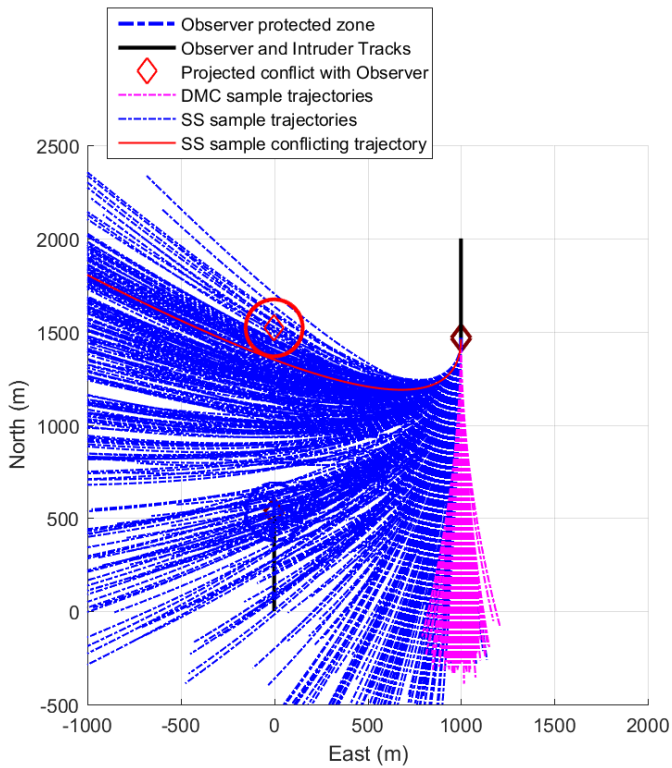}
	\label{fig:headonpass_SS_DMC_latsep_1000_before}}
	\\
	\subfloat[Head-on conflict scenario with 1000m Lateral separation after pass.]{%
	\includegraphics[width=\columnwidth]{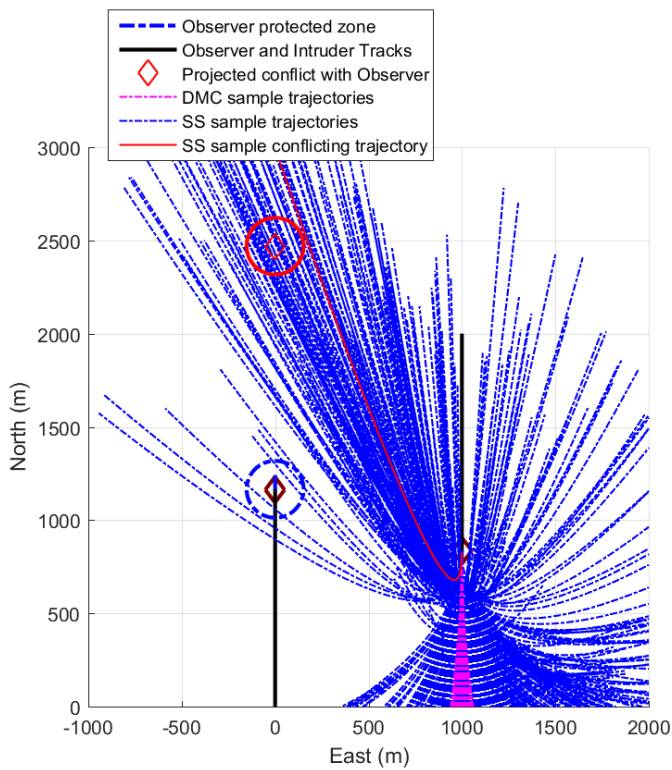}
	\label{fig:headonpass_SS_DMC_latsep_1000_after}}	
	\caption{SS and DMC trajectories for Head-on pass with lateral separation 1000m}
	\label{fig:headonpass_1000m_snapshot}	
\end{figure}

Figures~\ref{fig:headonpass_SS_DMC_latsep_1000_before} and~\ref{fig:headonpass_SS_DMC_latsep_1000_after} show the trajectories of the samples evaluated by SS and DMC methods at an instance before and after the Intruder and Observer pass each other respectively. The progressive nature of the SS method can be observed as a concentration of trajectories leading to the conflict trajectory. In contrast the DMC method has drawn the same number of samples (most are overlapping) without realizing any conflicts.

\subsection{Estimation of $P_{c}$ for Intruder Overtaking Observer}

The scenario parameters used are as follows: The Intruder speed is $300 \text{knots} = 154.3 \text{ms}^{-1}$ and the Observer speed is $150 \text{knots} = 77.17 \text{ms}^{-1}$. Both Intruder and Observer maintain a constant heading of a constant heading of $180^{\circ}$. The longitudinal distance $L_{o}$ between the Intruder and Observer is $L_{o} = 1000\text{m}$.

\begin{figure*}
	\centering
	\subfloat[$P_{c}$ during Intruder overtaking Observer conflict with 0m Lateral separation]{%
	\includegraphics[trim={0 0cm 0 0cm},clip,width=0.66\columnwidth]{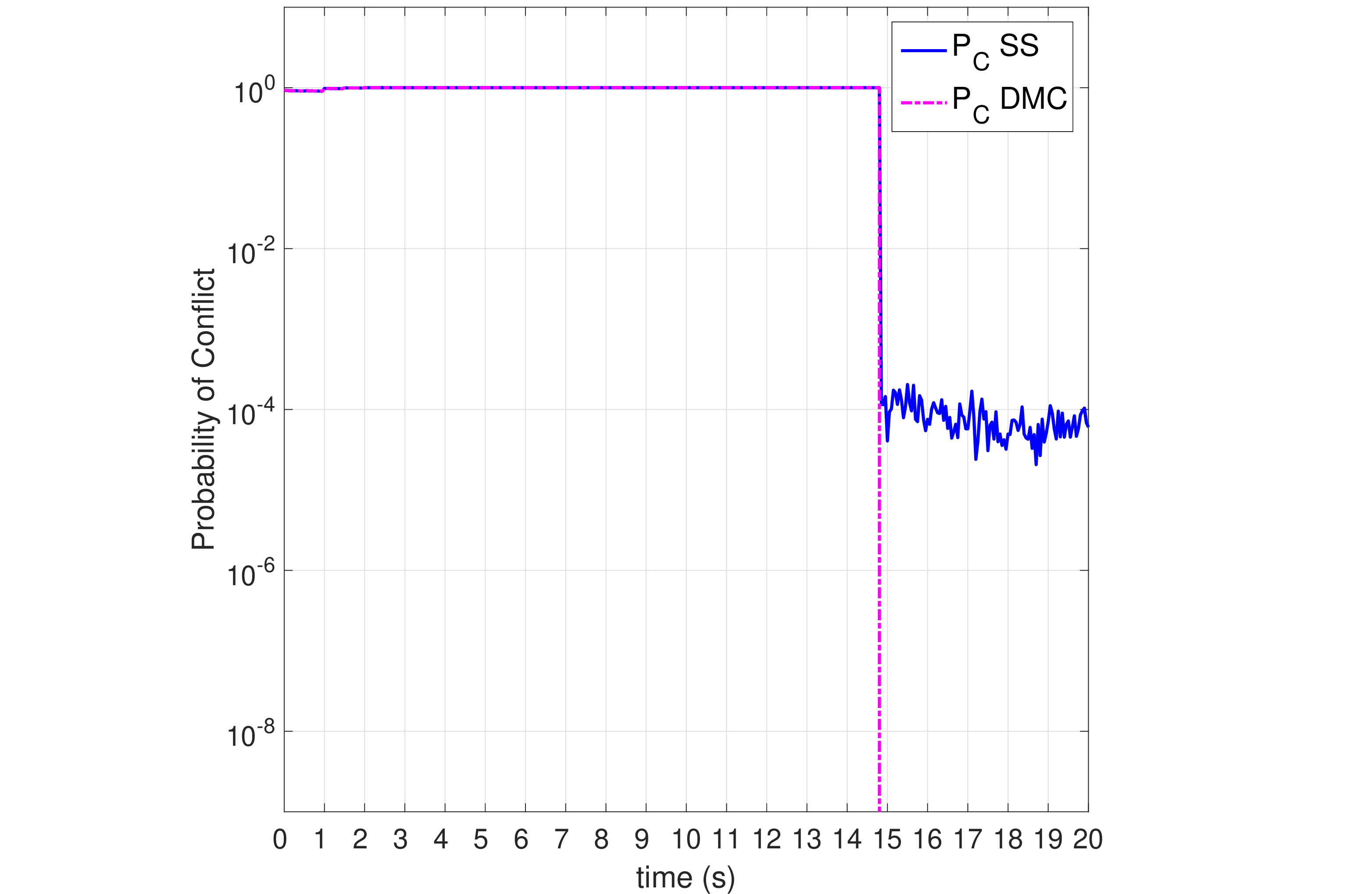}
	\label{fig:intruderOvertaking_0}}	
	\hfill
	\subfloat[$P_{c}$ during Intruder overtaking Observer conflict with 100m Lateral separation]{%
	\includegraphics[trim={0 0cm 0 0cm},clip,width=0.66\columnwidth]{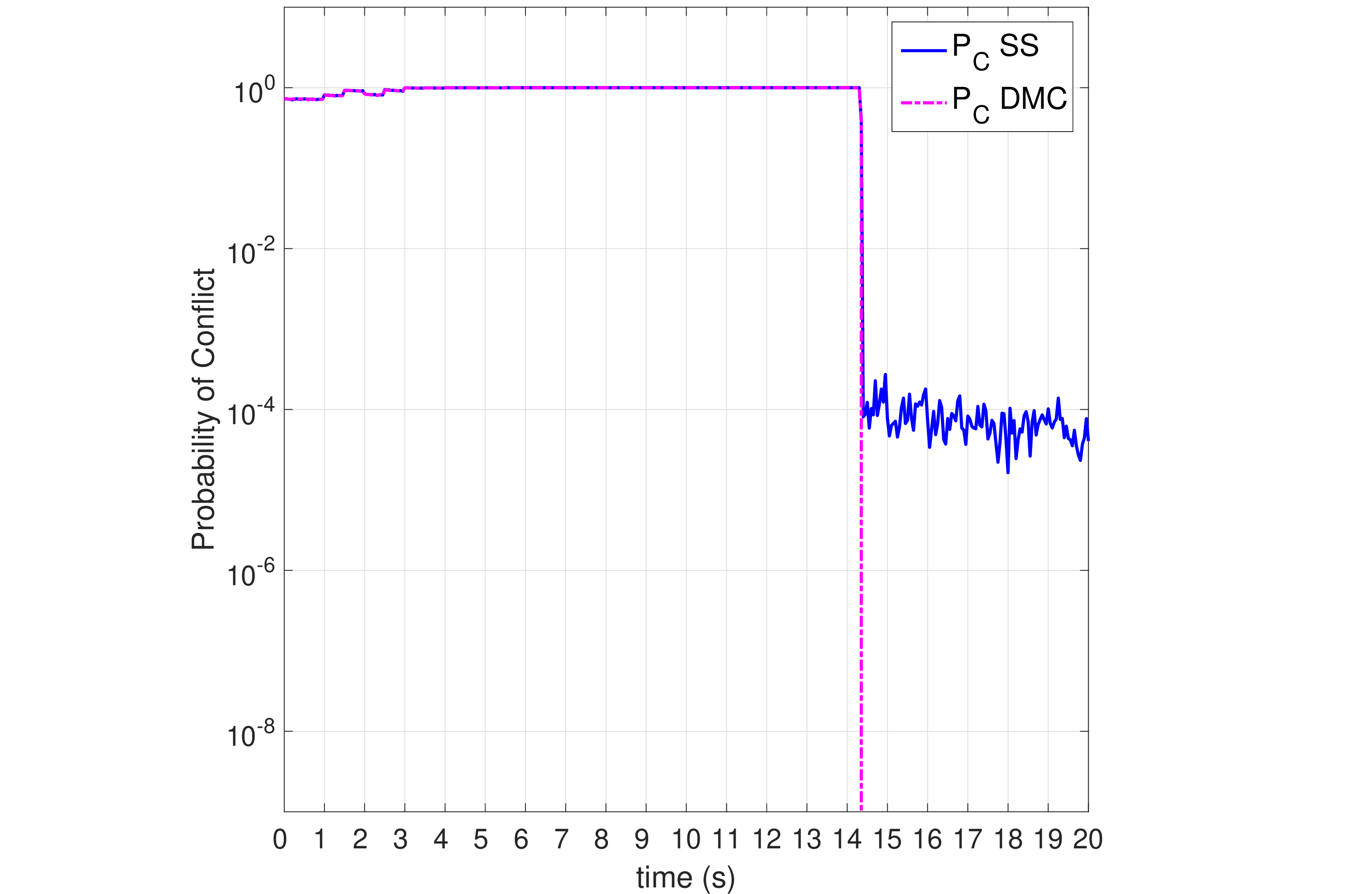}
	\label{fig:intruderOvertaking_100}}
	\hfill
	\subfloat[$P_{c}$ during Intruder overtaking Observer conflict with 152m Lateral separation]{%
	\includegraphics[trim={0 0cm 0 0cm},clip,width=0.66\columnwidth]{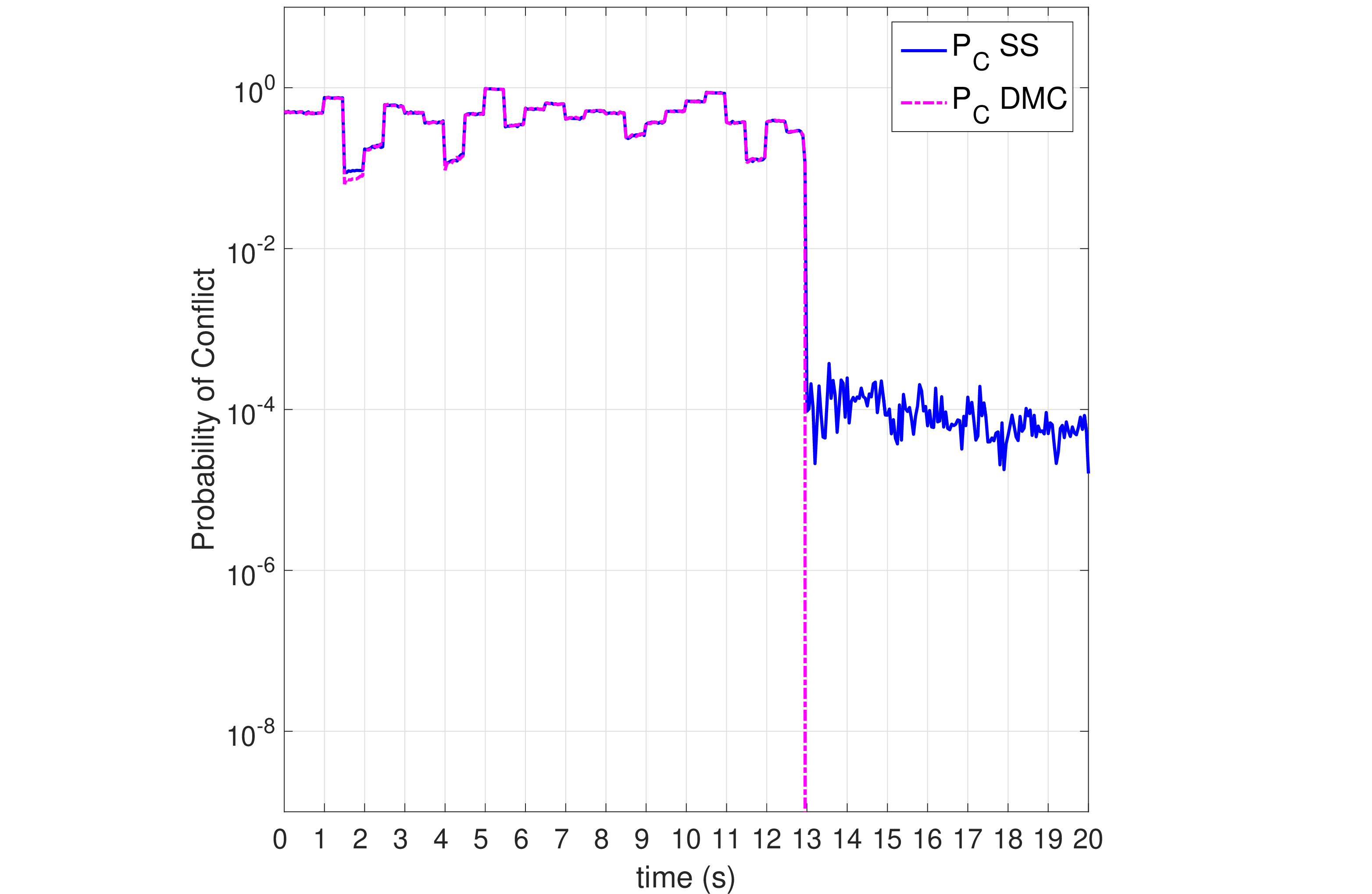}
	\label{fig:intruderOvertaking_152}}
	\hfill	
	\subfloat[$P_{c}$ during Intruder overtaking Observer with 500m Lateral separation]{%
	\includegraphics[trim={0 0cm 0 0cm},clip,width=0.66\columnwidth]{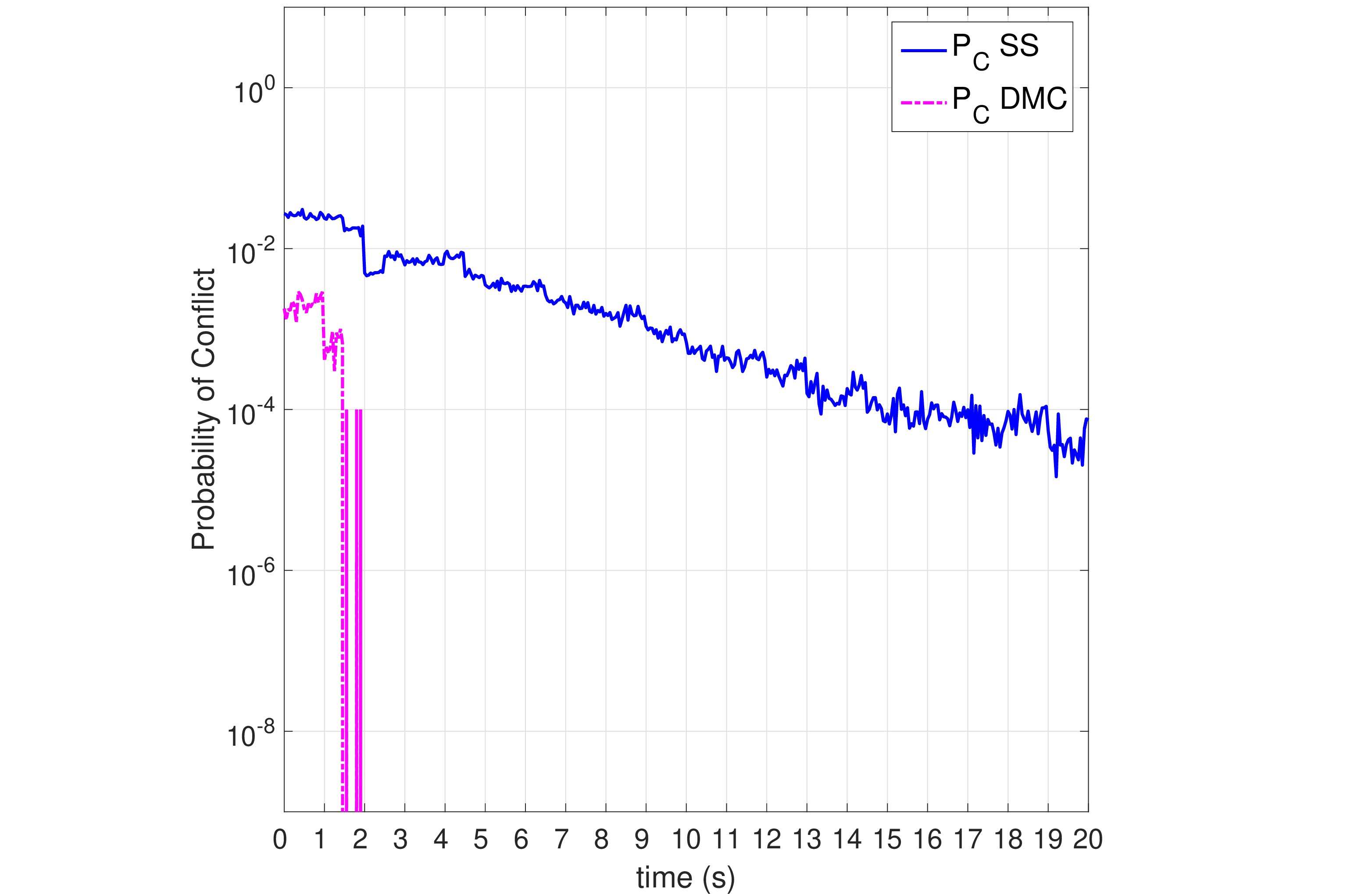}
	\label{fig:intruderOvertaking_500}}		
	\hfill
	\subfloat[$P_{c}$ during Intruder overtaking Observer with 1000m Lateral separation]{%
	\includegraphics[trim={0 0cm 0 0cm},clip,width=0.66\columnwidth]{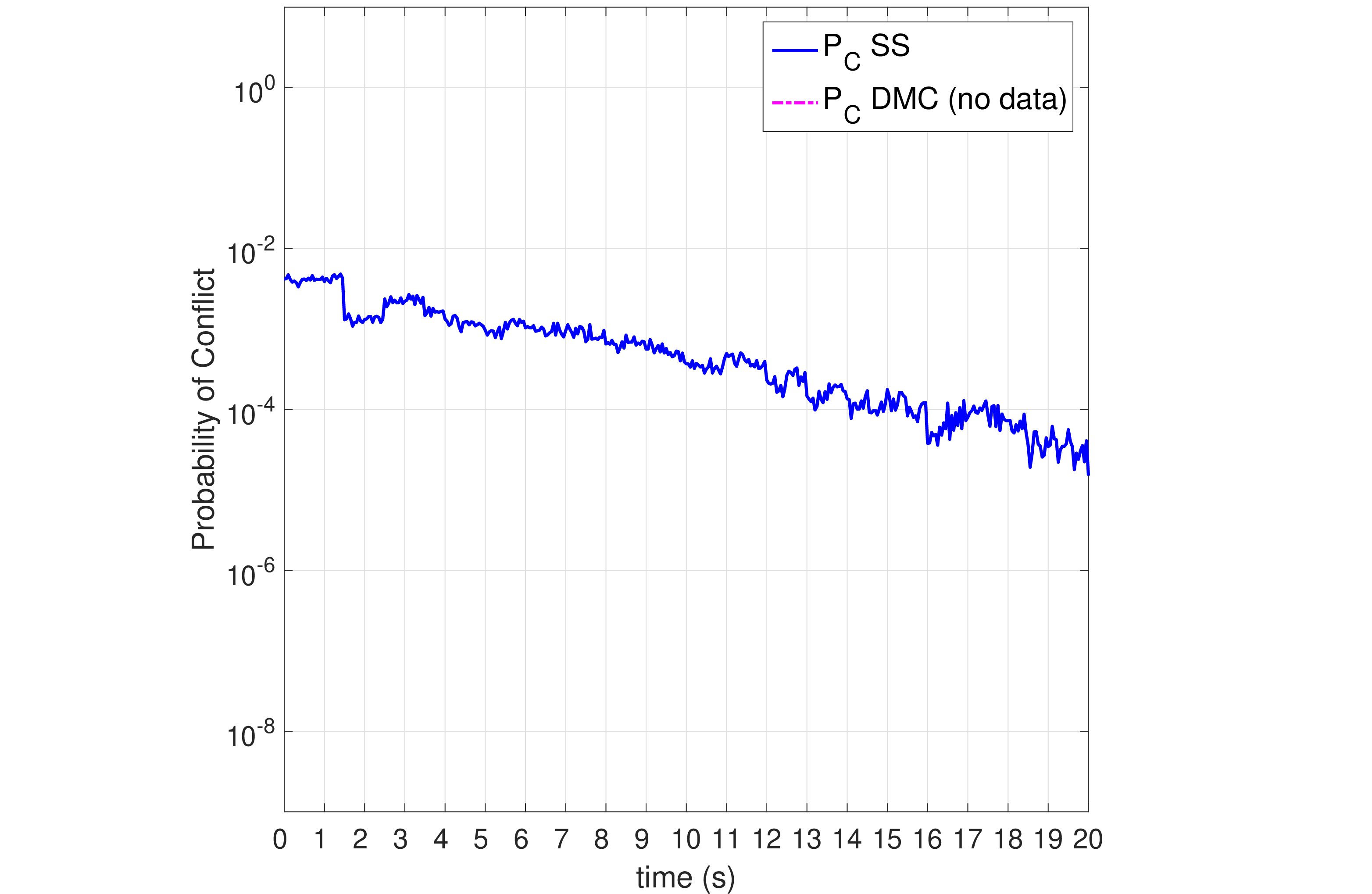}
	\label{fig:intruderOvertaking_1000}}	
	\hfill
	\subfloat[$P_{c}$ during Intruder overtaking Observer with 1100m Lateral separation]{%
	\includegraphics[trim={0 0cm 0 0cm},clip,width=0.66\columnwidth]{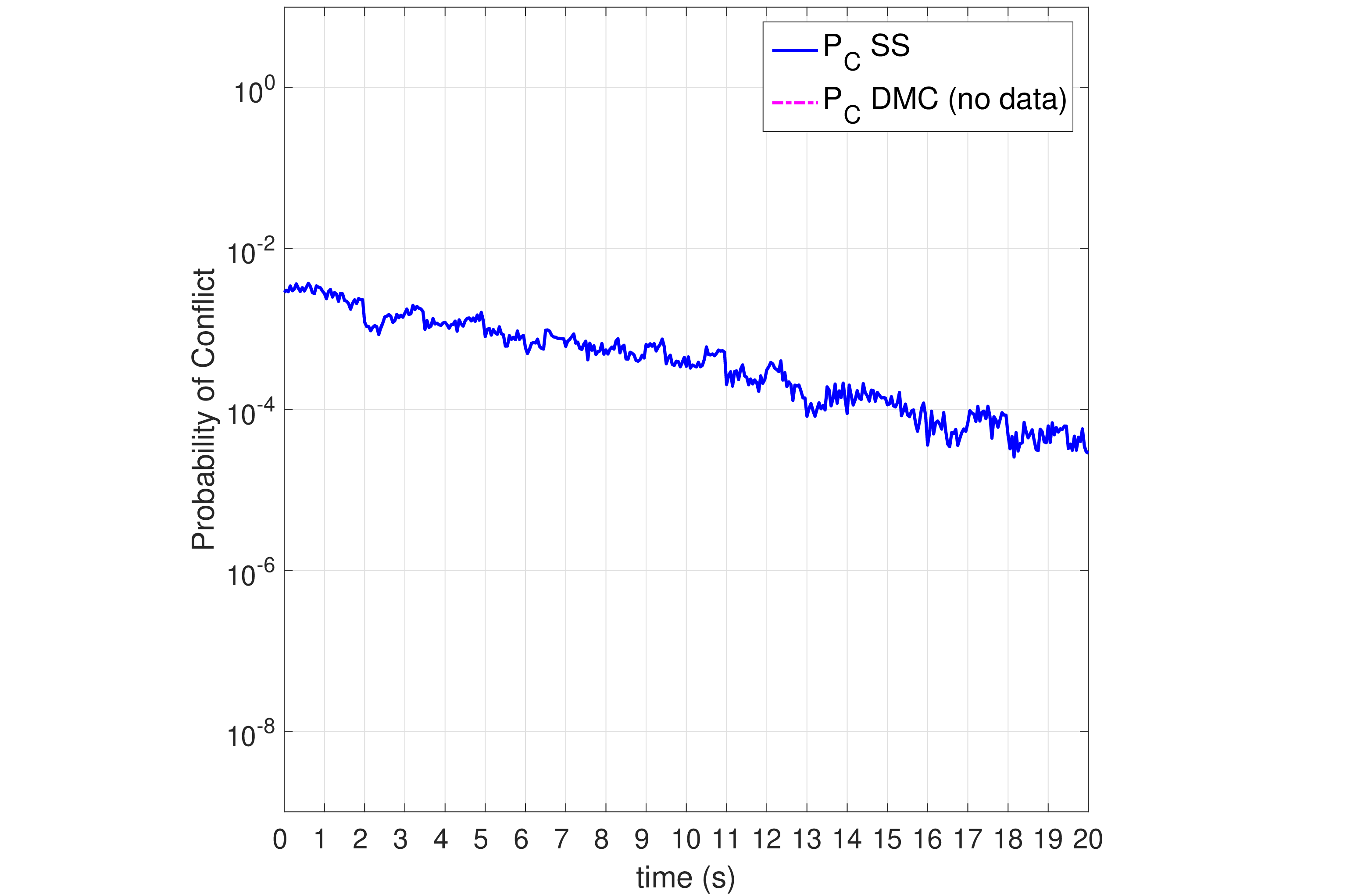}
	\label{fig:intruderOvertaking_1100}}
	\caption{The $P_{c}$ is estimated using the SS and DMC methods during the Intruder Overtaking the Observer scenario as shown in Fig.~\ref{fig:intruderovertaking_potential} with varying lateral separation $L_{a}=\{0,100,152,500,1000,1100\}\text{m}$.}
	\label{fig:intruderOvertaking_set1}	
\end{figure*}

Both SS and DMC methods have been applied to the Overtaking scenario as shown in Fig.~\ref{fig:intruderovertaking_potential}. Similar to the previous scenario, the SS method is able to obtain samples from the rare conflicting region of the pdf consistently throughout the duration of the simulation for this scenario. As the lateral separation increases, the $P_{c}$ decreases (as expected). Figures~\ref{fig:intruderOvertaking_1000} and~\ref{fig:intruderOvertaking_1100} show the $P_{c}$ when the lateral separation is 1000m and 1100m respectively. The change in $P_{c}$ is less abrupt compared to the 100m lateral separation after the Intruder as passed the Observer when $t > 13$s. The $P_{c}$ is approximately the same throughout the duration of the simulation. This is because the increased lateral separation includes samples with low turn rates in the conflict category and these are common enough to be drawn by the DMC method and SS method. With low lateral separation the conflicting samples will need high turn rates. These are rare and are realized by using SS method. In contrast the DMC method is unable to realize them. Also throughout the simulation, the relative change in angle of the Intruder from the Observer's perspective reduces as the lateral separation is increased. The conflicting samples can have lower turn rates despite the Intruder having passed the Observer. Such samples are common and can be realized by both methods.

\section{Accuracy and Efficiency of Subset Simulation}
\label{sec:acc_eff}

A range of magnitudes of probabilities have been evaluated within the simulated scenarios shown in the previous section. This section analyzes the accuracy and efficiency of using the Subset Simulation and Direct Monte Carlo methods to estimate probabilities at each of a number of orders of magnitude. In order for a fair comparison to be conducted -- a common phase within a simulation scenario must be found where both methods are able to realize conflicting samples and estimate the probability of conflict. 

The first order of magnitude considered for comparison is $P_{c_{1}} \approx 10^{-1}$. A suitable phase to conduct the comparison is at $t = 1\text{s}$ during the Head-on scenario with lateral separation $L_{a} = 152.4$ and longitudinal separation $L_{o} = 2000\text{m}$ where a conflict is inevitable. At this phase $p_{0} \leq P_{c_{1}} < 1$ and both methods estimate a similar probability of conflict. This is as expected since the probability is large enough to generate sufficient conflicting samples in the first level of Subset Simulation and it does not progress to higher levels of Subset Simulation. The first level of Subset Simulation is Direct Monte Carlo so the performance is the same.

The second order of magnitude considered is $P_{c_{2}}$. This probability needs to be lower than $P_{c_{1}}$ where $P_{c_{2}} < p_{0}$. Such phases occur frequently in the Head-on pass and Overtaking scenarios, typically when $t > 14 \text{s}$ as shown in figures~\ref{fig:headon_set1} and~\ref{fig:intruderOvertaking_set1} respectively. Note, during such phases the Subset Simulation method is able to obtain conflicting samples and provide a good estimate for $P_{c}$. However, the Direct Monte Carlo method fails to find conflicting samples and is unable to estimate the probability of conflict accurately (other than in a trivial case, $P_{c} = 0$ that is inaccurate). For example the Head-on pass scenarios in Fig.~\ref{fig:headon_set1} shows abrupt changes in $P_{c}$ in some cases from a magnitude of $10^{-1}$ to $10^{-8}$ at approximately $13 \text{s}$ as the Observer and Intruder pass each other. This change in magnitude of probability is very large and abrupt (steep). The magnitude $10^{-8}$ is very rare. For such probabilities the Subset Simulation method is able to obtain conflicting samples and estimate the $P_{c}$ but Direct Monte Carlo method fails to obtain conflicting samples and results in estimating $P_{c} = 0$. The Direct Monte Carlo method requires a large number of samples to estimate probabilities of such magnitude ($10^{-8}$). This might not be practical due to limited simulation resources. Therefore, this order of magnitude of probability is impractical for comparison since although the Subset Simulation method is able to find conflicting samples and estimate the $P_{c}$, the Direct Monte Carlo method is unable to find conflicting samples and fails to estimate the $P_{c}$.

\begin{figure}
\centering
	\includegraphics[width=\columnwidth]{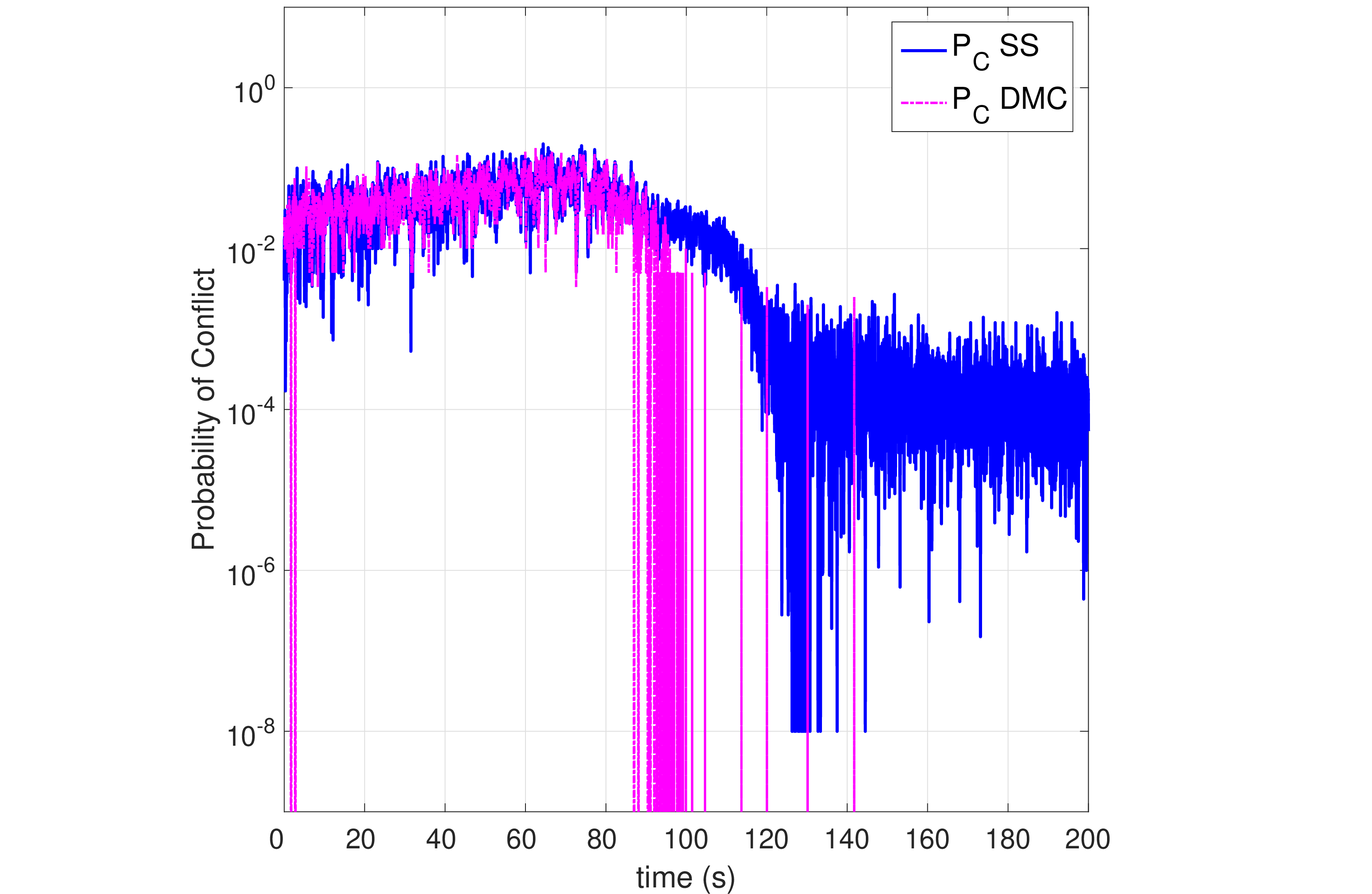}
	\caption{Head-on pass scenario with 1000m lateral separation and 20km longitudinal separation}
	\label{fig:headon_long}
\end{figure}

In order to find a phase where $P_{c_{2}}$ can be evaluated by both methods the simulation of the Head-on pass scenario with lateral separation of 1000m was repeated once with increased longitudinal separation $L_{o} = 20000\text{m}$ for an increased period of $t = 200\text{s}$. This allowed the change in $P_{c}$ to occur less abruptly. Fig.~\ref{fig:headon_long} shows $P_{c}$ estimated by Subset Simulation and Direct Monte Carlo methods during this scenario. Note, during the period $80\text{s} < t < 120\text{s}$, there are frequent abrupt variations in the $P_{c}$ estimated by the Direct Monte Carlo method as zero. These are phases where the method was unable to find a conflicting sample and estimated the probability of conflict as zero. A suitable phase for $P_{c_{2}}$ is at $t = 100\text{s}$ where the probability of conflict estimated by Subset Simulation has reduced to approximately $10^{-2}$; ($P_{c_{2}} \approx 10^{-2}$). This satisfies the $p_{0} > P_{c_{2}}$ criteria. Also, it is the last phase after which the frequency of the Direct Monte Carlo method finding conflicting samples to estimate the $P_{c}$ diminishes. In other words, it is the last phase where both methods are able to generate conflicting samples to estimate the probability of conflict for a comparison to be conducted.

The accuracy and efficiency are compared by calculating the coefficient of variance (c.o.v.) $\delta = \frac{\sigma}{\mu}$ for estimating the probabilities of conflict $P_{c_{1}}$ and $P_{c_{2}}$ using both Subset Simulation and Direct Monte Carlo methods for varying samples sizes $N$. The mean $\mu$ and standard deviation $\sigma$ is calculated over 50 Monte Carlo runs. The sample intervals for Direct Monte Carlo are $N_{\text{dmc}} = \{10^{2}, 10^{3}, 10^{4}, 10^{5}, 10^{6}\}$ and the sample intervals for Subset Simulation are $N_{\text{SS}} = \{100n: n = 1,...,100\}$. Note, that $N_{\text{SS}}$ is the number of samples at each level of Subset Simulation. The total number of levels can vary for each Monte Carlo run of Subset Simulation. This causes a total number of samples to vary for each Monte Carlo run. To allow a fair comparison an average of the total number of samples for each Monte Carlo run of Subset Simulation is used. 



\begin{figure*}
	\centering	
	\subfloat[Coefficient of variance for varying number of samples using SS and DMC methods for estimating $P_{c_{1}}$]{%
	\includegraphics[trim={0 0cm 0 0cm},clip,width=\columnwidth]{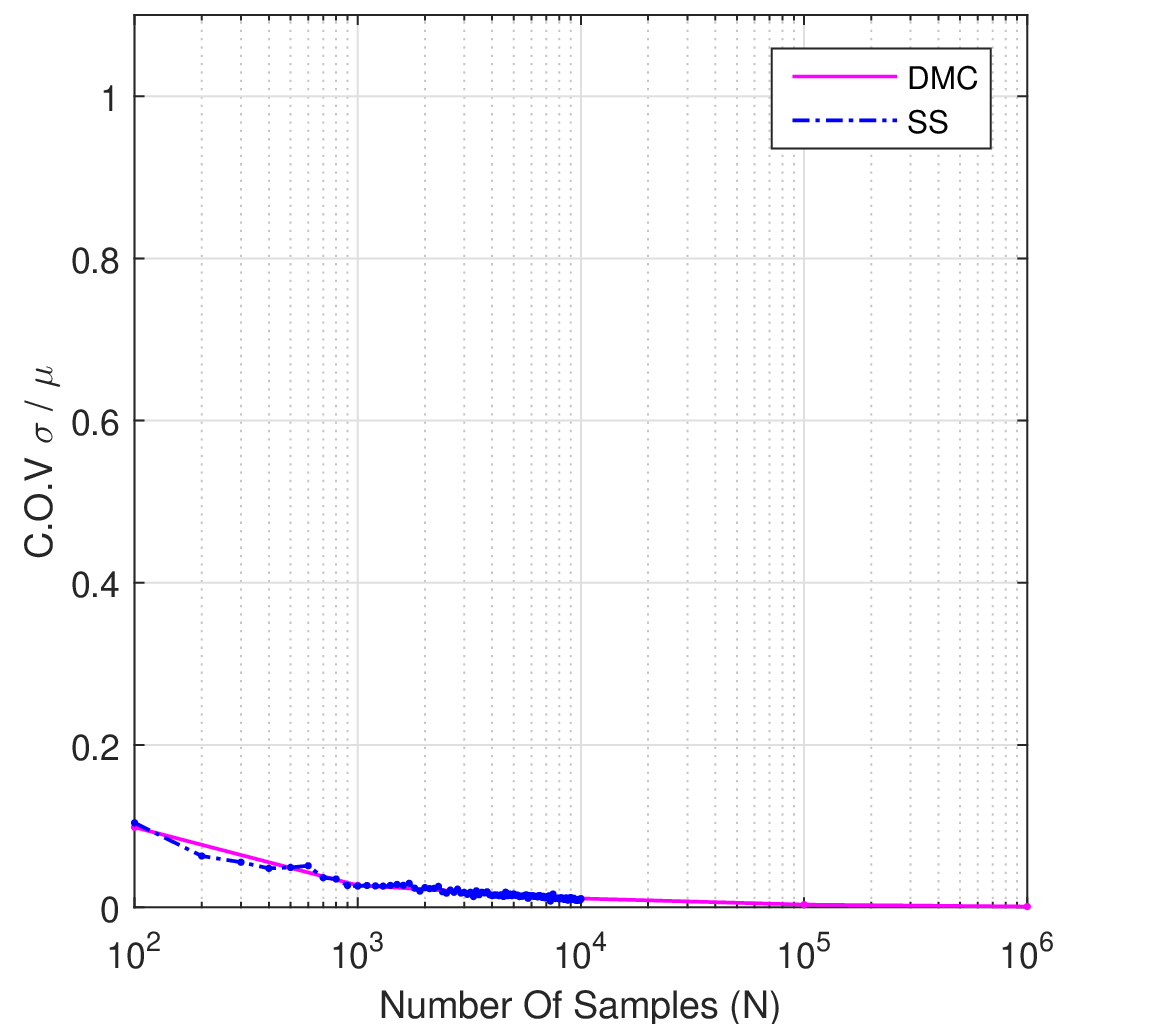}
	\label{fig:SS_vs_DMC_scenario_2}}	
	\hfill
	\subfloat[Coefficient of variance for varying number of samples using SS and DMC methods for estimating $P_{c_{2}}$]{%
	\includegraphics[trim={0 0cm 0 0cm},clip,width=\columnwidth]{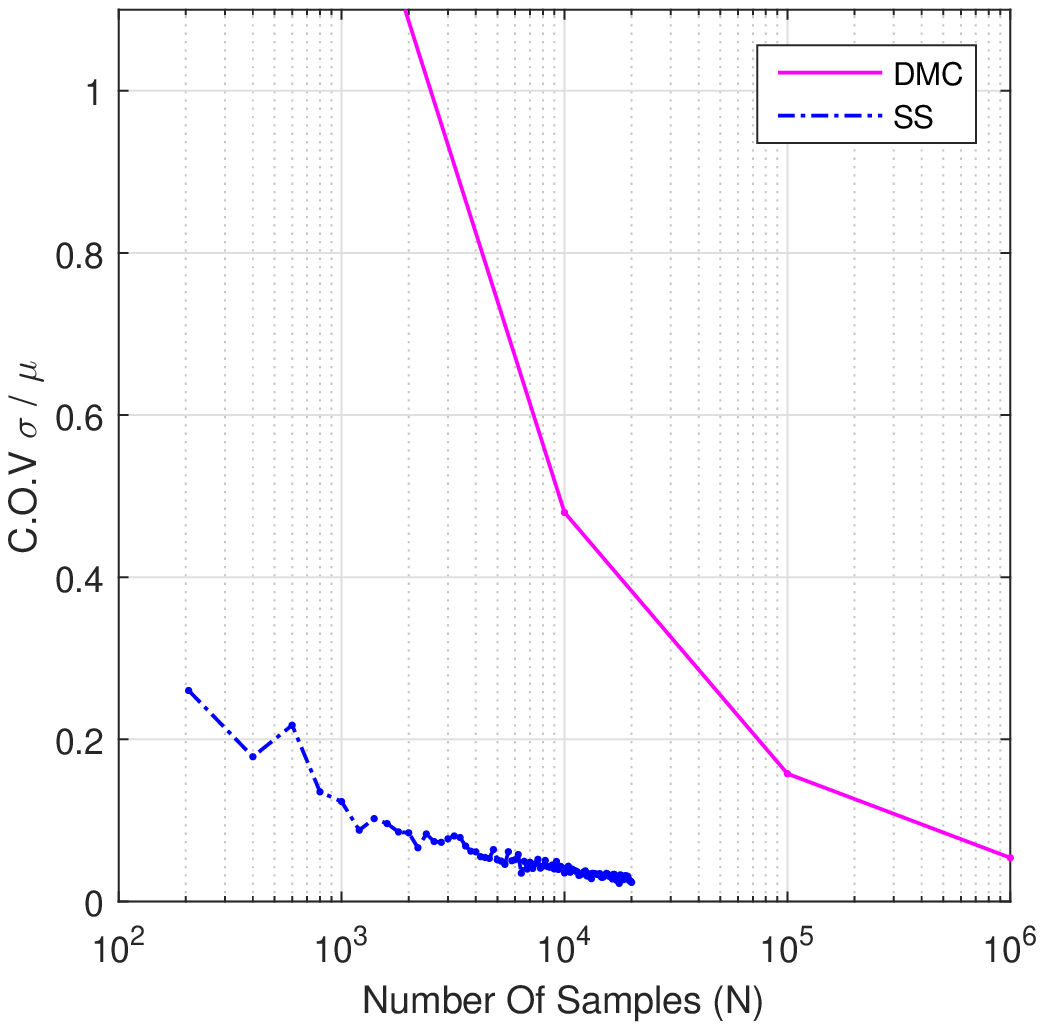}
	\label{fig:SS_vs_DMC_scenario_140}}	
	\hfill
	\caption{A comparison of accuracy and efficiency between DMC and SS methods for estimating the $P_{c}$ during the Head-on pass scenario.}
	\label{fig:headonpass_1100m_snapshot}	
\end{figure*}

The c.o.v. for estimating $P_{c_{1}}$ using Subset Simulation and Direct Monte Carlo methods at varying sample sizes $N$ is shown by Fig.~\ref{fig:SS_vs_DMC_scenario_2}. Note both methods have similar c.o.v. as the average sample size increases. This is expected since the probability is large enough to be realized in level 0 of Subset Simulation that is Direct Monte Carlo. In Fig.~\ref{fig:SS_vs_DMC_scenario_140} the c.o.v. of Subset Simulation for the lower probability of conflict $P_{c_{2}}$ becomes significantly lower than the c.o.v. of DMC as the average number of samples is increased. A point of comparison between both methods can be made where the number of samples $N = 10^{4}$. Note that the c.o.v for Direct Monte Carlo is approximately $0.48$ and the c.o.v for Subset Simulation is approximately $0.04$. Also note that in order for the DMC method to achieve similar c.o.v as the Subset Simulation method it must use $N = 10^{6}$ samples. Therefore the Subset Simulation estimates probabilities of magnitude $10^{-2}$ approximately 10 times more accurately than the Direct Monte Carlo method while using a fraction of the samples (approximately $\frac{1}{100}$) that are required by the Direct Monte Carlo method to achieve similar levels of accuracy. 

\section{Conclusion}
\label{sec:conclusion}

This paper has demonstrated the utility of the Subset Simulation method to estimate the Probability of Conflict ($P_{c}$) between air traffic within a block of airspace during conflicting and potentially conflicting scenarios based on the Rules of the Air defined by the International Civil Aviation Organization. These scenarios can be used to conduct benchmarks for comparing future algorithms. The Subset Simulation method has demonstrated the ability to seek samples from the rare conflict region of interest in an effort to estimate the probability of conflict with lower computational effort than Direct Monte Carlo method. For the equivalent number of samples, the Direct Monte Carlo method fails to consistently obtain samples from the region of interest within the probability distribution function. 

This paper has also demonstrated the ability of Subset Simulation to estimate low probability of conflict (of magnitude $10^{-2}$) approximately 10 times more accurately than the Direct Monte Carlo method while using approximately $\frac{1}{100}$ of the total samples used by the Direct Monte Carlo method to achieve the same level of accuracy as Subset Simulation. This has been demonstrated at a phase during a potentially conflicting scenario based on the Rules of the Air. This example situation has demonstrated that the Subset Simulation method is able to estimate low probabilities more accurately than Direct Monte Carlo method while using less samples than the Direct Monte Carlo method. We conclude that Subset Simulation method is more accurate and efficient than the Direct Monte Carlo method for estimating low probability of conflict between air traffic.  

The Subset Simulation method is scalable to involve multiple Intruders where the $P_{c}$ is estimated for each Intruder. This would be useful for the resolution stage, where Intruders can be prioritized based on the respective $P_{c}$ and an optimized resolution maneuver determined to minimize the new $P_{c}$ after the resolution maneuver. A more efficient method of estimating the $P_{c}$ would be to modify the SS method further to use Sequential Monte Carlo Samplers instead of Markov Chain Monte Carlo~\cite{del2006sequential}. This will allow the implementation to be parallelized in the seed selection stage and will give rise to improved statistical efficiency. We plan to investigate such improvements in future work.


\section*{Acknowledgments}
The authors would like to thank Matteo Fasiolo, Fl\'avio De Melo, Elias Griffith and James Wright for their contributions. This work was supported by the Engineering and Physical Sciences Research Council (EPSRC) Doctoral Training Grant.

\bibliography{conflict_detection_ref}{}
\bibliographystyle{ieeetr}

\end{document}